\documentclass[11pt]{article}
\usepackage[mathscr]{eucal}
\usepackage{epsfig,amsfonts}
\usepackage{amsmath}
\usepackage{amsthm,amssymb}
\usepackage{graphicx}
\usepackage{hhline}
\usepackage{cite}
\usepackage{psfrag}
\usepackage{mathrsfs} 
\usepackage{hyperref}
\usepackage{bm}
\usepackage{array}
\usepackage{tikz}
\usepackage{multirow}
\usepackage{textcomp}
\usepackage{hyperref}
\usepackage{mathrsfs} 
\usepackage{enumitem}
\DeclareMathOperator*{\slim}{slim}
\DeclareMathAlphabet{\mathpzc}{OT1}{pzc}{m}{it}

\makeatletter
\@addtoreset{equation}{section}
\makeatother

\def\bea{\begin{eqnarray}}
\def\eea{\end{eqnarray}}
\def\be{\begin{equation}}
\def\ee{\end{equation}}

\def\be{\begin{equation}}
\def\ee{\end{equation}}
\def\bdm{\begin{displaymath}}
\def\edm{\end{displaymath}}
\def\bea{\begin{eqnarray}}
\def\eea{\end{eqnarray}}

\def\sgn{{\rm sgn}}

\def\ri{{\rm i}}
\def\half{\textstyle\frac{1}{2}}
\def\XXint#1#2#3{{\setbox0=\hbox{$#1{#2#3}{\int}$}
    \vcenter{\hbox{$#2#3$}}\kern-.5\wd0}}

\newcommand{\rd}{\mbox{d}}
\newcommand{\re}{\mbox{e}}

\DeclareMathAlphabet{\mathpzc}{OT1}{pzc}{m}{it}

\begin{document}

%%%%%%%%%%%%%%%%%%%%%%%%%%%%

\begin{titlepage}
$\phantom{I}$
\vspace{2.8cm}

\begin{center}
\begin{LARGE}

{\bf  On the scaling behaviour of an integrable spin chain
with ${\cal Z}_r$ symmetry}

\end{LARGE}

\vspace{1.3cm}
\begin{large}

{\bf 

Gleb A. Kotousov$^{1}$ and Sergei  L. Lukyanov$^{2}$}

\end{large}

\vspace{1.cm}
${}^{1}$Institut f$\ddot{{\rm u}}$r Theoretische Physik, 
Leibniz Universit$\ddot{{\rm a}}$t Hannover\\
Appelstra\ss e 2, 30167 Hannover, Germany\\
\vspace{.4cm}
${}^{2}$NHETC, Department of Physics and Astronomy\\
     Rutgers University,
     Piscataway, NJ 08855-0849, USA\\
\vspace{1.0cm}

\end{center}

\begin{center}

\parbox{13cm}{%
\centerline{\bf Abstract} \vspace{.8cm}
The subject matter of this work is a  1D quantum 
spin\,-\,$\frac{1}{2}$ chain associated with the inhomogeneous 
six-vertex model possessing an additional  ${\cal Z}_r$ symmetry.
The model is studied in a certain parametric domain, where it is
critical. Within the ODE/IQFT approach, 
 a class of ordinary differential equations
and a quantization condition are proposed
which describe the scaling limit of the system.
Some remarkable features of the CFT underlying the critical behaviour are observed. 
Among them is an infinite degeneracy of the conformal
primary states and the presence of a continuous component in the spectrum
in the case of even $r$.}
\end{center}

\vfill

\end{titlepage}

\setcounter{page}{2}

%\tableofcontents
%\newpage

\section{Introduction}

Much of the early interest in exactly solvable models in 2D classical 
statistical mechanics  came from the study of 
critical phenomena. Starting from the Ising model, exact solutions
were crucial for the development of
key concepts such as the scaling hypothesis and universality. 
The latter was greatly informed by the six-vertex model \cite{Baxter:1982zz}.
Within the standard parameterization, its local Boltzmann weights depend on
$q$, which is sometimes referred to as the anisotropy parameter. 
As long as $q$ is a unimodular number, i.e., $q=\re^{\ri\gamma}$,
the statistical system is critical and it turns out that the corresponding critical exponents 
depend on $\gamma\in(0,\pi]$.  This  was in contradiction with the original
understanding of universality. The clarification came within the framework of
 QFT as an example of an exactly marginal
deformation of the critical point. It was realized that the
 universal behaviour of the six-vertex model
is governed by the massless Gaussian CFT, where the fundamental Bose field
is compactified to a circle with radius $\sqrt{2\gamma/\pi}$ \cite{Luther,Kadanoff,Barber}.
\smallskip

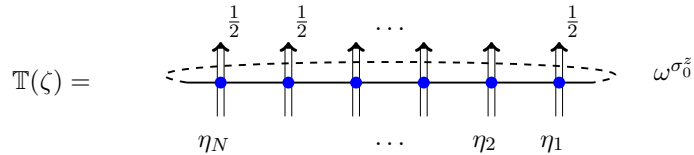
\begin{figure}
\begin{center}
\scalebox{0.9}{
\begin{tikzpicture}
\node at (0,0) {$\mathbb{T}(\zeta)=$};
\node at (2.4,-0.9) {$\eta_N$};
\node at (6.4,-0.9) {$\eta_2$};
\node at (7.4,-0.9) {$\eta_1$};
\node at (5,-0.9) {$\ldots $};
\draw[thick] (2,0) -- (8,0);
\draw (2.55,-0.5) -- (2.55,0.5);
\draw (2.45,-0.5) -- (2.45,0.5);
\draw (3.55,-0.5) -- (3.55,0.5);
\draw (3.45,-0.5) -- (3.45,0.5);
\draw (4.55,-0.5) -- (4.55,0.5);
\draw (4.45,-0.5) -- (4.45,0.5);
\draw (5.55,-0.5) -- (5.55,0.5);
\draw (5.45,-0.5) -- (5.45,0.5);
\draw (6.55,-0.5) -- (6.55,0.5);
\draw (6.45,-0.5) -- (6.45,0.5);
\draw (7.45,-0.5) -- (7.45,0.5);
\draw (7.55,-0.5) -- (7.55,0.5);
\draw [ultra thick,->] (2.5,0.5) -> (2.5,0.6);
\draw [ultra thick,->] (3.5,0.5) -> (3.5,0.6);
\draw [ultra thick,->] (4.5,0.5) -> (4.5,0.6);
\draw [ultra thick,->] (5.5,0.5) -> (5.5,0.6);
\draw [ultra thick,->] (6.5,0.5) -> (6.5,0.6);
\draw [ultra thick,->] (7.5,0.5) -> (7.5,0.6);
\draw[dashed,thick] (2,0)  to[out = 170, in = 10] (8,0);
\fill[blue] (2.5,0) circle (2.5pt);
\fill[blue] (3.5,0) circle (2.5pt);
\fill[blue] (4.5,0) circle (2.5pt);
\fill[blue] (5.5,0) circle (2.5pt);
\fill[blue] (6.5,0) circle (2.5pt);
\fill[blue] (7.5,0) circle (2.5pt);
\node at (2.7,0.9) {$\frac{1}{2}$};
\node at (3.7,0.9) {$\frac{1}{2}$};
\node at (7.7,0.9) {$\frac{1}{2}$};
\node at (5,0.8) {$\hdots$};
\node at (9.2,0.2) {$\omega^{\sigma^z_0}$};
%\node at (10.8,0) {$\phantom{asd}$};
\end{tikzpicture}
}
\end{center}
\caption{\label{fig1}\small
A graphical representation of the transfer-matrix for the inhomogeneous six-vertex model.
Twisted boundary conditions are imposed as indicated by the presence of the factor
 $\omega^{\sigma^z_0}$, where $\sigma^z_0$ is the Pauli matrix acting in the two dimensional auxiliary space.
}
\end{figure}

In the work \cite{Baxter:1971}, Baxter introduced a multiparametric statistical system,
which is a generalization of the  six-vertex model,
and showed that it is solvable via the Bethe Ansatz technique. The corresponding set
of algebraic equations takes the form
\be\label{bae}
\prod_{J=1}^{N}
\frac{\eta_J+q^{+1} \,\,\zeta_m}
{\eta_J+q^{-1}\,\zeta_m }
=-\omega^2\,q^{2S^z}\,
\prod_{j=1}^M\,
\frac{\zeta_j-q^{+2}\,\zeta_m }
{\zeta_j-q^{-2}\,\zeta_m }
\,\qquad\qquad (\,m=1,2,\ldots,M,\ \ S^z=\tfrac{N}{2}-M\ge 0\,)\, .
\ee
Having found a solution $\{\zeta_m\}_{m=1}^M$ one can compute the eigenvalues 
of all members of the commuting family of operators including the one-row transfer-matrix
schematically depicted in fig.\,\ref{fig1}. 
\smallskip

Among the parameters is the anisotropy  $q=\re^{\ri\gamma}$
 and the set of complex numbers $\{\eta_J\}_{J=1}^N$, referred to as the inhomogeneities, 
which are all assumed to be fixed. 
 Together with these, the Bethe Ansatz equations
 also involve the parameter 
 $\omega$. The latter comes about as a result of imposing so-called quasi-periodic boundary conditions
on the lattice and for the periodic case $\omega=1$. With the parameters obeying certain conditions, the
lattice system develops critical behaviour which can be described by a conformal field theory. 
 We'll assume that the number
of sites $N$ is divisible by $r$ and 
 the inhomogeneities  obey the $r$-site periodicity condition
\be\label{mn21bszz}
\eta_{J+r}=\eta_J\qquad\qquad\qquad (\,J=1,2,\ldots, N-r\,;\ N/r\in\mathbb{Z}\,)
\ee
in order to ensure the presence of translational invariance for the continuous theory.
In the scaling limit $N\to\infty$ while  $r$ is kept fixed.
\smallskip

For the investigation of the scaling limit, it is useful to switch to the Hamiltonian picture,
where a central r\^{o}le belongs to the  Hamiltonian $\mathbb{H}$. The latter is a member of the commuting family of operators and, 
furthermore,
is expressible as a logarithmic derivative of the $r$-row transfer-matrix
(see, e.g., section 6.4 of ref.\cite{Bazhanov:2020new} for details). In the case of the homogeneous model,
where $\eta_J$ are the same and without loss of generality may be set to one, 
$\mathbb{H}$ coincides with
the Hamiltonian for the Heisenberg XXZ spin\,-\,$\frac{1}{2}$ chain:
\bea\label{asiisaias}
\mathbb{H}\,|_{r=1}=-\frac{1}{2\sin(\gamma)}\,
\sum_{m=1}^N\Big( \sigma_m^x\sigma_{m+1}^x+\sigma_m^y\sigma_{m+1}^y+\cos(\gamma)
\, \big(\sigma_m^z\sigma_{m+1}^z-\hat{\bm{1}}\big)\Big)\ .
\eea
It should be supplemented with the quasi-periodic boundary conditions \cite{Barber}
\be\label{BC1b}
\sigma^x_{N+m}\pm\ri\,\sigma^y_{N+m}=\omega^{\pm 2}\,\big(\,\sigma^x_{m}\pm\ri\,\sigma^y_{m}\,\big)\,,
\qquad \qquad \sigma_{N+m}^z=\sigma^z_m \,.
\ee
For $r=2$ there are two inhomogeneities $\eta_1$, $\eta_2$, but one may always restrict to the case with
$\eta_2=\eta_1^{-1}$. In the parameterization $\eta_1=\re^{\ri\alpha}$, the Hamiltonian
reads as \cite{Ikhlef:2008zz,Frahm:2013cma}\footnote{% 
The Hamiltonian presented in \cite{Ikhlef:2008zz} corresponds to the case $\alpha=\frac{\pi}{2}$,
while the one from ref.\cite{Frahm:2013cma} is related to \eqref{asjh123}  by a similarity transformation.}
\bea\label{asjh123}
\mathbb{H}\,|_{r=2}&=&\frac{\cos(\gamma)}{2\sin(\gamma-\alpha)\,\sin(\gamma+\alpha)}\,
\sum_{m=1}^N\,\bigg[\,\frac{\sin^2(\alpha)}{\sin(\gamma)}\,\big(\,
\sigma^x_m\,\sigma^x_{m+2}+\sigma^y_m\,\sigma^y_{m+2}+
\sigma^z_m\,\sigma^z_{m+2}-\hat{\bf 1}\,\big)\nonumber\\[0.2cm]
&-&2\sin(\gamma)\,\bigg(\frac{\cos^2(\alpha)}{\cos(\gamma)}\,(
\sigma^x_m\,\sigma^x_{m+1}+\sigma^y_m\,\sigma^y_{m+1})+ \sigma^z_m\,\sigma^z_{m+1}-\hat{\bf 1} \bigg)-
\ri\sin^2(\alpha)\,(\sigma_{m+2}^z-\sigma_{m-1}^z)
\nonumber\\[0.2cm]
&\times&
(\sigma_m^x\sigma_{m+1}^x+\sigma_m^y\sigma_{m+1}^y)
+(-1)^m\sin(2\alpha)\ \Big(\tan(\gamma)\, (\sigma^x_m\sigma^y_{m+1}-\sigma^y_m\sigma^x_{m+1})
\\[0.2cm]&+&
\frac{\ri}{2\cos(\gamma)}\,
(\sigma_m^x\,\sigma^y_{m+2}-\sigma_m^y\sigma_{m+2}^x)\,\sigma_{m+1}^z-
\frac{\ri}{2}\,(\sigma_m^x\,\sigma^y_{m+1}-\sigma_m^y\sigma_{m+1}^x)\,
(\sigma_{m+2}^z-\sigma^z_{m-1})\Big)\bigg]
\nonumber
\eea
and is subject to the quasi-periodic boundary conditions as above.
When $r>2$, the explicit formula for $\mathbb{H}$ becomes cumbersome and not particularly transparent. 
It involves a sum over terms describing interactions of up to $r+1$ adjacent  spins.
All such Hamiltonians commute with the $z$-projection of the total spin
operator
\be\label{mnbn211221}
\mathbb{S}^z=\frac{1}{2}\sum_{m=1}^N \sigma_m^z\, : \ \ \ \ \ \ \ \ \ \ \ [\,\mathbb{S}^z,\mathbb{H}\,]=0\,.
\ee

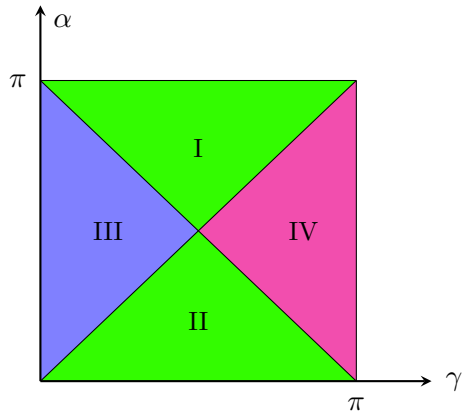
\begin{figure}
\begin{center}
\scalebox{1}{
\begin{tikzpicture}
\path[fill =green, opacity = 1] (0,0) -- (4.2,0) -- (2.1,2) -- cycle;
\path[fill =yellow, opacity = 0.2] (0,0) -- (4.2,0) -- (2.1,2) -- cycle;
\path[fill = green, opacity = 1] (0,4) -- (4.2,4) -- (2.1,2) -- cycle;
\path[fill = yellow, opacity = 0.2] (0,4) -- (4.2,4) -- (2.1,2) -- cycle;
\path[fill = magenta, opacity = 0.7] (4.2,0) -- (2.1,2) -- (4.2,4) -- cycle;
\path[fill = blue, opacity = 0.5] (0,0) -- (2.1,2) -- (0,4) -- cycle;
\draw[-stealth,thick] (0,0) -> (5.2,0); 
\draw[-stealth,thick] (0,0) -> (0,5); 
\draw (4.2,0) -- (4.2,4);
\draw (0,4) -- (4.2,4);
\draw (0,0) -- (4.2,4);
\draw (0,4) -- (4.2,0);
\node at (-0.3,4) {$\pi$};
\node at (4.2,-0.3) {$\pi$};
\node at (3.5,2) {\small IV};
\node at (2.1,0.8) {\small II};
\node at (2.1,3.1) {\small I};
\node at (0.9,2) {\small III};
\node at (5.5,0) {$\gamma$};
\node at (0.3,4.8) {$\alpha$};
\end{tikzpicture}
}
\end{center}
\caption{\small
The phase diagram for the model \eqref{asjh123}
with $\gamma,\alpha$ being real
numbers lying within the segment $(0,\pi]$. Phases I and II correspond to the massless compact Bose field, while
III is related to the 2D black hole sigma models \cite{Jacobsen:2005xz,Ikhlef:2008zz,Ikhlef:2011ay,Frahm:2012eb,Frahm:2013cma,
Candu:2013fva,Bazhanov:2019xvy,Bazhanov:2019xvyA}. 
In the parametric domain IV the critical behaviour of the
statistical system is governed by a CFT consisting of a massless compact boson and two Majorana fermions 
\cite{IJS2,Kotousov:2021vih}.\label{fig2}}
\end{figure}

The field theory provides a description of the excitations whose energy counted from that of the ground state is sufficiently low. 
The full phase diagram for the general lattice system, including the identification
of the critical surfaces, is currently not clear. The examples of critical models that have been studied so far,
since the pioneering paper of Jacobsen and Saleur \cite{Jacobsen:2005xz},
all correspond
to the case when the anisotropy is unimodular or, in other words, $\gamma$ is real and may be taken
from the domain
\be
q=\re^{\ri\gamma}\,:\qquad\qquad \gamma\in(0,\pi]\,.
\ee
Likewise, the inhomogeneities are also unimodular, $|\eta_J|=1$,  and, without loss of generality, one may always assume that
\be
 \prod_{J=1}^r\eta_J=1\ .
\ee
As was already mentioned, for $r=1$ the critical behaviour of the model, with quasi-periodic boundary conditions imposed,
is described by a massless compact Bose field. Already for
$r=2$ the model exhibits different types of critical behaviour and
 the phase diagram in the $(\gamma,\alpha)$ plane (the parameters of the lattice Hamiltonian \eqref{asjh123}) is depicted in fig.\,\ref{fig2}.
\smallskip

The different phases are already manifest  in the general pattern of  Bethe roots for the ground state.
For the homogeneous model, all the $\{\zeta_m\}$ are real and positive. In fact, this requires
the condition on the twist parameter:
\be\label{m1nbn3212}
\omega=\re^{\ri\pi{\tt k}}\,,\qquad\qquad {\tt k}\in(-\tfrac{1}{2},\tfrac{1}{2}]
\ee
with sufficiently small $|{\tt k}|$. In fig.\,\ref{fig3a} depicted are the typical pattern of Bethe roots for the
case $r=2$ in phases III and IV in the complex plane of $\beta\equiv-\frac{1}{2}\log(\zeta)$.
An important difference is that for phase III the roots accumulate exactly on the lines
$\Im m(\beta)=0,\frac{\pi}{2}$ independently of the value of the inhomogeneity parameter $\alpha$ (see, e.g.,
ref.\cite{Frahm:2013cma}).
For phase IV, they also accumulate along two lines. However these depend on $\alpha$ 
as $\Im m(\beta)=\pm\frac{1}{2}\,\alpha$  (see, e.g.,
ref.\cite{Kotousov:2021vih}).

\smallskip

The phase diagram for $r>2$ is yet to be mapped out in general. 
 In this work we will study the universal behaviour of the model in  the domain $0<\gamma<\frac{\pi}{r}$.
It turns out to exhibit qualitatively new features compared to the cases $r=1,2$. In particular, the pattern
of Bethe roots for the ground state do not accumulate on lines with fixed value of $\Im m(\beta)$,
 but  on a certain locus in the
complex $\beta$ plane that depends on the values of the inhomogeneities (for an illustration, see fig.\,\ref{fig3}). 
The exception is
\be\label{zsym1}
\eta_{J}=(-1)^r\,\re^{\frac{\ri\pi}{r} (2J-1)}\,,
\ee
where the lattice model possesses an extra ${\cal Z}_r$ invariance. Among other things, the symmetry manifests itself
in that the Bethe roots accumulate along the lines 
\be\label{asjjh1hg}
\Im m(\beta_j)=0\,,\ \frac{\pi}{r}\,,\ \frac{2\pi}{r}\,,\ldots,\ \frac{(r-1)\pi}{r}\qquad\qquad ({\rm mod}\ \pi)\, .
\ee

The most effective technique for studying the critical behaviour of an integrable lattice system is based
on the ODE/IQFT correspondence  \cite{Voros:1994,Dorey:1998pt,Bazhanov:1998wj,Bazhanov:2003ni}. This is demonstrated for the XXZ spin\,-\,$\frac{1}{2}$ chain  in    ref. \cite{Kotousov:2019ygw}  and
for the case $r=2$ in  phase III from fig.\,\ref{fig2} in the work \cite{Bazhanov:2019xvyA}. 
In ref.\cite{Kotousov:2021vih}, the ODE/IQFT correspondence was put forward
for the inhomogeneous six-vertex model with  $(1-\frac{1}{r})\pi<\gamma<\pi$.
\smallskip

In this paper we propose a set of differential equations that describe the scaling limit
of the ${\cal Z}_r$  invariant spin chain  subject to quasi-periodic boundary conditions
\eqref{BC1b},\,\eqref{m1nbn3212}, 
when the inhomogeneities
take the value \eqref{zsym1} and
\be\label{87hgvbasbv}
q=\re^{\ri\gamma}\ :\ \ \ \ \ 0<\gamma<\frac{\pi}{r}\ .
\ee
The  analysis turns out to be more complicated than for the ${\cal Z}_2$ case and a
detailed study of the critical behaviour is left for future work.

\begin{figure}
\centering
\scalebox{0.9}{
\begin{tikzpicture}
\node at (8,0.9) {\small $\beta$};
\draw (8,0.93) circle (0.25cm);
\node at (-1.7,0.9) {\small $\beta$};
\draw (-1.7,0.93) circle (0.25cm);
\node at (-4.8,0) {\includegraphics[width=7.5cm]{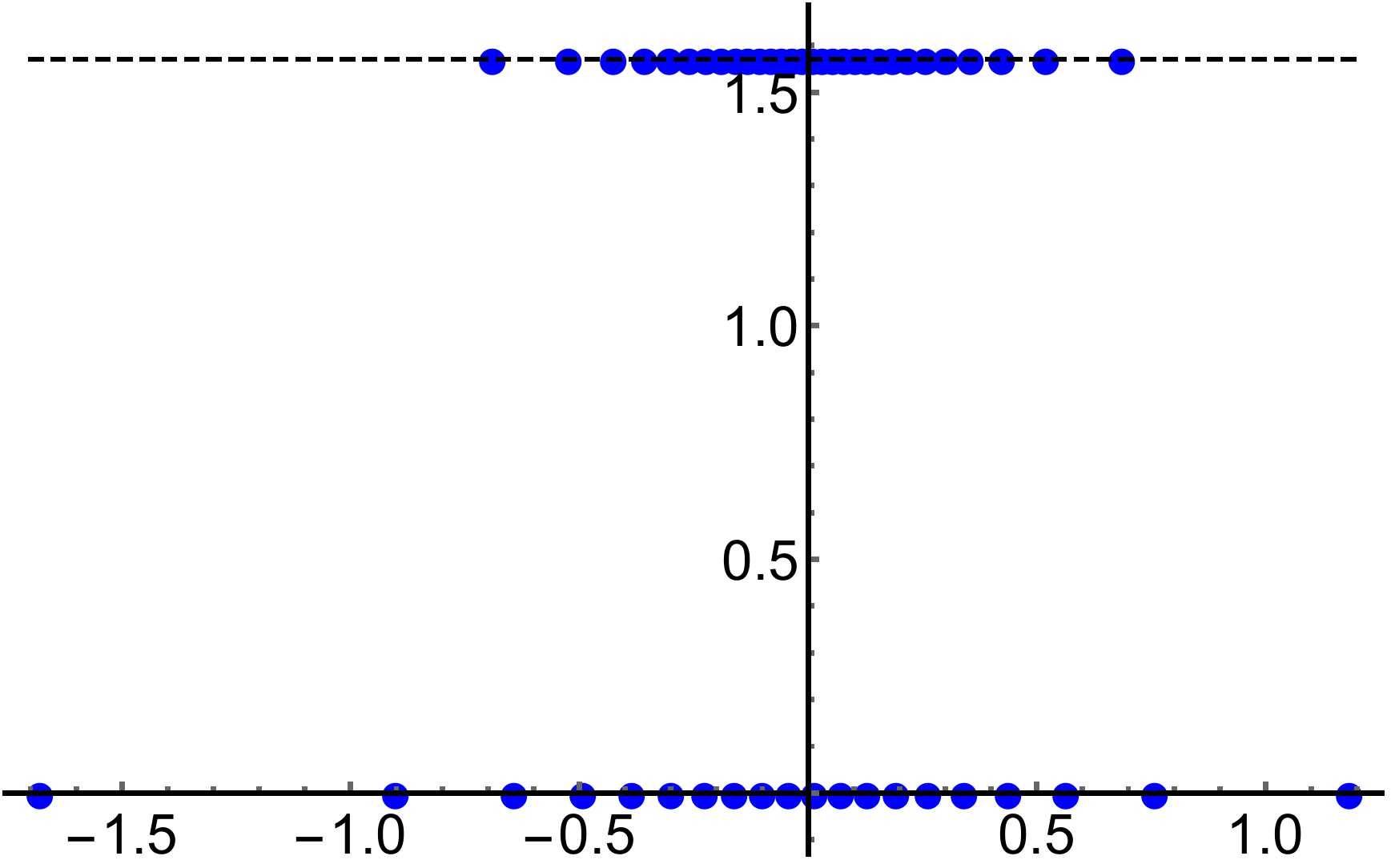}};
\node at (4.8,0) {\includegraphics[width=7.8cm]{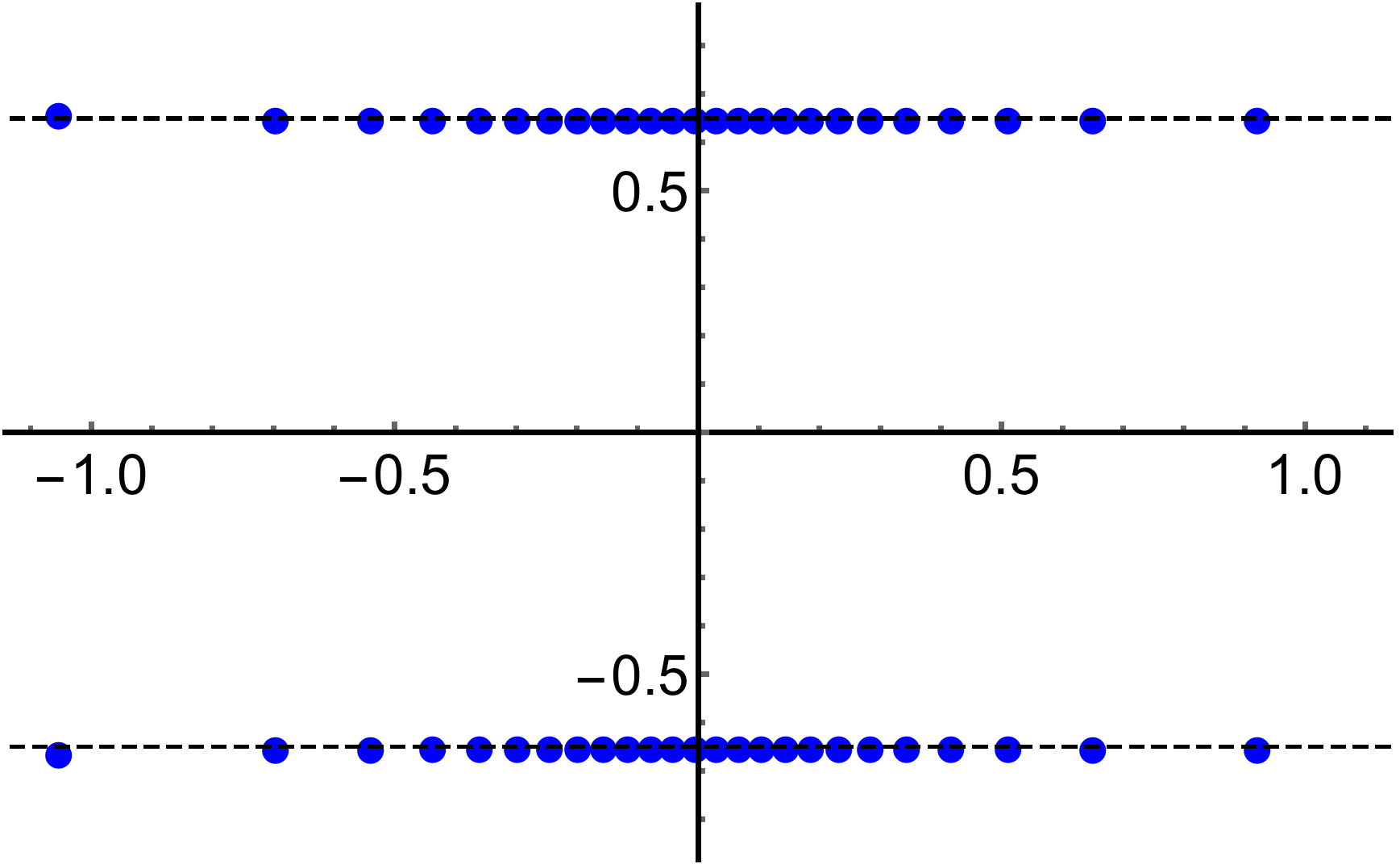}};
\node at (2.1,2.15) {$\Im m(\beta)=+\frac{1}{2}\,\alpha$};
\node at (2.1,-1.35) {$\Im m(\beta)=-\frac{1}{2}\,\alpha$};
\node at (-7.4,2.4) {$\Im m(\beta)=\frac{\pi}{2}$};
\end{tikzpicture}
}
\bigskip

\caption{\small
The typical pattern of Bethe roots for the ground state ($S^z=0$) of the Hamiltonian
\eqref{asjh123} are shown in the complex $\beta$ plane with $\beta=-\frac{1}{2}\log(\zeta)$.
For the left panel, the parameters were taken to be
 $(\gamma,\alpha)=(\frac{43\pi}{200},\frac{13}{10})$, so that the model is in the critical phase III from fig.\,\ref{fig2}. 
For the right panel $(\gamma,\alpha)=(\frac{143\pi}{200},\frac{13}{10})$, which corresponds to phase IV.
The number of sites $N=100$ and the twist parameter ${\tt k}$, which enters into the boundary conditions \eqref{BC1b} via
$\omega=\re^{\ri\pi{\tt k}}$ was set to ${\tt k}=\frac{1}{10}$.
\label{fig3a}}
\end{figure}
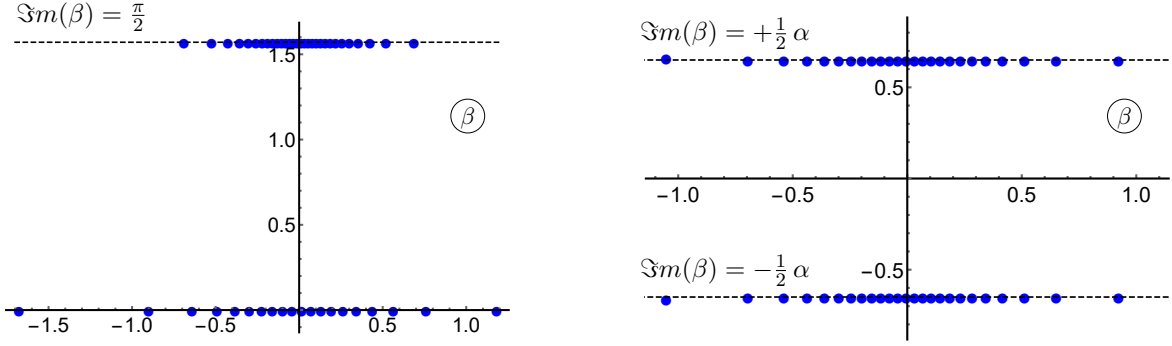
\begin{figure}
\begin{center}
\scalebox{0.9}{
\begin{tikzpicture}
\node at (0,0) {\includegraphics[width=10cm]{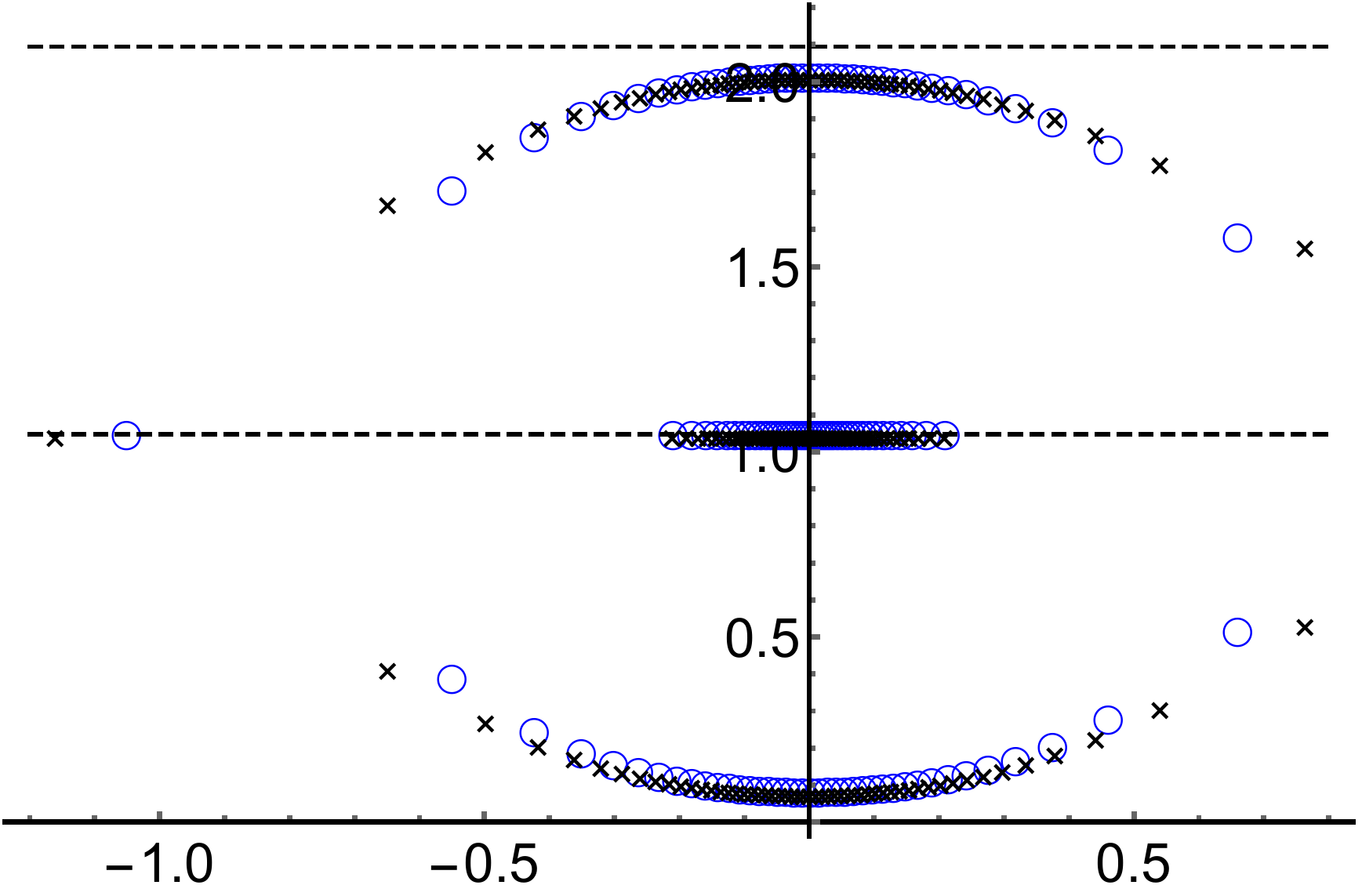}};
\node at (6.05,2.95)   {$\Im m(\beta)=\frac{2\pi}{3}$};
\node at (6.0,0.05)   {$\Im m(\beta)=\frac{\pi}{3}$};
\node at (-4.3,1.6) {\small $\beta$};
\draw (-4.3,1.63) circle (0.25cm);
\end{tikzpicture}
}
\end{center}
\bigskip

\caption{\label{fig3}\small
The pattern of  Bethe roots  $\beta_j=-\frac{1}{2}\log(\zeta_j)$  for the ground state of the spin chain Hamiltonian $\mathbb{H}$ with $r=3$ in the sector $S^z=0$. 
The  inhomogeneities are set to be $\eta_1=\re^{-\frac{2\pi\ri}{3}}$,
  $\eta_2=\re^{\frac{2\pi\ri}{3}\epsilon}$ and
$\eta_3=\re^{\frac{2\pi\ri}{3}(1-\epsilon)}$
where $\epsilon=\frac{3}{40}$. 
For the open blue circles, the number of lattice sites is $N=240$, while for the black crosses
$N=324$. The remaining parameters were taken as
$\gamma=\frac{\pi}{5},\ {\tt k}=\frac{1}{20}$.
}
\end{figure}
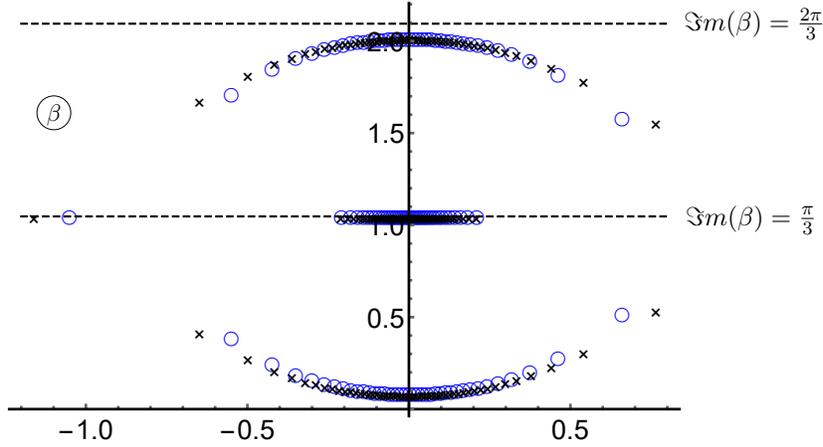

\section{Output from numerical work}
Described here are some results of our numerical analysis of
the spin chain with ${\cal Z}_r$ symmetry in the regime \eqref{87hgvbasbv}.

\bigskip

The study of the scaling limit requires one to assign an $N$ dependence to the low energy states, i.e.,
to form the RG trajectories $|\Psi_N\rangle$.
It is clear how to do this for the ground state or, for that matter, the lowest energy states in the disjoint
sectors of the Hilbert space, say, in the sector with given value of $S^z$
(eigenvalue of $\mathbb{S}^z$ from eq.\,\eqref{mnbn211221}). However forming individual
RG 
flow trajectories for low energy stationary states is not
a trivial task. In the case at hand, the procedure is facilitated by the existence of
the Bethe Ansatz equations.
\bigskip

One starts by diagonalizing the Hamiltonian for a small number of lattice sites $N=N_{\rm in}\lesssim 20$.\footnote{%
The explicit formula for the Hamiltonian and the other members of the commuting family 
may be found in sec.\,6 of ref.\cite{Bazhanov:2020new}.}   
Each stationary state may be chosen to be a Bethe state, which is an eigenvector of the full
family of commuting operators. An important member of this family is the $Q$-operator.
Its eigenvalues are polynomials in $\zeta$, whose zeroes $\{\zeta_m\}_{m=1}^M$ 
obey the Bethe Ansatz equations \eqref{bae}.  
For a given low energy state $|\Psi_{N_{\rm in}}\rangle$, one can compute the eigenvalue of the $Q$-operator
and thus determine the corresponding set of Bethe roots.
With these at hand,  the state $|\Psi_{N}\rangle$ where $N=N_{\rm in}+2r$ may be specified
without an explicit diagonalization of the Hamiltonian. One
finds the solution to the Bethe Ansatz equations, whose pattern qualitatively resembles the
Bethe roots for $|\Psi_{N_{\rm in}}\rangle$ (for technical details, see ref.\cite{Bazhanov:2019xvy}).
 By iterating this procedure, the RG trajectory 
$|\Psi_N\rangle$ 
for increasing $N$ is obtained. The corresponding energy is computed
 from the Bethe roots labeling the Bethe state as
\be\label{aoisd9819821A}
{\cal E}=\sum_{\ell=1}^r{\cal E}^{(\ell)}\,,\qquad\qquad
{\cal E}^{(\ell)}=2\ri\sum_{m=1}^{M}\,\bigg(\frac{1}{1+\zeta_m\,q^{-1}\,/\eta_\ell}
-\frac{1}{1+\zeta_m\,q/\eta_\ell}\bigg)\ ,
\ee
while the eigenvalue of the lattice translation operator $\mathbb{K}$ is given by\footnote{%
 The formulae for the eigenvalues of the energy and lattice shift operator are valid for any values of the
 inhomogeneities provided the $r$-site periodicity condition \eqref{mn21bszz} is satisfied. Clearly, 
for the ${\cal Z}_r$ symmetric case
 \eqref{zsym1} they may  be simplified, but we prefer to keep them as they are.}
\be\label{asj21gh}
{\cal K}=\prod_{\ell=1}^r{\cal K}^{(\ell)}\,,\qquad\qquad\qquad
{\cal K}^{(\ell)}=\re^{\ri \pi{\tt k}}\,q^{-M}\,
\prod_{m=1}^{M}
\frac{\zeta_m+\eta_{\ell}\,q}{\zeta_m+\eta_{\ell}\,q^{-1}}\ .
\ee
Recall that the number of Bethe roots in the sector with given value
$S^z$ coincides with $M=\frac{N}{2}-S^z$.

\subsection{Low energy spectrum for odd $r$}
We performed a study of the large $N$ behaviour of the low energy spectrum and found that the states organize into
conformal towers as predicted from conformal field theory \cite{Cardy:1986ie}. For the case of odd $r$, the low 
energy-momentum
spectrum is described as
\bea\label{aiosd9012221}
{\cal E}&=&N\,e_{\infty}+\frac{2\pi r v_{{\rm F}}}{N}\,\Big(P^2+\bar{P}^2
-\frac{r}{12}\,+{\tt L}+\bar{{\tt L}}\,\Big)+o(N^{-1}) \nonumber \\[0.4cm]
{\cal K}&=&(-1)^{r{\tt w}}\,\exp\bigg(\frac{2\pi\ri r }{N}\ \big(\,P^2-\bar{P}^2+{\tt L}-\bar{\tt L}\,\big)\bigg)\, .
\eea
Here the specific bulk energy and Fermi velocity are given explicitly by
 \bea\label{uassaysa}
e_{\infty}= -\frac{2v_{{\rm F}}}{\pi}\,\int_0^\infty{\rm d}t\ \frac{\sinh\big(\frac{r t}{n}\big)}
{\sinh\big(\frac{(n+r)\,t}{n}\big)\,\cosh(t)}\,,\qquad \qquad
v_{{\rm F}}=\frac{r\,(n+r)}{n}
\eea
and the anisotropy parameter $q=\re^{\ri\gamma}$ has been swapped for $n$ such that 
\be
\gamma=\frac{\pi}{n+r}\qquad\qquad (n>0)\, .
\ee
The $P$ and $\bar{P}$ that enter into the $1/N$ corrections take the discrete set of values, 
\be\label{askjhg21}
P=\frac{1}{2\sqrt{n+r}}\,\big(S^z+
(n+r)\,({\tt k}+{\tt w})\big)\,,\qquad\qquad
\bar{P}=\frac{1}{2\sqrt{n+r}}\,\,\big(S^z-
(n+r)\,({\tt k}+{\tt w})\big)\,,
\ee
which
are characterized by the eigenvalue of $\mathbb{S}^z$ and the so-called winding number ${\tt w}=0,\pm 1,\pm2,\ldots\ $.
Finally, the pair of non-negative integers $({\tt L},\bar{\tt L})$ give the chiral levels of the state in the conformal tower.
\medskip

The Bethe states  for which the asymptotic behaviour \eqref{aiosd9012221} is satisfied with
${\tt L}=\bar{\tt L}=0$ will be referred to as the primary Bethe states. We observe the surprising feature that
for given $S^z$ and ${\tt w}$, the number of primary Bethe states grows for increasing $N$ and in all likelihood
becomes infinite in the limit $N\to\infty$.
This implies the existence of an infinite number of conformal towers
labeled by the same pair of conformal dimensions. To resolve such a degeneracy,
one should consider the scaling limit of the spectrum of other operators from the commuting family. We used
the so-called quasi-shift operators,
\be
\mathbb{K}^{(\ell)}\ \ 
(\ell=1,2,\ldots,r)\ :\qquad\qquad  \big[\mathbb{K}^{(\ell)},\mathbb{H}\big]=
 \big[\mathbb{K}^{(\ell)},\mathbb{K}^{(\ell')}\big]=0\,,
\ee
whose definition is given in sec.\,6.2
of the work \cite{Bazhanov:2020new}. Note that these
 operators are reshuffled under the action of the ${\cal Z}_r$ symmetry:
\be
\hat{{\cal D}}^{-1}\mathbb{K}^{(\ell)}\hat{{\cal D}}=\mathbb{K}^{(\ell+1)}\qquad\qquad
\big(\mathbb{K}^{(r+1)}\equiv\mathbb{K}^{(1)}\big)\,,
\ee
where the notation $\hat{{\cal D}}$ from ref.\cite{Bazhanov:2020new} for the generator of the  symmetry is being used.
Also their product coincides with the $r$-site translation operator:
\be
\mathbb{K}=\prod_{\ell=1}^r\mathbb{K}^{(\ell)}\ .
\ee
The eigenvalues of $\mathbb{K}^{(\ell)}$ are denoted as ${\cal K}^{(\ell)}$ and
are given in terms of the Bethe roots as in eq.\,\eqref{asj21gh}.
\medskip

Keeping in mind the ${\cal Z}_r$ symmetry, we studied the discrete Fourier transform of the logarithm 
of ${\cal K}^{(\ell)}$:\footnote{% 
For the low energy states ${\cal K}^{(\ell)}\to (-1)^{{\tt w}}$ as $N\to\infty$, where ${\tt w}\in\mathbb{Z}$
is the winding number. The leading term in the asymptotics $\log ({\cal K}^{(\ell)})=\ri\pi+\ldots$  for ${\tt w}$ odd
gives no contribution to the definition of $b_a$
since it is canceled when the sum over $\ell$ in \eqref{ajs12v} is taken.}
\bea\label{ajs12v}
b_a\equiv \frac{ N^{1-\frac{2|a|}{r}}}{2\pi \ri r  }\
  \sum_{\ell=1}^r\re^{\frac{\ri\pi}{r} a(r+1-2\ell)}\ \log\big({\cal K}^{(\ell)}\big)\ \ \ \  \ \
  \qquad  \big(1\le|a|\leq[\tfrac{r-1}{2}]\big)\, .
\eea
The factor $ N^{1-\frac{2|a|}{r}}$ was introduced for the following reason;
we observed that for the low energy states
$b_a=b_a(N)$, defined as above, tend to finite limits as $N\to \infty$ which,
in general, are non-vanishing (complex) numbers.
Also, in \eqref{ajs12v}, the symbol $[\cdots]$ stands for the integer part. 
It's placed here in view of the later discussion of even $r$, where this same
 definition of $b_a$ will be utilized.

\medskip

The limiting values of $b_a$ \eqref{ajs12v}  as $N\to\infty$ are important characteristics of the
RG trajectories. It turns out that they, together with $P$ and $\bar{P}$ \eqref{askjhg21}, are sufficient to
 distinguish the primary Bethe states. 
Also, we found that $b_a$ 
appear in the corrections to the scaling of the energy in eq.\,\eqref{aiosd9012221}.
In particular,  for any low energy Bethe state
\bea\label{k21vgghsd}
{\cal E}&=& N\,e_{\infty}+\frac{2\pi r v_{{\rm F}}}{N}\,\Big(P^2+\bar{P}^2
-\frac{r}{12}\,+{\tt L}+\bar{{\tt L}}\,\Big)
\\[0.0cm]
&-&\frac{4\pi^2 r v_{{\rm F}}}{N^{1+\frac{2}{r}}}\,\cot\Big(\frac{\pi}{2n}\Big)\
b_{\frac{r-1}{2}}\,b_{\frac{1-r}{2}}+
O\Big(N^{-1-\frac{6}{r}},N^{-1-\frac{4n}{r}}\Big)\qquad\qquad 
(r=3,5,\ldots) \, .\nonumber\vphantom{\Bigg(}
\eea

\subsection{Low energy spectrum for $r$ even}
For even $r$ special attention needs to be paid to the operator
\be
\mathbb{B}=\prod_{m=1}^{\frac{r}{2}} \mathbb{K}^{(2m-1)}\,\big(\mathbb{K}^{(2m)}\big)^{-1} \ .
\ee
A numerical study of its eigenvalues  shows there are low energy states in the lattice model such that
\bea\label{ajs12vAA}
{b}_{\frac{r}{2}}\equiv \frac{(-1)^{\frac{r}{2}+1}}{2\pi  r  }\,\log({\cal B})=
 \frac{1}{2\pi  r  }\
  \sum_{\ell=1}^r(-1)^{\ell+\frac{r}{2}}\, \log\big({\cal K}^{(\ell)}\big)
\eea 
either tends to a non-zero value or decays to zero as
\be
{b}_{\frac{r}{2}}\sim \frac{1}{\log(N)}\qquad\qquad {\rm for}\qquad\qquad
N\to\infty\, .
\ee
Such a phenomenon was already observed and extensively studied  in the context of
 the ${\cal Z}_2$ invariant spin chain ($r=2$) \cite{Ikhlef:2008zz,Ikhlef:2011ay,Candu:2013fva,Frahm:2013cma,Bazhanov:2019xvy,Bazhanov:2019xvyA}. 
The scaling limit should be taken such 
 that the value of ${b}_{\frac{r}{2}}$
is kept fixed as $N\to\infty$.
This leads to the presence of 
a continuous component in the spectrum of the underlying conformal field theory.
 In the case $r=4,6,\ldots$ and for the low energy states, $b_a$ defined through \eqref{ajs12v}   remain
finite and (generically) non-vanishing as $N\to\infty$.
 This way,  for both
$r$ odd and $r$ even,
the RG trajectories are   labeled by
the limiting values of $b_a$.
For even $r$ we observed that the large $N$ asymptotic behaviour of the energy follows a similar
formula to \eqref{k21vgghsd}:
\bea\label{k21vgghsdAA}
{\cal E}&=& N\,e_{\infty}+\frac{2\pi r v_{{\rm F}}}{N}\,\Big(P^2+\bar{P}^2
+2n\ ({b}_{\frac{r}{2}})^2
-\frac{r}{12}\,+{\tt L}+\bar{{\tt L}}\,\Big)
\\[0.0cm]
&-&\frac{8\pi^2 r v_{{\rm F}}}{N^{1+\frac{4}{r}}}\,\cot\Big(\frac{\pi}{n}\Big)\
b_{\frac{r}{2}-1}\,b_{1-\frac{r}{2}}+
O\Big(N^{-1-\frac{8}{r}},N^{-1-\frac{4n}{r}}\Big)\qquad\qquad 
(r=4,6,\ldots)\,. \nonumber
\eea

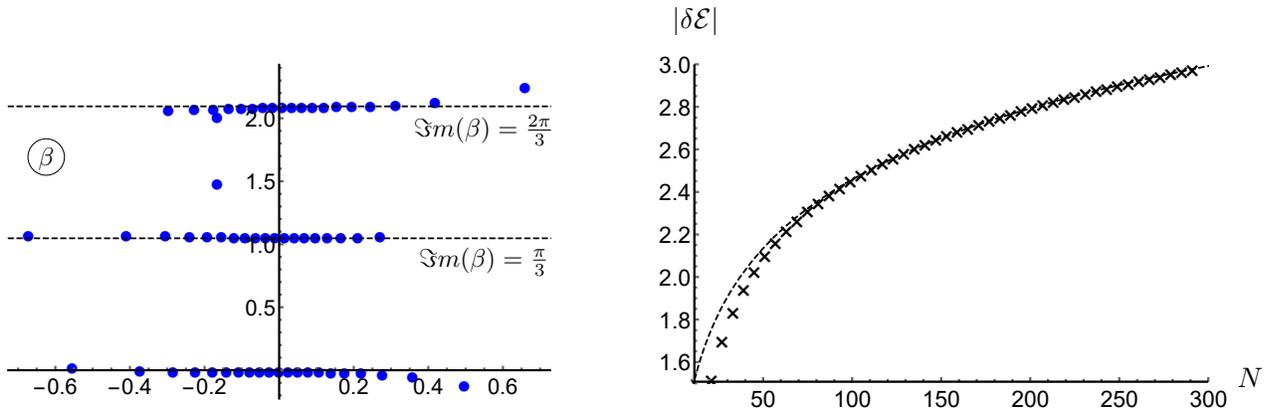
\begin{figure}
\centering
\scalebox{0.97}{
\begin{tikzpicture}
\node at (-8,1.0) {\small $\beta$};
\draw (-8,1.03) circle (0.25cm);
\node at (-4.8,0) {\includegraphics[width=7.5cm]{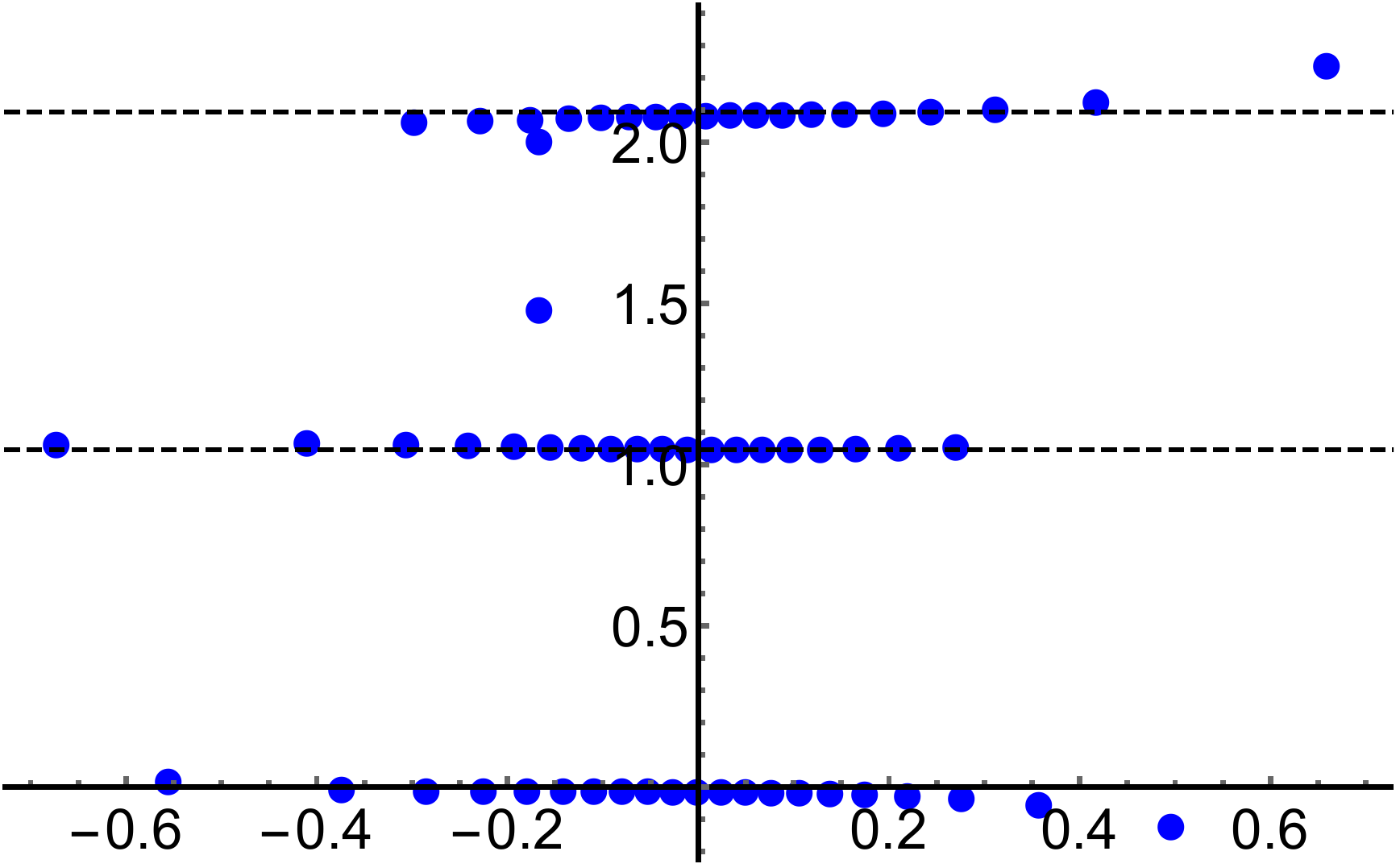}};
\node at (4.3,0) {\includegraphics[width=7.8cm]{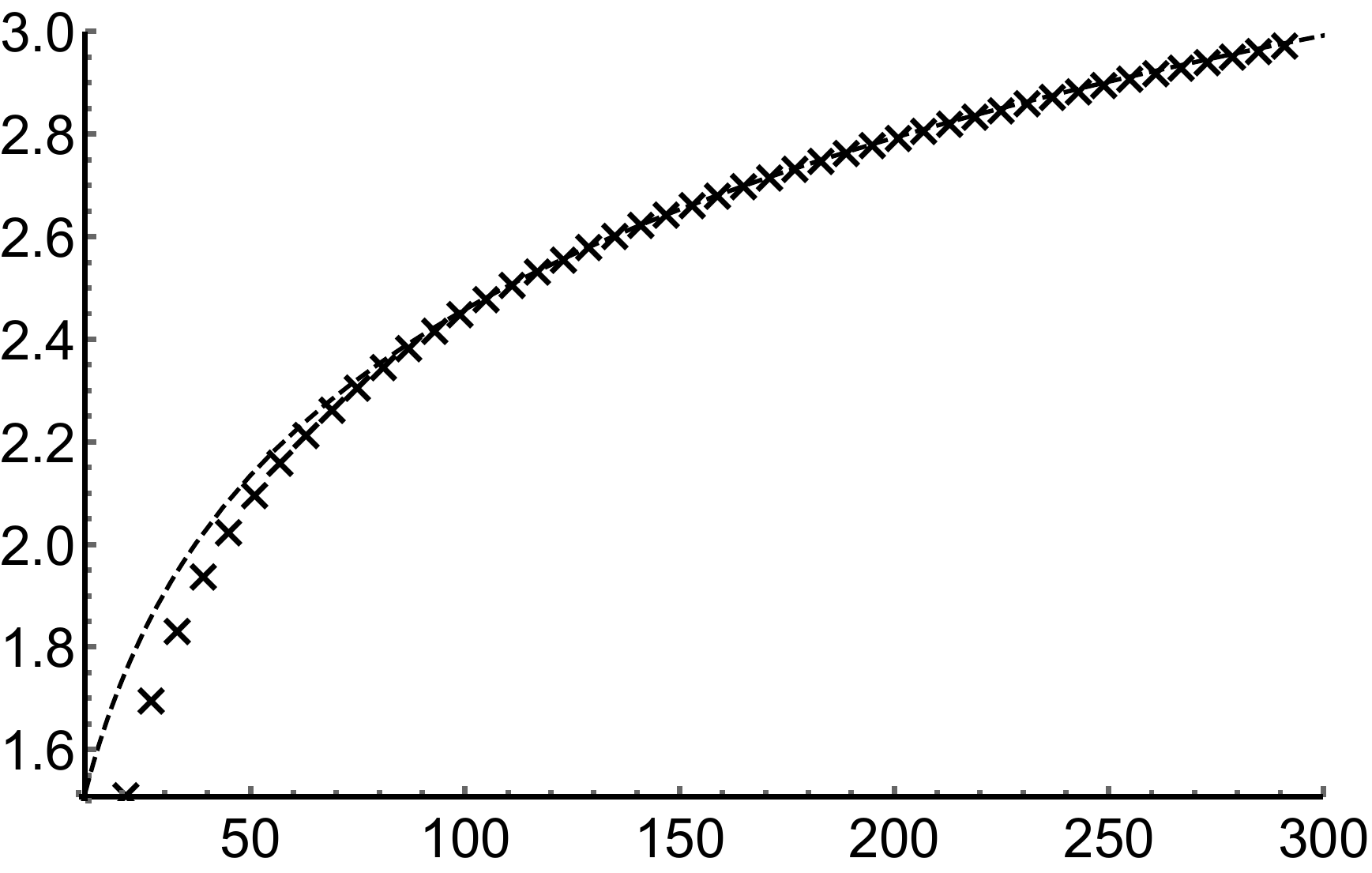}};
\node at (-2,1.4) {\small $\Im m(\beta)=\frac{2\pi}{3}$};
\node at (-2,-0.4) {\small $\Im m(\beta)=\frac{\pi}{3}$};
\node at (0.9,2.9) {$|\delta {\cal E}|$};
\node at (8.5,-2.0) {$N$};
\end{tikzpicture}
}

\caption{\small Presented is numerical data for an RG trajectory of the ${\cal Z}_3$ invariant
spin chain in the sector $S^z=\frac{1}{2}$ whose scaled energy 
$\delta{\cal E}=\frac{N}{2\pi r v_{\rm F}}\,\big({\cal E}-N e_\infty\big)$ grows logarithmically
as $N\to\infty$. 
 The typical pattern of Bethe roots  ($N=123$) is depicted  in the complex $\beta$ plane on the left panel.
On the right panel, the crosses correspond to the numerical values of $|\delta {\cal E}|$ obtained via
the solution of the Bethe Ansatz equations. The dashed line is a plot of the absolute value of the
fit  $\delta{\cal E}\approx -0.03-1.11\,\ri+(0.521+0.099\,\ri)\log(N)$. The  parameters are
$n=3$ and ${\tt k}=\frac{1}{20}$.
\label{fig4a}}
\end{figure}

The following  comments are in order here. In our numerical work we constructed a variety
of RG trajectories for $r=2,3,\ldots,7$. In all the cases our  data was in full agreement 
with the asymptotic formula
\bea
{\cal E}&=& N\,e_{\infty}+\frac{2\pi r v_{{\rm F}}}{N}\,\bigg[P^2+\bar{P}^2+2n\ ({b}_{\frac{r}{2}})^2
-\frac{r}{12}\,+{\tt L}+\bar{{\tt L}}
\\[0.0cm]
&-&\sum_{a=1}^{[\frac{r-1}{2}]}\,2\pi\,(r-2a)\,\cot\big(\tfrac{\pi\,(r-2a)}{2n}\big)\,
\frac{{b}_a\,{b}_{-a}}{N^{2-\frac{4a}{r}}}+
O\Big(N^{-2},N^{-\frac{4n}{r}}\Big)\ \bigg]\,,
\nonumber
\eea
where $b_{\frac{r}{2}}\equiv 0$ for odd $r$ and $n$ is assumed to be generic.
Also, it is worth mentioning that the summands ${\cal E}^{(\ell)}$ in \eqref{aoisd9819821A}
coincide with the eigenvalues of the mutually commuting operators
\be
\mathbb{H}^{(\ell)}\ \ :\qquad 
 \big[\mathbb{H}^{(\ell)},\mathbb{H}^{(\ell')}\big]=0\,,\qquad\qquad
 \mathbb{H}=\sum_{\ell=1}^r\mathbb{H}^{(\ell)}\, .
\ee
Similar to $\mathbb{H}$ these are  given by a sum of terms describing interactions of
up to $r+1$ adjacent spins. However, they are not ${\cal Z}_r$ invariant, but rather
\be
\hat{{\cal D}}^{-1}\mathbb{H}^{(\ell)}\hat{{\cal D}}=\mathbb{H}^{(\ell+1)}\qquad\qquad
\big(\mathbb{H}^{(r+1)}\equiv\mathbb{H}^{(1)}\big)\,.
\ee
For the
 leading large $N$ behaviour of their eigenvalues we found that for $r\geq 2$
\bea\label{ajs12vAAAA}
  \sum_{\ell=1}^r\re^{\frac{\ri\pi}{r} a(r+1-2\ell)}\ {\cal E}^{(\ell)}=2\pi v_{\rm F}\,\sgn(a)\,(r-2|a|)\,
 N^{-1+\frac{2|a|}{r}}\ \big({b}_a+o(1)\big)\ \ \ \  \ \  \big(1\le|a|\leq[\tfrac{r}{2}]\big)\,,
\eea
where $\sgn(a)$ denotes the sign of the integer $a$.
Finally, Bethe states were observed for which   $\delta{\cal E}=\frac{N}{2\pi r v_{\rm F}}\,\big({\cal E}-N e_\infty\big)$
grows logarithmically as $N\to\infty$, see fig.\,\ref{fig4a}. We do not count  these states
as being part of the low energy spectrum.
A similar phenomenon was discussed in the case of the ${\cal Z}_2$ invariant spin chain in the original works
 \cite{Ikhlef:2008zz,Jacobsen:2005xz}.

\section{ODEs for the scaling limit of the  ${\cal Z}_r$ invariant spin chain}
As was mentioned in the Introduction, the most effective way of studying
 the universal behaviour of integrable lattice systems is based on the ODE/IQFT correspondence.
 In this section, we propose the set of Ordinary Differential Equations which  describe the 
scaling limit of the ${\cal Z}_r$ invariant  spin
 chain where the anisotropy parameter is taken to be $q=\re^{\frac{\ri\pi}{n+r}}$ with $n>0$.
Also, some numerical verifications of the proposal are discussed.

\subsection{Differential equations for the primary Bethe states}
The limiting values of $b_a$ \eqref{ajs12v},\,\eqref{ajs12vAA} as $N\to\infty$ are important characteristics
of an individual RG trajectory.
For our purposes, it will be  convenient to switch from these to
\bea\label{asnb2vb}
s_a= C_a \ \slim_{N\to\infty}b_a\,,
\qquad\qquad \qquad\bar{s}_a=-C_a \ \slim_{N\to\infty}b_{-a}\,,\qquad\qquad 1\le a\le[\tfrac{r}{2}]
\eea
with the proportionality coefficients being given by
\be\label{asnb2vbAA}
C_a=(-1)^{a(r-1)}\,2n\ \frac{\sqrt{\pi}\,\Gamma(1+\frac{r-2a}{2n})}{\Gamma(\frac{1}{2}+\frac{r-2a}{2n})}\ 
 \Bigg[\frac{\Gamma(\frac{3}{2}+\frac{r}{2n})}{\sqrt{\pi}
\Gamma(1+\frac{r}{2n})}\Bigg]^{\frac{r-2a}{r}}\, .
\ee
The symbol ``$\slim$'' is used as a reminder that the limit \eqref{asnb2vb}
exists only for the class of low energy
states and for even $r$ should be 
taken such that   $b_{\frac{r}{2}}$ is kept fixed as $N\to\infty$.
Note that, by definition
$b_{-\frac{r}{2}}\equiv-b_{\frac{r}{2}}$ and therefore
\be
\bar{s}_{\frac{r}{2}}\equiv s_{\frac{r}{2}}\ .
\ee
The $s_a$ and $\bar{s}_a$ will be combined into the two sets
\be\label{kjassshg21}
\bm{s}=\big({s}_{1},\ldots, {s}_{[\frac{r}{2}]}\big)\,,\qquad
\ \ \ \ \ \ {\bar {\bm{s}}}=\big(\bar{s}_{1},\ldots, \bar{s}_{[\frac{r}{2}]}\big)\ .
\ee
Also we'll use the notations
\bea\label{asnb2vbAAA}
p=\frac{n+r}{2}\ \bigg(\frac{S^z}{n+r}+{\tt k}+{\tt w}\bigg)\ ,\ \ \ \  \ \ \ 
{\bar{p}}=\frac{n+r}{2}\ \bigg(\frac{S^z}{n+r}-{\tt k}-{\tt w}\bigg)\, .
\eea

Among other things, the ODE/IQFT correspondence implies a relation between the scaling limit of the eigenvalues
of the $Q$-operator for the low energy Bethe states and the spectral determinants of  a certain set of ODEs. 
In the case at hand,
the primary Bethe states of the ${\cal Z}_r$ invariant spin chain  are fully characterized by 
$p$, $\bar{p}$ as well as  $\bm{s}$, ${\bar {\bm{s}}}$ and
the corresponding pair of differential equations  are proposed to be
\bea\label{hassay}
&&\bigg[-\partial_z^2+\frac{1}{z^{2}}\ \Big( E^{-n-r}\ z^{n+r}- z^{r}+p^2-\frac{1}{4}+
\sum_{a=1}^{[\frac{r}{2}]} s_a\, (-z)^{a}\Big)
\bigg]\, \Psi=0\\[0.2cm]\label{hassayA}
&&\bigg[-\partial_{\bar{z}}^2+\frac{1}{\bar{z}^{2}}\ \Big( {\bar E}^{-n-r}\ 
\bar{z}^{n+r}-\bar{z}^{r}+{\bar p}^2-\frac{1}{4}+
\sum_{a=1}^{[\frac{r}{2}]}{\bar { s}}_a\, (-\bar{z})^{a}\Big)
\bigg]\,{\bar  \Psi}=0\ .
\eea
The cases $r=1$ and $r=2$
have been extensively explored  in the work \cite{Bazhanov:2019xvyA} (see also Appendix \ref{AppA}). In order to
not overburden the text with technical details,  the similar analysis for
general $r$ will not be repeated here. Instead we give a brief reminder of the construction of the spectral determinant,
which is numerically efficient, focusing on  eq.\,\eqref{hassay}.
\bigskip

It is convenient to re-write the ODE using the variables
\be\label{akskj21981298}
z=E\,\re^y\,,\qquad\qquad \Psi=\re^{\frac{y}{2}}\,\psi\ .
\ee
The differential equation \eqref{hassay} then becomes
\be\label{kkjj2198}
\Big[-\partial^2_y+p^2+\re^{(n+r)y}-E^r\,\re^{ry}+
\sum_{a=1}^{[\frac{r}{2}]} s_a\, (-E\,\re^y)^{a}\Big]\psi=0\ .
\ee
Assuming that $\Re e(p)\ge 0$, there exists a solution which is uniquely defined by the asymptotic condition
\be\label{askjjh2143}
\psi_p(y)\to\re^{py}\qquad\qquad {\rm as}\qquad\qquad y\to-\infty\ .
\ee
Introduce another solution, $\chi=\chi(y)$, by means of the WKB asymptotic:
\bea
\chi\asymp
\exp\bigg[-\frac{n+r}{4}\, y-\frac{2}{n+r}\,\re^{(n+r)\frac{y}{2}}\ {}_{2}F_1\Big(-\frac{1}{2},
-\frac{1}{2}
-\frac{r}{2n},
\frac{1}{2}
-\frac{r}{2n}\,\Big|\, E^r \re^{-ny}\Big)\bigg]\ \ \ \ {\rm as}\ \  y\to+\infty
\eea
Here ${}_{2}F_1$ is the conventional  hypergeometric function and
 it is  assumed that $\frac{r}{n}\not=1,3,\ldots\ $. 
The spectral determinant is given in terms of the Wronskian of the two solutions:
\be\label{aklsjhj1298}
D_p(E)=\frac{\sqrt{\pi}}{\Gamma(1+\frac{2p}{n+r})}\ (n+r)^{-\frac{1}{2}-\frac{2p}{n+r}}\ 
\big(\chi\partial_y\psi_p-\psi_p\partial_y\chi\big)
\ee
with the factor out the front being chosen so that $D_p(0)=1$ for generic values of $p$.
\medskip

The spectral determinant
is an entire function of $E$ and hence the series
\be\label{asnbv21}
\log D_p(E)=-\sum_{j=1}^\infty J_j\,E^j
\ee
has a finite radius of convergence. 
The expansion coefficients depend on $p$ and ${\boldsymbol s}=(s_1,\ldots,s_{[\frac{r}{2}]})$
(also, of course, on $n>0$),
\bea
J_j= J_j(p,{\boldsymbol s})\ ,
\eea
which are parameters of the
differential equation.
By applying perturbation theory  in $E$ to the ODE \eqref{kkjj2198}
it is possible to obtain explicit analytic formulae for the first two of them. The result
for the cases $r=1,2$ is quoted in ref.\cite{Bazhanov:2019xvyA}. It possesses a straightforward generalization for
$r\ge 3$:
\bea\label{akjshgj873287}
J_1&=&s_1\ f_1\big(\tfrac{p}{n+r},\tfrac{1}{n+r}\big)\ C
\\[0.2in]
J_2&=&
\bigg(s_1^2\ f_2\big(\tfrac{p}{n+r},\tfrac{1}{n+r}\big)-
s_2\ 2^{\frac{4}{n+r}}\ \frac{\pi\Gamma^2(-\frac{1}{n+r})}{\Gamma^2(\frac{1}{2}-\frac{1}{n+r})}\ 
f_1\big(\tfrac{p}{n+r},\tfrac{2}{n+r}\big)\,\bigg)\  C^2
\nonumber
\eea
with
\bea
C=\frac{(n+r)^{\frac{2}{n+r}}}{\Gamma^2(-\frac{1}{n+r})}
\eea
(for $r=3$, one should set $s_2$ identically to zero).
The explicit form of the functions $f_1$ and $f_2$ are presented in Appendix \ref{AppB}. 
\bigskip

The eigenvalues of the lattice $Q$-operator are polynomials in the spectral parameter $\zeta$, whose roots
satisfy the Bethe Ansatz equations. In the notations of ref.\cite{Bazhanov:2020new},\footnote{%
As explained in that work, the Baxter $TQ$-relation possesses two solutions $\mathbb{Q}_\pm$.
The operator $\mathbb{A}_+$   coincides with $\mathbb{Q}_+$ up to a simple factor involving
fractional powers of $\zeta$ and has the advantage that its eigenvalues $A_+(\zeta)$ are  polynomials in $\zeta$
of order $M\le N/2$\,.}
\be\label{sajhg1278}
A_+(\zeta)=\prod_{m=1}^M\bigg(1-\frac{\zeta}{\zeta_m}\bigg)\qquad\qquad\qquad\qquad (M\le N/2\,)\, .
\ee
 The coefficients of the polynomial 
are simply expressed in terms of the sums
\be\label{mnbnw8129}
h^{(j)}_N=j^{-1}\sum_{m=1}^M(\zeta_m)^{-j}\,,
\ee
for instance,
\be
\log A_+(\zeta)=-h^{(1)}_N\,\zeta-h^{(2)}_N\,\zeta^2+\ldots\ .
\ee
For the low energy Bethe states, the scaling behaviour of $h^{(j)}_N$ is described
by $J_j$ from the expansion \eqref{asnbv21} of the spectral determinant.
In particular,  
\bea\label{askjh12}
&&\slim_{N\to\infty}\ \bigg(\frac{N}{rN_0}\bigg)^{-\frac{2n}{r(n+r)}}\, h_N^{(1)}=
J_1
\\
&&
\slim_{N\to\infty}\ \bigg(\frac{N}{rN_0}\bigg) ^{-\frac{4n}{r(n+r)}}\, h_N^{(2)}=
    J_2
\ ,\nonumber
\eea
where
\bea\label{iaisissu}
N_0=
\frac{\sqrt{\pi}\Gamma(1+\frac{r}{2n})}{r\Gamma(\frac{3}{2}+\frac{r}{2n})}\ .
\eea
Having at hand a low energy Bethe state at fixed $N\gg 1$, the l.h.s. of eqs.\,\eqref{askjh12} may be computed  from 
the corresponding Bethe roots by means of \eqref{mnbnw8129}. 
Also, from 
the eigenvalues of the quasi-shift operators ${\cal K}^{(\ell)}$ \eqref{asj21gh} 
one extracts the complex numbers $b_a$ via the definitions \eqref{ajs12v} and \eqref{ajs12vAA}.
These determine
the parameters $s_a$  appearing on the ODE side  through formula
\eqref{asnb2vb}. 
For the primary Bethe states, the r.h.s. may be calculated using 
the explicit analytic expressions \eqref{akjshgj873287} for the coefficients $J_1, J_2$
as functions of $s_a $ and $p$.
\medskip

Similar relations hold true for the sums
\be\label{mnbnw8129vvv}
\bar{h}^{(j)}_N=j^{-1}\sum_{m=1}^M(\zeta_m)^j\, .
\ee
They take the same form as \eqref{askjh12} with $h_N^{(j)}\mapsto  \bar{h}_N^{(j)}$, while 
$J_j(p,\bm{s})$ are replaced by ${J}_j(\bar{p},\bar{\bm{s}})$. The latter occur in the Taylor series
of the logarithm of the spectral determinant for the second ODE \eqref{hassayA}.
 Note that $\bar{h}^{(j)}_N$
are the expansion coefficients of 
$-\log\,\big(\prod_{m=1}^M(1-\zeta_m/\zeta)\big)$ in the variable $\zeta^{-1}$:
\be
\bar{A}_+(\zeta)\equiv
\prod_{m=1}^M\bigg(1-\frac{\zeta_m}{\zeta}\bigg)=\exp\bigg(-\sum_{j=1}^\infty \bar{h}_N^{(j)}\zeta^{-j}\bigg)\, .
\ee
The polynomials  $\bar{A}_+(\zeta)$  are  eigenvalues of 
\be\label{kjsajh1289}
\bar{\mathbb{A}}_+(\zeta)=\zeta^{{\mathbb S}^z-N/2}\,\mathbb{A}_+(\zeta)\,\big(
\mathbb{A}_+^{(\infty)}\big)^{-1}\,,
\ee
where the operator $\mathbb{A}_+^{(\infty)}$ belongs to the commuting family and its eigenvalue
for a Bethe state with corresponding Bethe roots $\{\zeta_m\}$
is given by the product:
\be
A_+^{(\infty)}=\prod_{m=1}^M\,(-\zeta_m)^{-1}\ .
\ee

\bigskip

\begin{figure}
\begin{center}
\scalebox{0.95}{
\begin{tikzpicture}
\node at (-6.2,2.7) {$\big(\frac{N}{rN_0}\big)^{-\frac{2n}{r(n+r)}}\,h_N^{(1)}$};
\node at (2.4,2.7) {$\big(\frac{N}{rN_0}\big)^{-\frac{4n}{r(n+r)}}\,\bar{h}_N^{(2)}$};
\node at (-4.3,0) {\includegraphics[width=7.25cm]{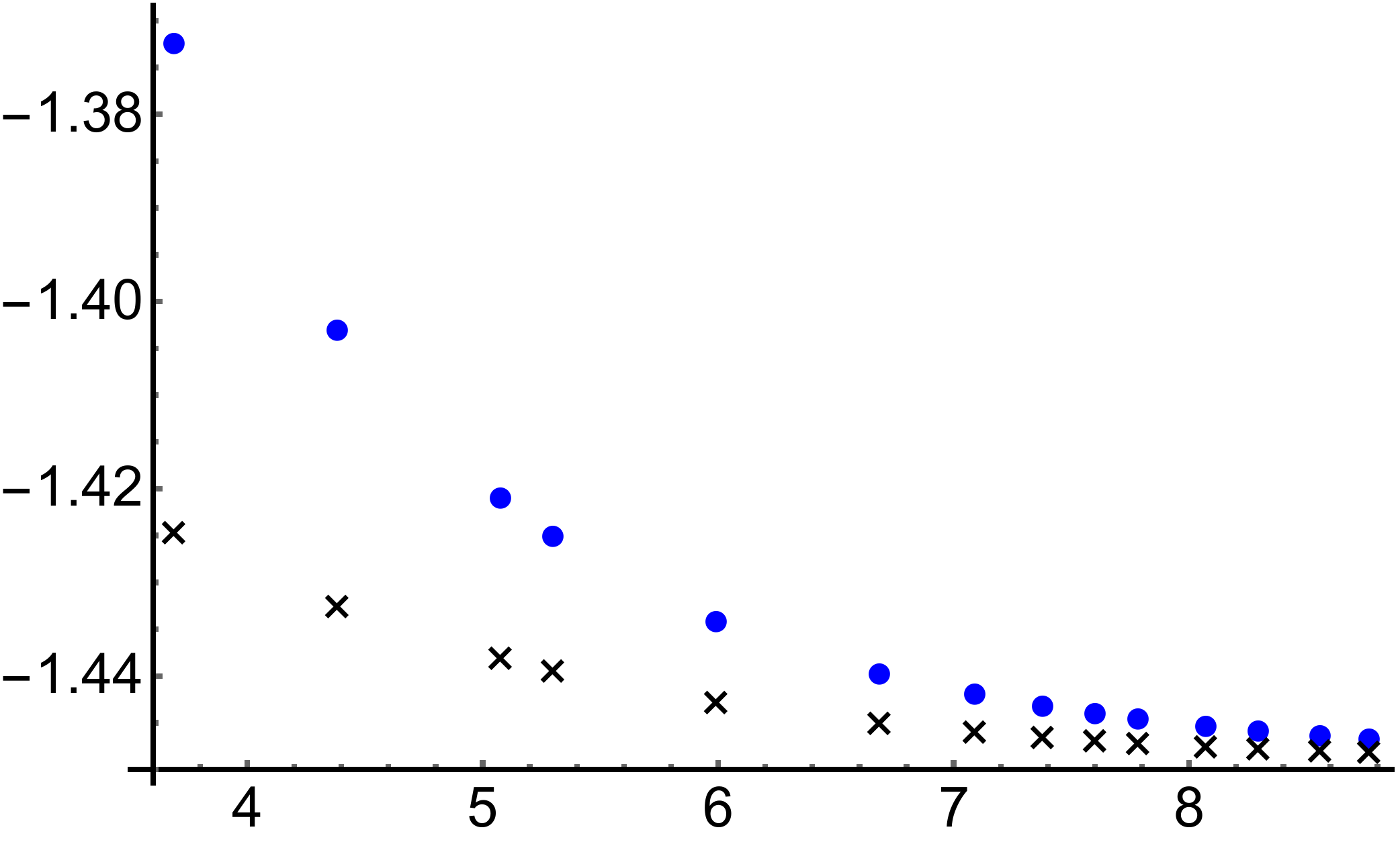}};
\node at (4.5,0) {\includegraphics[width=7.0cm]{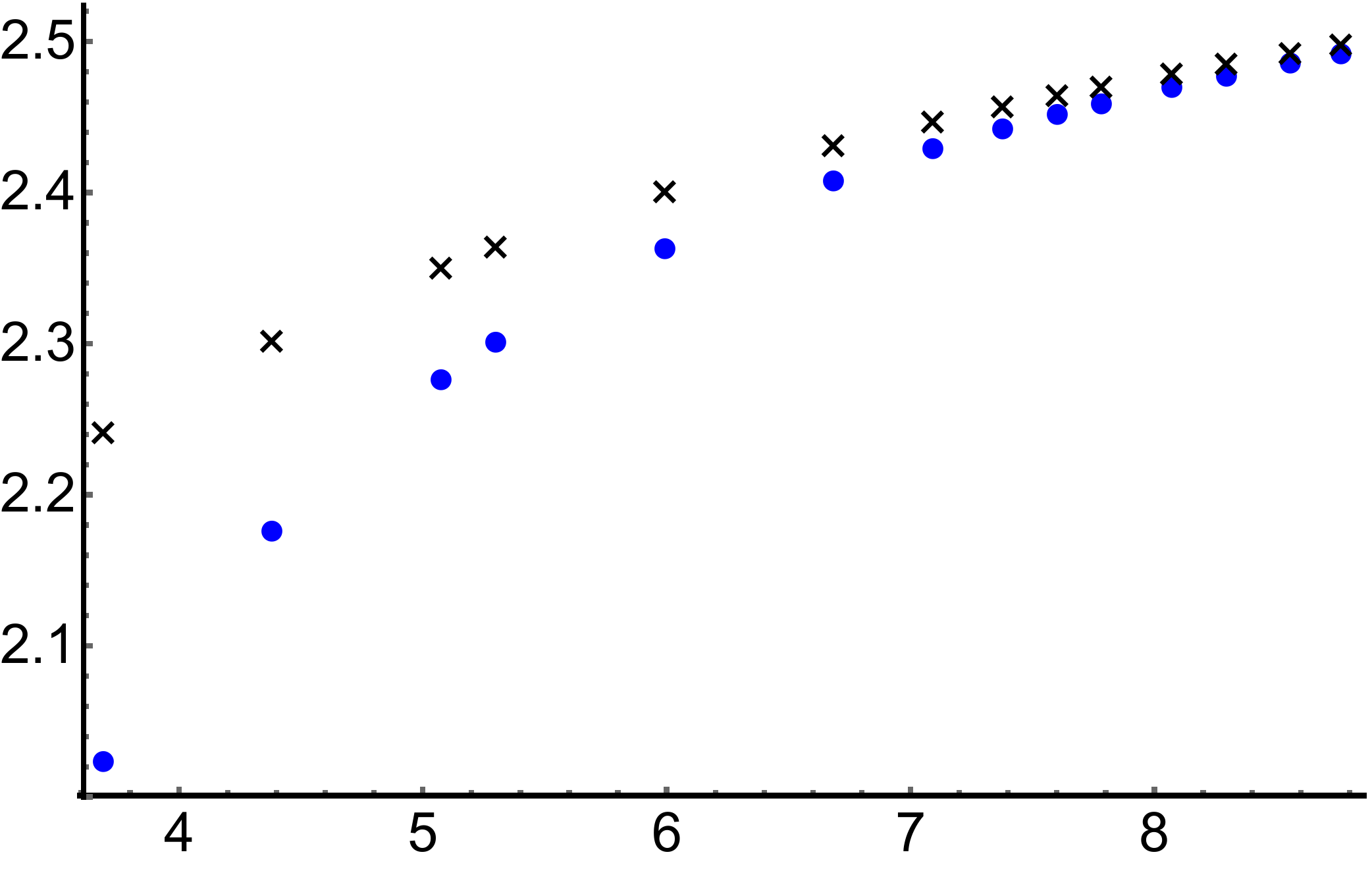}};
\node at (0,-1.75) {\small $\log(N)$};
\node at (8.7,-1.8) {\small $\log(N)$};
\end{tikzpicture}
}
\end{center}
\caption{\label{fig5} \small
Depicted is numerical data for $h_N^{(1)}$ (left panel) and $\bar{h}_N^{(2)}$ (right panel),
scaled by the appropriate $N$ dependent factor,  as a function of $\log(N)$ for the ${\cal Z}_r$
symmetric model with $r=4$. The Bethe state was chosen to be a primary one, and the black crosses were obtained from
the corresponding Bethe roots via the definitions \eqref{mnbnw8129} and \eqref{mnbnw8129vvv}. The blue dots depict
the values of $J_1(p,s_1,s_2)$ and $J_2(\bar{p},\bar{s}_1,\bar{s}_2)$ for the left and right panels, respectively, which
are given by formula \eqref{akjshgj873287}. Here
$s_1$ ($\bar{s}_1$) was substituted by $C_1\,b_1(N)$ ($-C_1\,b_{-1}(N)$), 
see eq.\,\eqref{asnb2vb}, where $b_{\pm 1}(N)$ were obtained 
by means of \eqref{ajs12v} from the eigenvalues of the  quasi-shift operators.
Similarly, $s_2\equiv\bar{s}_2\mapsto C_2\,b_2(N)$ with
$b_2(N)$ being  computed via \eqref{ajs12vAA}. 
 The parameters were set to be
$n=\frac{1}{2}$, ${\tt k}=\frac{1}{10}$, while $S^z={\tt w}=0$. Note that,  for computing 
 $J_2(\bar{p},\bar{s}_1,\bar{s}_2)$ in the case when $\bar{p}=-(n+r)\,\frac{{\tt k}}{2}=-\frac{9}{40}<0$,
one needs to evaluate the function $f_2$ entering into \eqref{akjshgj873287} when the first argument is
negative. Formulae \eqref{jassusau000}-\eqref{jassusau} remain applicable provided the integration contour
over the real line
is locally deformed such that it goes below the pole at $x=\ri h$.
}
\end{figure}
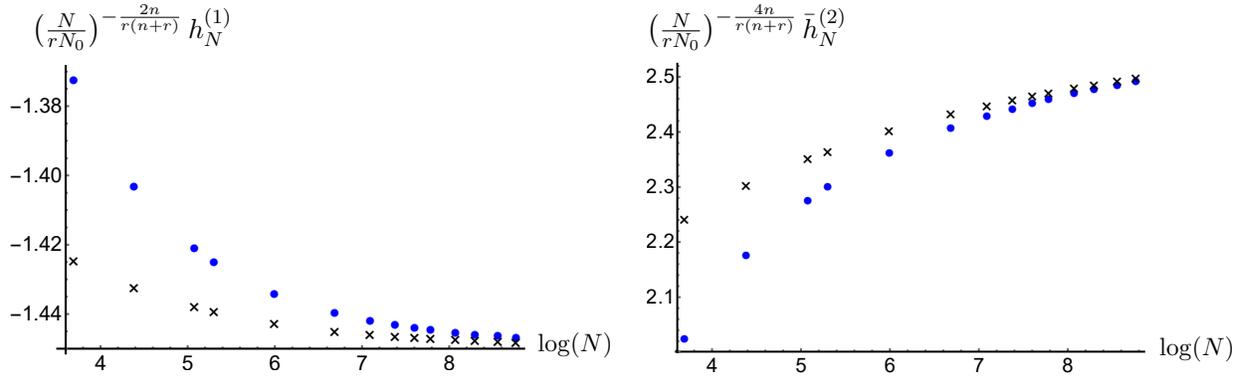

We performed  extensive numerical checks of the
scaling relation
\eqref{askjh12} and its barred counterpart. Some of the data is presented in fig.\,\ref{fig5}.

\subsection{Eigenvalues of the $Q$-operator in the scaling limit}
For the excited states, the ODEs are obtained from \eqref{hassay} and \eqref{hassayA} following
the well known procedure \cite{Bazhanov:2003ni}. It was discussed in refs.\cite{Bazhanov:2019xvy,Bazhanov:2019xvyA} 
for the case of the ${\cal Z}_2$ invariant spin chain.
With the same line of arguments, we propose that the
 generalization of the ODE  \eqref{hassay} that  describes
 the scaling limit  of any low energy Bethe state of the ${\cal Z}_r$ invariant model has the form 
\bea\label{hasas}
\Big(-\partial_z^2+t_0(z)+t_1(z)+E^{-n-r}\ z^{n+r-2}
\Big)\, \Psi=0\ ,
\eea
where 
\bea\label{n1n9832hjds}
t_0(z)&=&\frac{1}{z^{2}}\ \bigg(- z^{r}+p^2-\frac{1}{4}+
\sum_{a=1}^{[\frac{r}{2}]} s_a\, (-z)^{a}\bigg)\\
t_1(z)&=&\sum_{\alpha=1}^{{\tt L}}\bigg(\frac{2}{(z-w_\alpha)^2}+\frac{n_\alpha}{z(z-w_\alpha)}\,\bigg)\ .\nonumber
\eea
The position of the poles $w_\alpha$ and the residues $n_\alpha$ must be taken in such a way that any solution
of the ODE is single  valued in the vicinity
of $z=w_\alpha$ for any $\alpha=1,\ldots, {\tt L}\,$. 
This condition leads to the system of algebraic equations on $w_\alpha$
and $n_\alpha$: 
\bea\label{aosaasiBBB}
&&n_\alpha\, \big(\, \tfrac{1}{4}\, n_\alpha^2- w_\alpha^2\,t_0^{(\alpha)}\big)+w^3_\alpha\, t_1^{(\alpha)}=0\\[0.2cm]
&&n_\alpha=n+r-2 \qquad \qquad\qquad\qquad\qquad (\alpha=1,2,\ldots,{\tt L})\ .\nonumber
\eea
Here
\bea\label{aosaasiCCC}
t_0^{(\alpha)}&=&t_0(w_\alpha)-\frac{n_\alpha}{w_\alpha^2}+ \sum_{\beta\not=\alpha}
\bigg(\frac{2}{(w_\alpha-w_\beta)^2}+\frac{n_\beta}{w_\alpha(w_\alpha-w_\beta)}\bigg)\\[0.1in]
t_1^{(\alpha)}&=&t'_0(w_\alpha)+\frac{n_\alpha}{w_\alpha^3}- \sum_{\beta\not=\alpha}\bigg(\frac{4}{(w_\alpha-w_\beta)^3}
+\frac{n_\beta\,(2w_\alpha-w_\beta)}{w_\alpha^2\,(w_\alpha-w_\beta)^2}\bigg)\nonumber
\eea
and  the prime stands for the derivative w.r.t. the argument. The ODE  that would generalize \eqref{hassayA}
takes an analogous form. It involves the additional term
\be
\bar{t}_1(\bar{z})=\sum_{\alpha=1}^{\bar{\tt L}}\Big(\frac{2}{(\bar{z}-\bar{w}_\alpha)^2}+
\frac{\bar{n}_\alpha}{\bar{z}(\bar{z}-\bar{w}_\alpha)}\,\Big)\,,
\ee
where the set $\bar{n}_\alpha$ and $\{\bar{w}_\alpha\}_{\alpha=1}^{\tt \bar{L}}$ satisfy the system obtained from 
eqs.\,\eqref{aosaasiBBB} and \eqref{aosaasiCCC} by the formal substitutions
$(p,s_a,n_\alpha,w_\alpha,{\tt L})\mapsto (\bar{p},\bar{s}_a,\bar{n}_\alpha,\bar{w}_\alpha,\bar{{\tt L}})$.
\medskip

We expect that any low energy Bethe state
 $|\Psi_N\rangle$  flows to the state in the conformal tower labeled by, together with the parameters $p$, $\bar{p}$,
 $\bm{s}$, $\bar{\bm{s}}$, also  two sets of ``apparent'' singularities:
 \be
 \bm{w}=\{w_\alpha\}_{\alpha=1}^{\tt L}\,,\qquad\qquad 
 \bar{\bm{w}}=\{\bar{w}_\alpha\}_{\alpha=1}^{\tt \bar{L}}
 \,.
 \ee
The scaling behaviour of the eigenvalue ${A}_+(\zeta)$ \eqref{sajhg1278} of the $Q$-operator 
for the state $|\Psi_N\rangle$ is then  described
in terms of the spectral determinant $D_p(E\,|\,\bm{w})$ of the ODE \eqref{hasas} and 
$\bar{D}_{\bar{p}}(\bar{E}\,|\,\bar{\bm{w}})$
of the ``barred'' differential equation. For the excited states,  formulae \eqref{akskj21981298},\,
\eqref{askjjh2143}-\eqref{aklsjhj1298} involved in the definition of $D_p(E\,|\,\bm{w})$
 remain the same.\footnote{%
The additional term $t_1(z)$ in the ODE does not change the leading asymptotic behaviour
of the solutions $\psi_p$ and $\chi$ in the vicinity of the  singularities at $z=0$ and $z=\infty$.}  The
corresponding equations for $\bar{D}_{\bar{p}}(\bar{E}\,|\,\bar{\bm{w}})$ are only notationally different.
Similar to the case of the ${\cal Z}_2$ invariant spin chain studied in \cite{Bazhanov:2019xvyA} 
we expect that
\bea\label{iisis1a}
 &&\slim_{N\to\infty} G^{(N/r)}\big(E^r\,|\,\tfrac{r}{n+r}\big)\ 
A_+\Big(\, \big(N/(rN_0)\big)^{-\frac{2n}{r(n+r)}}
 E \,\Big)=D_p({ E}\,|\,\bm{ w})\,.
\eea
Here the function 
\bea\label{saysysa}
G^{(N)}(E\,|\,g)=\exp\Bigg(\
{\displaystyle \sum_{m=1}^{\big[\frac{1}{2(1-g)}\big]}}\
\dfrac{(-1)^{m}\,N}{2m\cos(\pi m g)}\  (N/N_0)^{2m(g-1)}\  E^{m}\Bigg)\,, \qquad \ \ \ \  g\ne 1-\tfrac{1}{2k}
\eea
(which is the same for all the states) has been introduced to ensure the existence of the limit.
Note that the result is valid for any $n>0$ provided that $\frac{r}{n}\ne 1,3,5,\ldots\ $. At these special values additional
terms $\propto \log(N)$ need to be included in the exponent in the definition of $G^{(N)}(E\,|\,g)$
for the limits \eqref{iisis1a} to exist, see ref.\cite{Bazhanov:2019xvyA} for details. Similarly, the scaling limit of the eigenvalues
of the operator $\bar{\mathbb{A}}_+(\zeta)$ \eqref{kjsajh1289} is described as
\be\label{iisis1aaa}
\slim_{N\to\infty} G^{(N/r)}\big(\bar{E}^r\,|\,\tfrac{r}{n+r}\big)\ 
\bar{A}_+\Big(\, \big(N/(rN_0)\big)^{\frac{2n}{r(n+r)}}
 \bar{E}^{-1} \,\Big)=\bar{D}_{\bar{p}}(\bar{ E}\,|\,\bar{\bm{ w}}) \,.
\ee

\bigskip

\begin{figure}
\begin{center}
\scalebox{0.95}{
\begin{tikzpicture}
\node at (-6.1,2.7) {$\big(\frac{N}{rN_0}\big)^{-\frac{2n}{r(n+r)}}\,\Re e\big(\bar{h}_N^{(2)}\big)$};
\node at (2.7,2.7) {$\big(\frac{N}{rN_0}\big)^{-\frac{4n}{r(n+r)}}\,\Im m\big(\bar{h}_N^{(2)}\big)$};
\node at (-4.3,0) {\includegraphics[width=7.25cm]{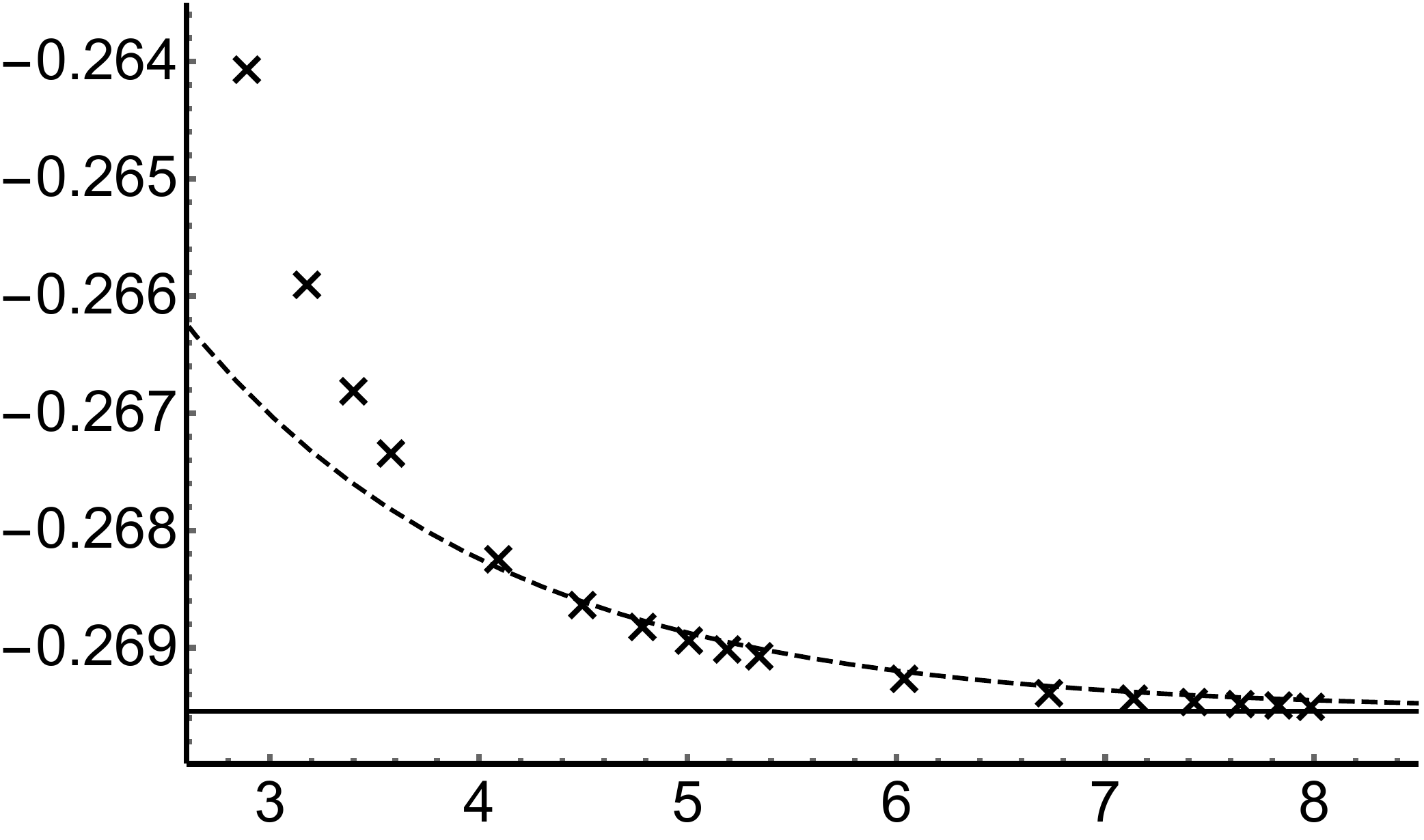}};
\node at (4.5,0) {\includegraphics[width=7.0cm]{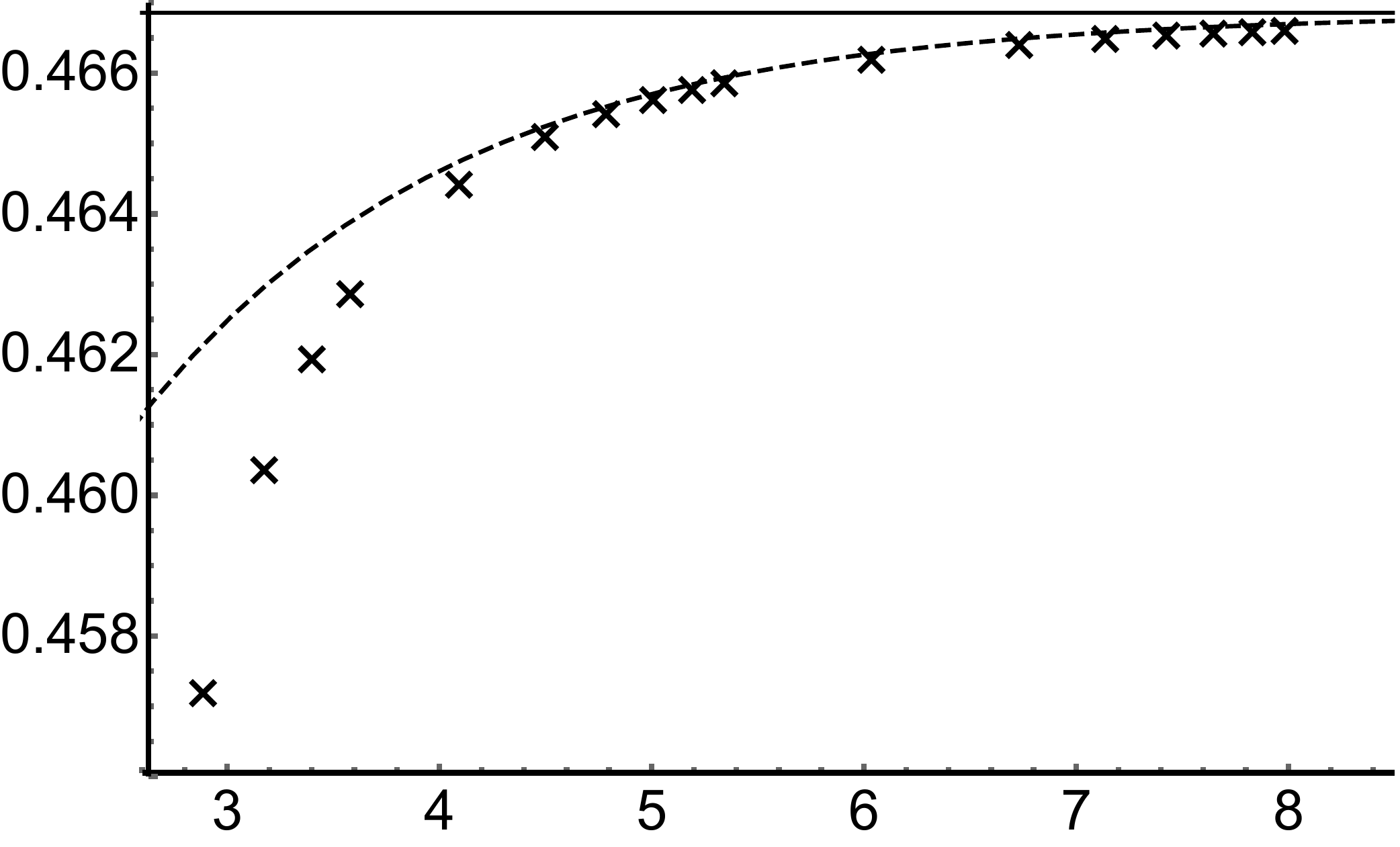}};
\node at (0,-1.75) {\small $\log(N)$};
\node at (8.7,-1.8) {\small $\log(N)$};
\end{tikzpicture}
}
\end{center}
\caption{\label{fig6} \small
Presented is numerical data for an
 RG trajectory of the ${\cal Z}_3$ invariant spin chain with $n=2$, ${\tt k}=\frac{1}{20}$,
characterized by $S^z=1$, ${\tt w}=0$ and levels $({\tt L},\bar{\tt L})=(0,1)$. 
The values of $s_1=-0.942033-1.631649\,\ri$ and $\bar{s}_1=0.085024 - 0.147266\,\ri$
were extracted by means of \eqref{asnb2vb} using a certain
interpolation procedure to take the limit $N\to\infty$.
The  crosses
stand for the real and imaginary parts of 
$\big(\frac{N}{rN_0}\big)^{-\frac{2n}{r(n+r)}}\,\bar{h}_N^{(2)}$, where $\bar{h}_N^{(2)}$
denotes the sum over the Bethe roots \eqref{mnbnw8129vvv}. The solid  lines represent
the prediction of the barred counterpart to the scaling formula \eqref{askjh12}, namely,
$J_2(\bar{p},\bar{\bm{s}}\,|\,\bar{\bm{w}})=-0.269543 + 0.466857\,\ri$. The latter
was obtained via a numerical
integration of the ODE \eqref{hasas},\,\eqref{n1n9832hjds} with the parameters replaced as
$p\mapsto \bar{p}=\frac{3}{8}$,\, ${\tt L}\mapsto \bar{\tt L}=1$, 
$s_1\mapsto \bar{s}_1=-0.942033-1.631649\,\ri$
and $\{w_\alpha\}\mapsto \{\bar{w}_\alpha\}$ with $\bar{w}_1= 0.778812 - 1.348942\,\ri$, therein.
The dashed lines depicts the fit $-0.269539 + 0.466854\,\ri + (0.0186614 - 0.0323224\,\ri)\,N^{-\frac{2}{3}}$.
}
\end{figure}
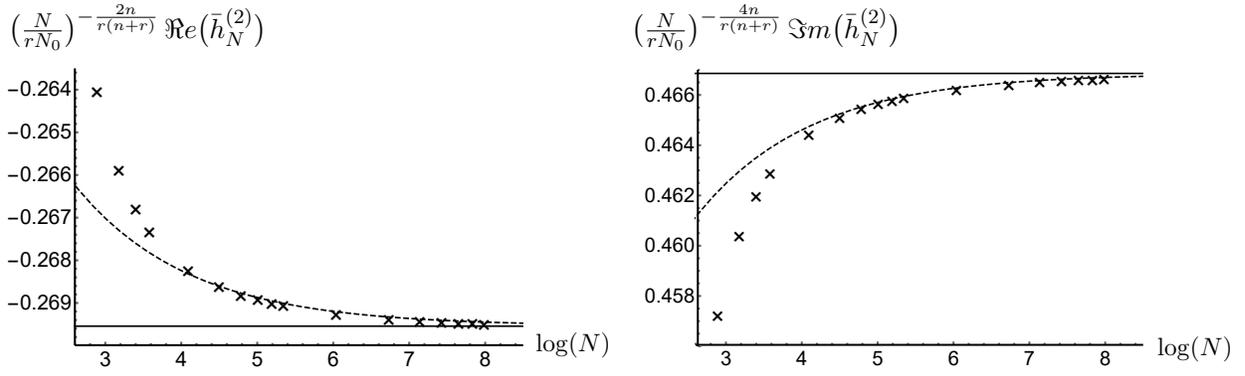

A verification of the conjectured relations \eqref{iisis1a} and \eqref{iisis1aaa} 
 may be carried out along the same lines as for the primary Bethe states. 
The scaling relations \eqref{askjh12} for the sums $h^{(j)}_N$ \eqref{mnbnw8129} and their barred
versions still hold true provided  the $J_j$ in the r.h.s. are taken to be the
 expansion coefficients \eqref{asnbv21} of the spectral
determinant for eq.\eqref{hasas}.
Apart from $p$ and $s$, they depend now on
the set $\{w_\alpha\}_{\alpha=1}^{\tt L}$, which solves the algebraic system \eqref{aosaasiBBB},\,\eqref{aosaasiCCC}, i.e.,
\be
J_j=J_j(p,\bm{s}\,|\,\bm{w})\,,\qquad\qquad\bar{J}_j=J_j(\bar{p},\bar{\bm{s}}\,|\,\bar{\bm{w}})\, .
\ee
For the case of the excited states  no explicit analytical expressions are available for the expansion coefficients
$J_j$. Nevertheless, they may be obtained via a numerical integration of the differential equations. Together
with the primary states,  a verification of \eqref{iisis1a}-\eqref{iisis1aaa} for the excited states
was performed, see fig.\,\ref{fig6}. We also analyzed the algebraic system \eqref{aosaasiBBB},\,\eqref{aosaasiCCC}.
It was observed that for given ${\tt L}$ and generic values of the parameters $n$, $p$ and $\bm{s}$ 
the number of solutions is equal to the $r$ colored partitions of ${\tt L}$:
\be
\sum_{{\tt L}=0}^{\infty}{\rm par}_r({\tt L})\,{ \tt q}^{\tt L}=\prod_{m=1}^\infty\frac{1}{(1-{\tt q}^m)^r}=
1+r{\tt q}+\frac{1}{2}\,r\,(r+3)\,{\tt q}^2+\frac{1}{6}\,r\,(r+1)(r+8)\,{\tt q}^3+\ldots\ .
\ee

\subsection{Product rules}

Being an entire function of $E$,  the spectral determinant  admits a convergent series expansion
near zero.  At infinity, it possesses an asymptotic expansion, whose description 
requires one to  identify the Stokes lines, i.e., the lines which split the complex $E$ plane into domains where the
asymptotic behaviour is described differently. In the case at hand, the Stokes lines
are the lines of accumulation of zeroes. 
Their location  may be deduced from formula \eqref{iisis1a}
for $D_p(E\,|\,\bm{w})$ and \eqref{iisis1aaa} for $\bar{D}_{\bar{p}}(\bar{E}\,|\,\bar{\bm{w}})$
from the  knowledge of the pattern of Bethe roots 
for the low energy Bethe states
of the ${\cal Z}_r$ invariant spin chain.
According to  \eqref{asjjh1hg}, the 
Bethe roots 
 accumulate along   lines parallel to the real axis
 in the complex $\beta$ plane with $\beta=-\frac{1}{2}\log(\zeta)$. 
At the edges of the distribution, the roots develop a scaling behaviour.
Namely, ordering $\{\zeta_m\}_{m=1}^M$ w.r.t.  their absolute value,
$$|\zeta_1|\le|\zeta_2|\le\ldots\le|\zeta_{M-1}|\le |\zeta_{M}|\,,$$
 formulae
\eqref{iisis1a} and \eqref{iisis1aaa} imply the existence of the limits
\be
\slim_{N\to\infty\atop j\ {\rm fixed}} \ (N/(rN_0))^{\frac{2n}{r(n+r)}}\,\zeta_j=E_j\, ,
\qquad\qquad
\slim_{N\to\infty\atop M-j\ {\rm fixed}} \ (N/(rN_0))^{\frac{2n}{r(n+r)}}\,\zeta_{M-j}^{-1}=\bar{E}_j\, .
\ee
Evidently, $E_j$  coincide with the zeroes of the spectral determinant
$D_p(E\,|\,\bm{w})$ and accumulate along the rays
\bea\label{askb21j}
\arg(E)=0,\, \frac{2\pi}{r},\, \frac{4\pi}{r},\ldots\,, \frac{2\pi(r-1)}{r}\qquad\qquad \qquad ({\rm mod}\,2\pi)
\eea  
and similarly for $\bar{E}_j$.
\bigskip

The rays \eqref{askb21j} divide the complex plane into wedges, in   which the
spectral determinants exhibit a different large $E$ ($\bar{E}$)
 asymptotic behaviour.
To write down the asymptotic formulae for the different wedges, 
which will be labeled by $\ell=1,2,\ldots, r$, we
 swap $E$ and $\bar{E}$ for $\theta$ as
\bea\label{askb1287ashhg}
E=(-1)^{r-1}\ \re^{+\frac{\ri\pi }{r}(2\ell-1)} \ \re^{\frac{2n\theta}{r(n+r)}}\,,\qquad\qquad
\bar{E}=(-1)^{r-1}\ \re^{-\frac{\ri\pi }{r}(2\ell-1)} \ \re^{\frac{2n\theta}{r(n+r)}}\, .
\eea
Then, from an analysis of the ODE \eqref{hasas}
and its barred counterpart,
one can show that as 
 \bea
\Re e(\theta)\to+\infty\ \ \ \ \qquad {\rm and}\ \ \ \ \qquad |\Im m(\theta)|<\frac{\pi(n+r)}{2n}
\eea
the following asymptotics hold true
\bea\label{askj38921}
D_p(E\,|\,\bm{w})&=& {\mathfrak C}^{(\ell)}_{p,{\boldsymbol s}}({\boldsymbol w})\ \exp\bigg[\, \bigg(\,
(-1)^{\ell -1}\, \ri s
 -\frac{2 n p}{n+r}\bigg)
\, \frac{\theta}{r}+
\frac{N_0}{\cos(\frac{\pi r}{2n})}\ \re^\theta+O\Big(\re^{-\frac{\theta}{r}}\,\Big)\,\bigg]\nonumber\\[-0.2cm]
\\[-0.2cm]
\bar{D}_{\bar{p}}(\bar{E}\,|\,\bar{\bm{w}})&=& 
\bar{\mathfrak C}^{(\ell)}_{\bar{p},\bar{\boldsymbol s}}(\bar{\boldsymbol w})\ \exp\bigg[\, \bigg(\,
(-1)^{\ell -1}\, \ri s
 -\frac{2 n \bar{p}}{n+r}\bigg)
\, \frac{\theta}{r}+
\frac{N_0}{\cos(\frac{\pi r}{2n})}\ \re^\theta+O\Big(\re^{-\frac{\theta}{r}}\,\Big)\,\bigg]\, .\nonumber
\eea
Here, by definition,
\bea
{s}\equiv \begin{cases}
\,0\ \ \ \ &{\rm for\ \ odd}\ \ r\\
\, {s}_{\frac{r}{2}}\ \ \ \ &{\rm for\ \ even}\ \ r
 \end{cases}
\eea
and the numerical constant $N_0$  is given by \eqref{iaisissu}. For $r=1,2$ 
the coefficients ${\mathfrak C}^{(\ell)}_{p,{\boldsymbol s}}({\boldsymbol w})$
and $\bar{\mathfrak C}^{(\ell)}_{\bar{p},\bar{\boldsymbol s}}(\bar{\boldsymbol w})$
are available in closed analytic form \cite{Kotousov:2019nvt}. 
An efficient numerical procedure for their
computation for any $r$ is described in
 Appendix  \ref{AppC}.
\begin{figure}
\begin{center}
\scalebox{0.95}{
\begin{tikzpicture}
\node at (-7.3,2.6) {$R$};
\node at (-7.3,-2.9) {$R$};
\node at (-7.3,-8.4) {$R$};
\node at (1.5,2.6) {$R$};
\node at (1.5,-2.9) {$R$};
\node at (1.5,-8.4) {$R$};
\node at (-4.3,0) {\includegraphics[width = 7cm]{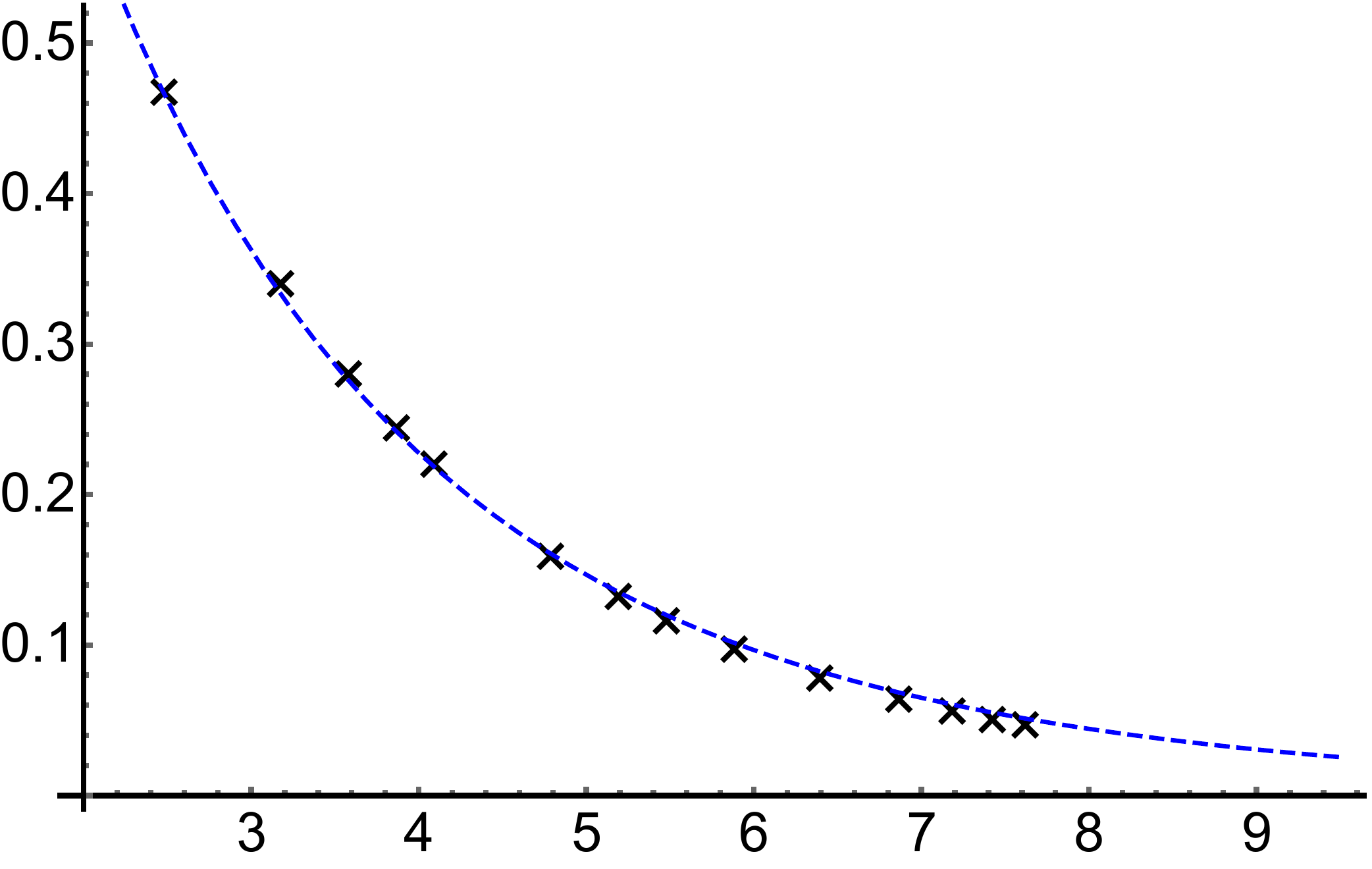}};
\node at (4.5,0) {\includegraphics[width = 7cm]{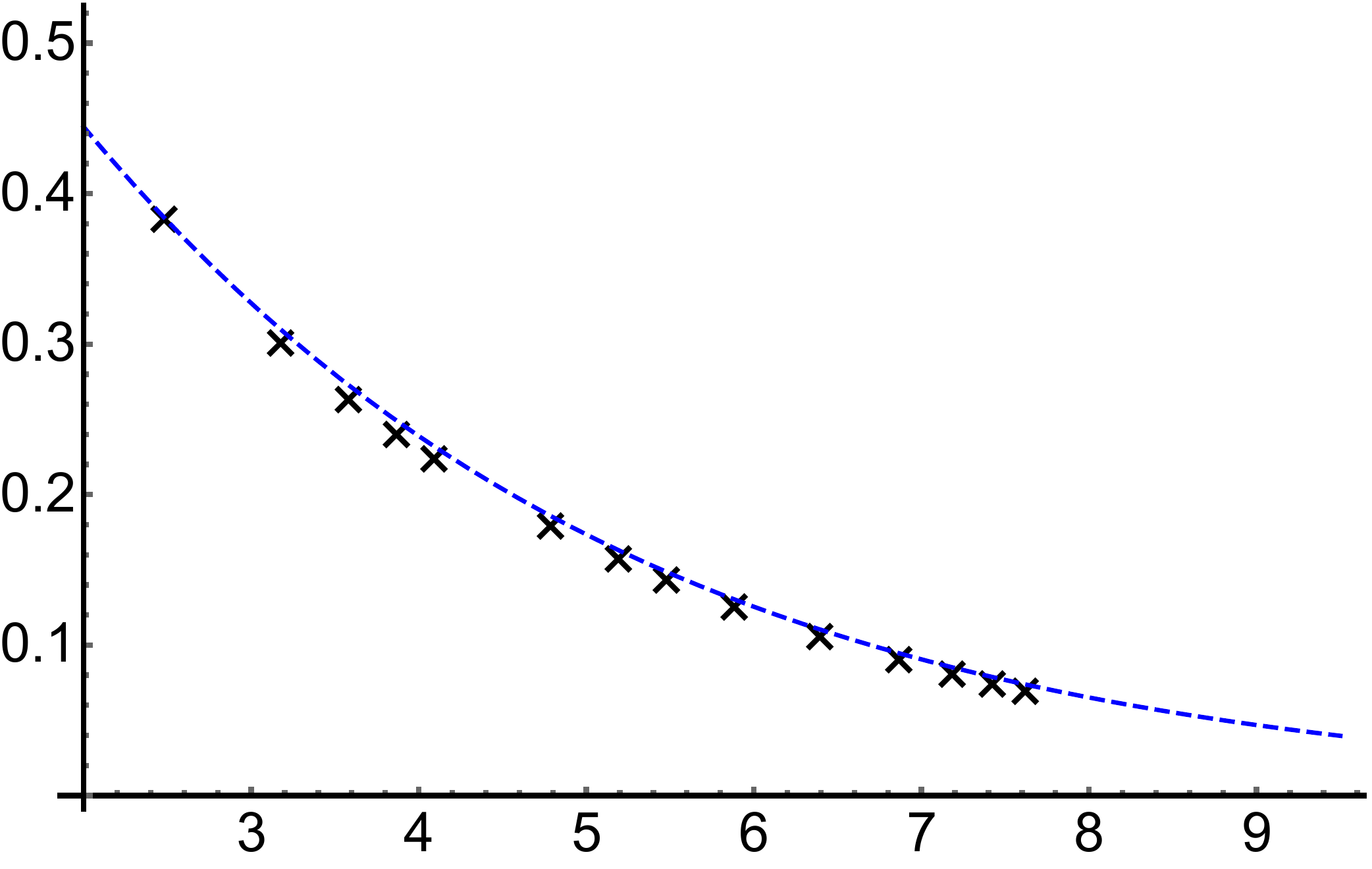}};
\node at (-4.3,-5.5) {\includegraphics[width = 7cm]{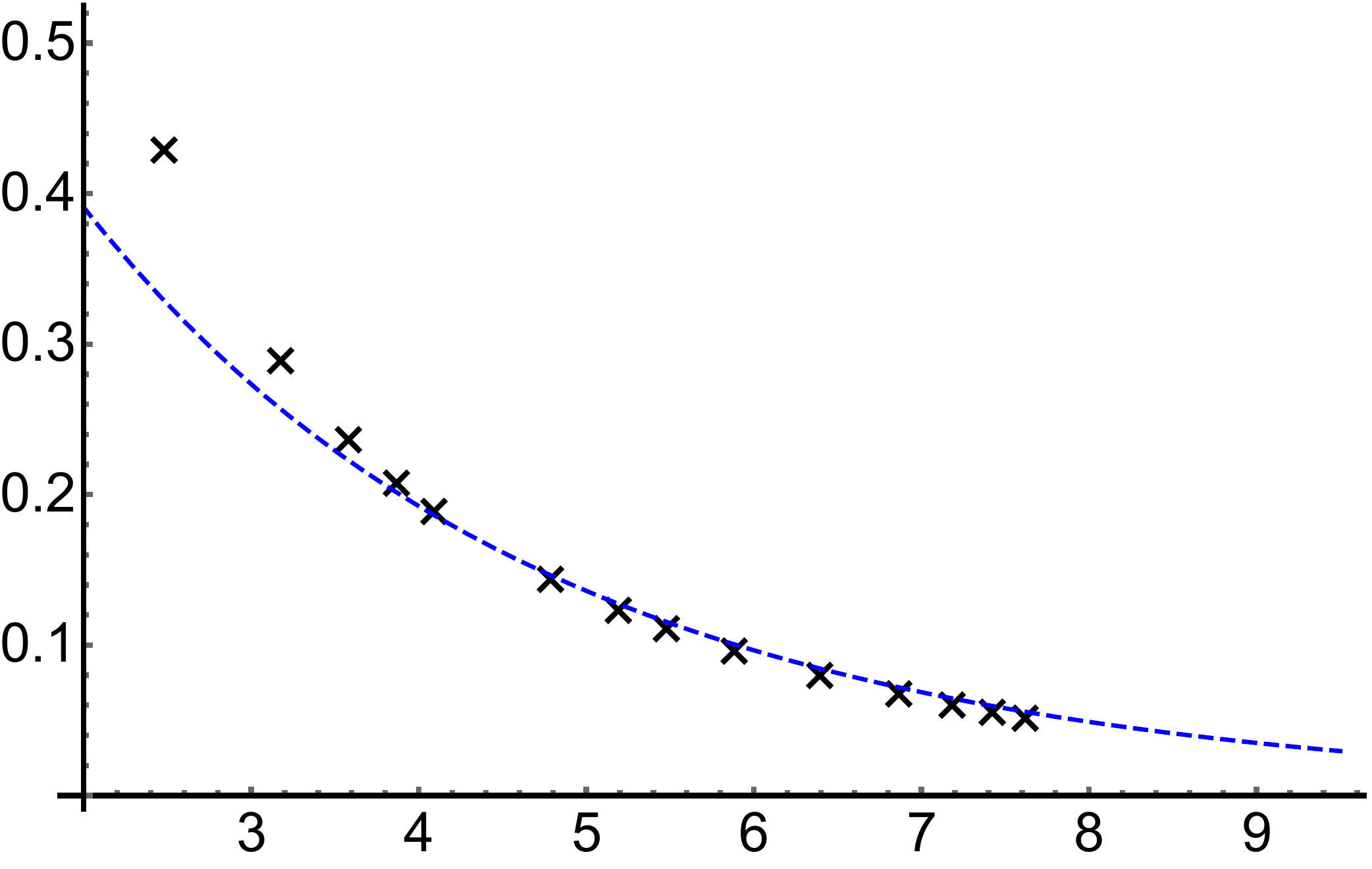}};
\node at (4.5,-5.5) {\includegraphics[width = 7cm]{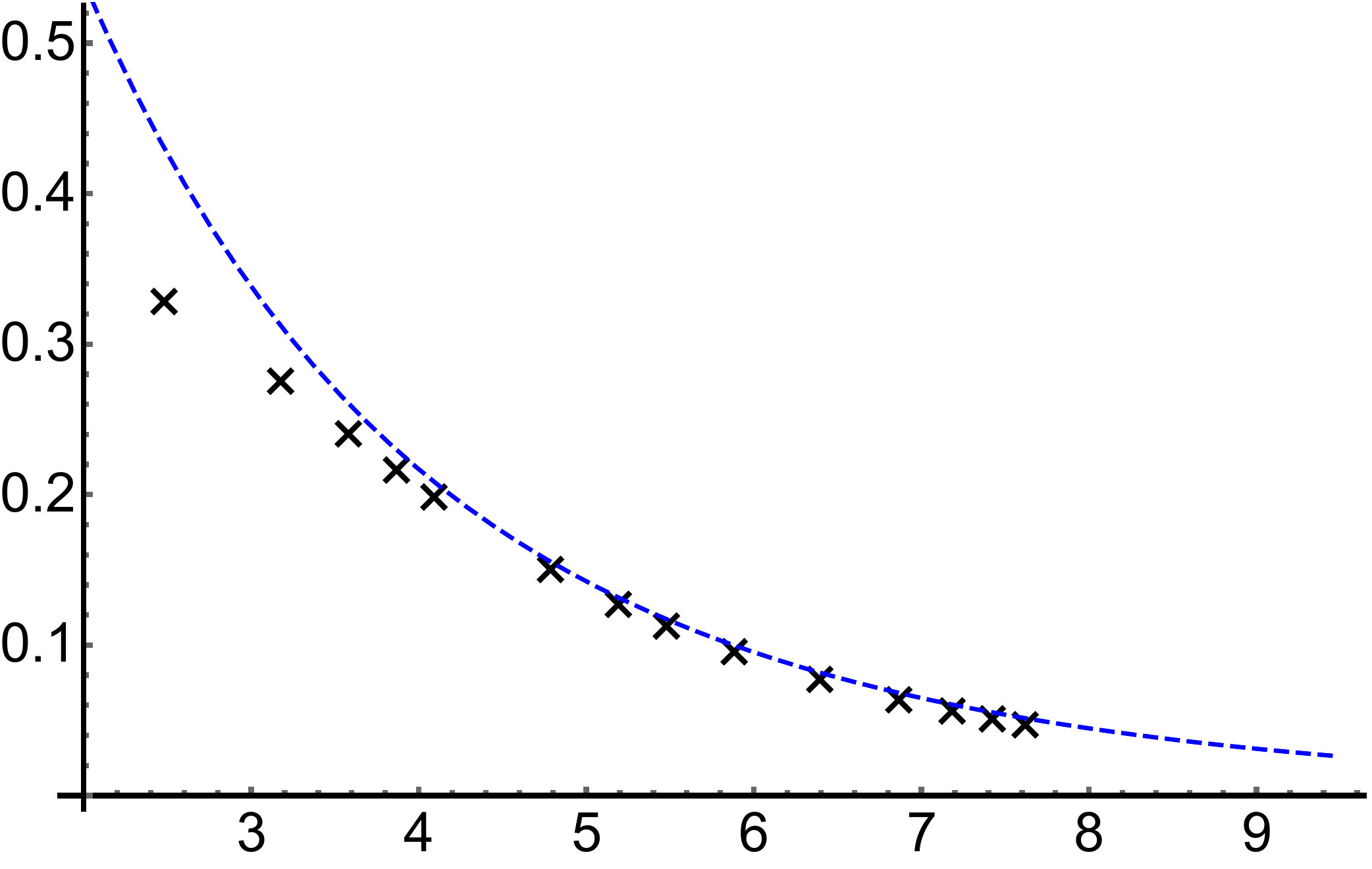}};
\node at (-4.3,-11) {\includegraphics[width = 7cm]{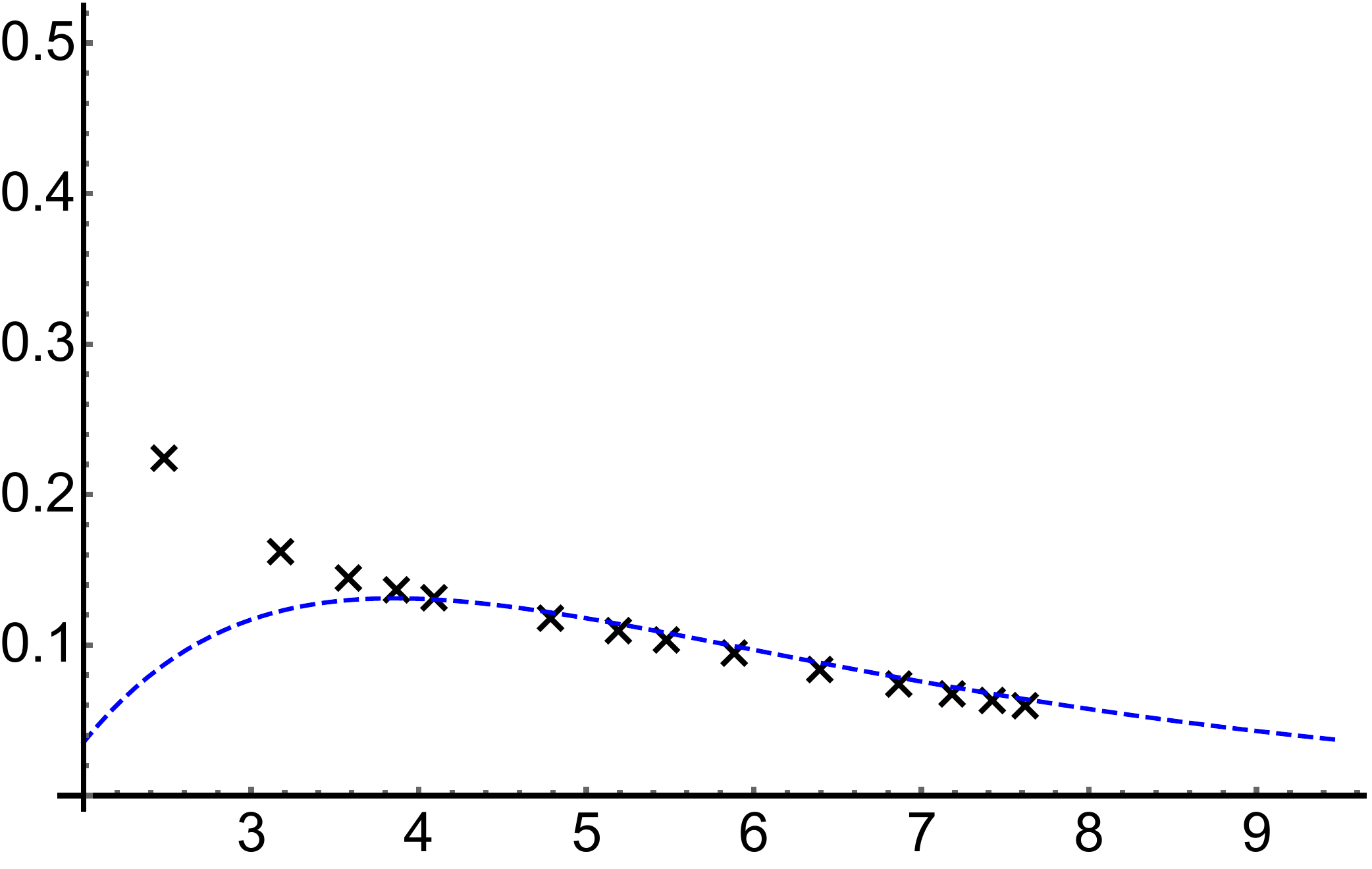}};
\node at (4.5,-11) {\includegraphics[width = 7cm]{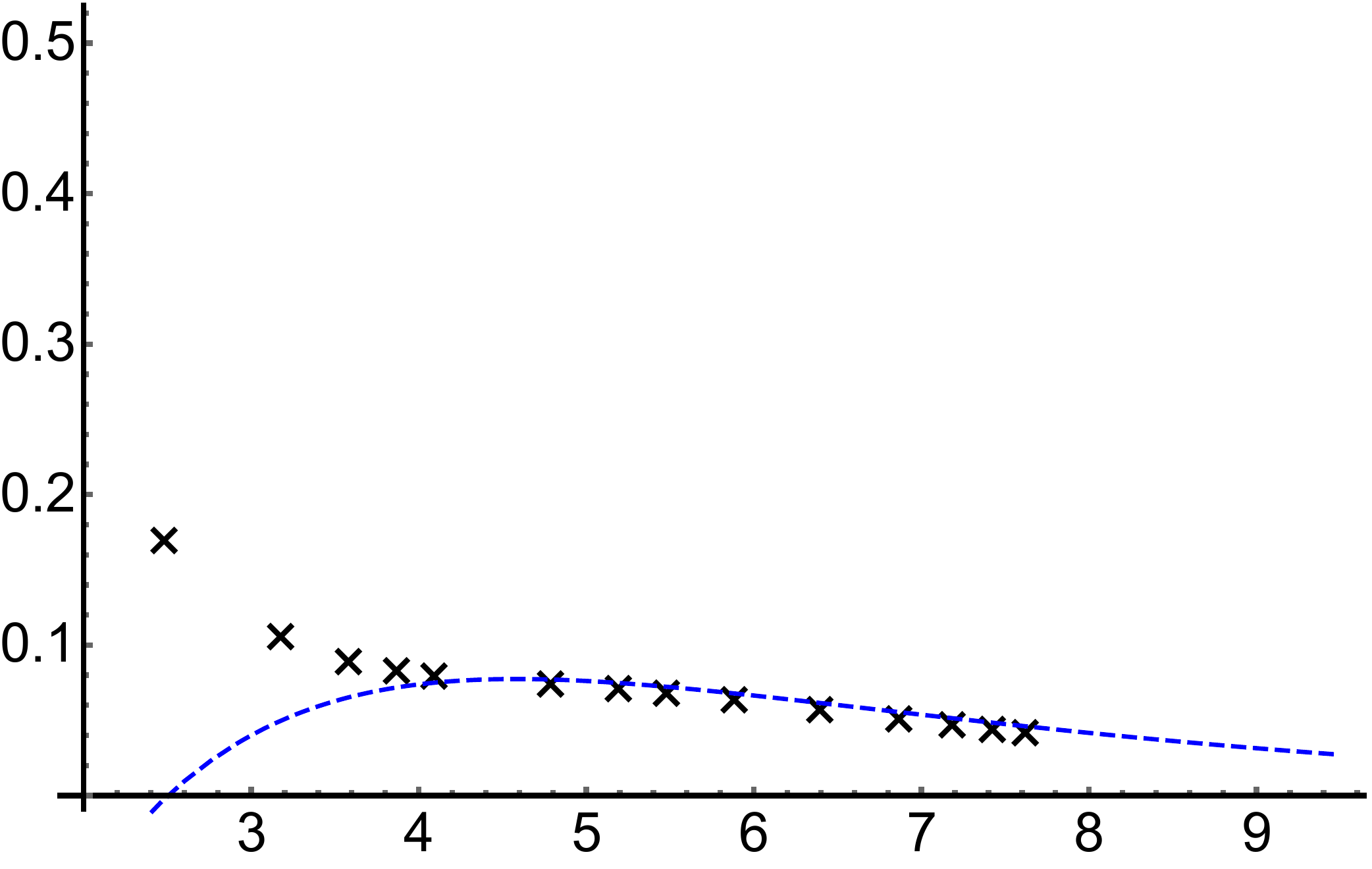}};
\node at (-0.1,-1.8) {\small $\log(N)$};
\node at (8.7,-1.8) {\small $\log(N)$};
\node at (-0.1,-7.3) {\small $\log(N)$};
\node at (8.7,-7.3) {\small $\log(N)$};
\node at (-0.1,-12.8) {\small $\log(N)$};
\node at (8.7,-12.8) {\small $\log(N)$};
\node at (-4.3,2.1) {$\ell=\ell'=1$};
\node at (-4.3,-3.4) {$\ell=1,\,\ell'=3$};
\node at (-4.3,-8.9) {$\ell=1,\,\ell'=5$};
\node at (4.5,2.1) {$\ell=1,\,\ell'=2$};
\node at (4.5,-3.4) {$\ell=1,\,\ell'=4$};
\node at (4.5,-8.9) {$\ell=1,\,\ell'=6$};
\end{tikzpicture}
}
\end{center}
\caption{\label{fig7}\small%
Depicted is  the absolute value of the logarithm of the ratio
of the l.h.s. to the r.h.s. of the relation \eqref{prod1a}, $R=|\log({\rm l.h.s.\ of\ \eqref{prod1a}}/{\rm r.h.s.\ of\ \eqref{prod1a}})|$,  for $\ell=1$  and  $\ell'=1,2,\ldots,6$.
This was computed for an RG trajectory with ${\tt w}={\tt L}=\bar{\tt L}=0$ for the ${\cal Z}_r$ invariant spin chain with $r=6$, $n=3.5$ and ${\tt k}=0.1$
in the sector $S^z=0$.
The l.h.s. is obtained from the Bethe roots corresponding to the trajectory.
For the r.h.s. 
 $\bm{s}=(s_1,s_2,s_3)$ and $\bar{\bm{s}}=(\bar{s}_1,\bar{s}_2,\bar{s}_3)$ 
 were swapped in favour of the ``running couplings'' 
$s_a\mapsto C_a\,b_a(N)$ and $\bar{s}_a\mapsto -C_a\,b_{-a}(N)$ in accordance with eq.\,\eqref{asnb2vb}. 
The $b_a(N)$ themselves are extracted from the eigenvalues of the quasi-shift operators via formulae 
\eqref{ajs12v}  and \eqref{ajs12vAA}. Also, in the computation of the r.h.s., the correction term $O(N^{-\epsilon})$ was ignored.
The dashed lines  are  linear-log plots
of $c_1\,N^{-\frac{1}{3}}+c_2\,N^{-\frac{2}{3}}$, where for each value of $\ell'$ the coefficients 
$c_1$ and $c_2$
were determined  via a fit.
}
\end{figure}

\bigskip

Focusing on the cases with $r=1$ and $r=2$, it was observed in refs.\cite{Kotousov:2019ygw} and \cite{Bazhanov:2019xvyA}, respectively, that 
${\mathfrak C}^{(\ell)}_{p,{\boldsymbol s}}({\boldsymbol w})$ and $\bar{{\mathfrak C}}^{(\ell)}_{\bar{p},\bar{\boldsymbol s}}(\bar{\boldsymbol w})$ 
appear in the scaling limit of certain products over the Bethe roots.
We found that the similar relations hold true for $r\ge 3$, namely:
\bea\label{prod1a}
\prod_{m=1}^Mq\,\big(\zeta_m+\eta_\ell\, q^{-1}\big)\big(\zeta^{-1}_m+\eta_{\ell'}^{-1} \, q^{-1}\big)&=&
 \re^{ 
((-1)^{\ell'}-(-1)^\ell) \frac{\pi s}{2n}}\ 
{\mathfrak C}^{(\ell)}_{p,{\boldsymbol s}}({\boldsymbol w})\,
{\bar  {\mathfrak C}}^{(\ell')}_{{\bar p},{\bar {\boldsymbol s}}}({\bar {\boldsymbol w}})\,\\
&\times& 
\bigg(\frac{N}{rN_0}\bigg)^{ 
\ri\,((-1)^{\ell'}-(-1)^\ell) \frac{s}{r}
 -\frac{ 2n (p+{\bar p})}{r(n+r)}}\ \bigg(\frac{4n}{n+r}\bigg)^{\frac{N}{r}}\ 
 \, \Big(1+O\big(N^{-\epsilon}\big)\Big)
 \nonumber
\eea
Here the remainder terms fall off as a power of $N$ with some positive exponent $\epsilon$
and $\eta_\ell$ stand for the inhomogeneities as in eq.\,\eqref{zsym1}, i.e.,
\bea
\eta_\ell=(-1)^{r}\ \re^{\frac{\ri\pi }{r}(2\ell-1)}\qquad\qquad\qquad (\ell=1,2,\ldots, r)\, .
\eea
In addition to the relations \eqref{prod1a},  we also established that
\bea\label{prod1b}
\prod_{m=1}^M(\zeta_m)^r=
\bigg(\frac{N}{rN_0}\bigg)^{ 
 \frac{2 n (p-{\bar p})}{n+r}}\ 
\prod_{\ell=1}^r\frac{{\bar  {\mathfrak C}}^{(\ell)}_{{\bar p},{\bar {\boldsymbol s}}}({\bar {\boldsymbol w}})}
{{\mathfrak C}^{(\ell)}_{p,{\boldsymbol s}}({\boldsymbol w})}\ \  \Big(1+O\big(N^{-\epsilon}\big)\Big)\, .
\eea
By taking the product of the l.h.s. of eq.\,\eqref{prod1a} over $\ell=\ell'=1,2,\ldots,r$ and dividing/multiplying
the result by $\prod_{m=1}^M(\zeta_m)^{r}$ one obtains 
\bea
 \prod_{m=1}^M\big(\zeta^{-r}_m+q^{r}\big)\big(\zeta^{-r}_m+q^{-r}\big)&=&
\prod_{\ell=1}^r\Big({\mathfrak C}^{(\ell)}_{p,{\boldsymbol s}}({ {\boldsymbol w}})\Big)^2\ \  \bigg(\frac{N}{rN_0}\bigg)^{
 -\frac{ 4n p}{n+r}}\ 
 \bigg(\frac{4n}{n+r}\bigg)^{N}\ 
 \, \Big(1+O\big(N^{-\epsilon}\big)\Big) \nonumber \\[-0.2cm]
&& \\[-0.2cm]
\prod_{m=1}^M\big(\zeta^{+r}_m+q^{r}\big)\big(\zeta^{+r}_m+q^{-r}\big)&=&
\prod_{\ell=1}^r\Big({\bar {\mathfrak C}}^{(\ell)}_{{\bar p},{\bar {\boldsymbol s}}}({\bar {\boldsymbol w}})\Big)^2\ \  \bigg(\frac{N}{rN_0}\bigg)^{
 -\frac{ 4n {\bar p}}{n+r}}\ 
 \bigg(\frac{4n}{n+r}\bigg)^{N}\ 
 \, \Big(1+O\big(N^{-\epsilon}\big)\Big)\, .\nonumber
\eea
The advantage of the last two product rules 
is that they involve  the 
asymptotic coefficients for the spectral determinant of the ODE \eqref{hasas} and its barred variant separately.
\medskip

The above product rules provide additional support for the proposal for the scaling limit
of the  eigenvalues of the
$Q$-operator \eqref{iisis1a} and \eqref{iisis1aaa}.
We performed numerical checks for a range of cases with some  of the data being presented in fig.\,\ref{fig7}.

\section{Quantization condition}

A complete description of the conformal field theory underlying the critical behaviour of the inhomogeneous
six-vertex model in the regime $q=\re^{\frac{\ri\pi}{n+r}}$ with $n>0$ 
is a difficult task. The proposed ODEs provide 
some guidance that allows one to make progress in this problem. Our analysis suggests that
under a suitable normalization,
the low energy Bethe states $|\Psi_N\rangle$ for the ${\cal Z}_r$ invariant spin chain
 possess a scaling limit such that
\be\label{kjasnaaab3}
\slim_{N\to\infty}|\Psi_N\rangle=
|\bm{\psi}_{p,\bm{s}}(\bm{w})\rangle\otimes\overline{|{\bm{\psi}}_{\bar{p},\bar{\bm{s}}}(\bar{\bm{w}})\rangle}\  .
\ee
We expect that the states $|\bm{\psi}_{p,\bm{s}}(\bm{w})\rangle$ 
organize into highest weight representations of
a certain chiral algebra of extended conformal symmetry and similar for 
$\overline{|{\bm{\psi}}_{\bar{p},\bar{\bm{s}}}(\bar{\bm{w}})\rangle}$. 
In particular, the  chiral 
components that appear in the scaling limit of the primary Bethe states would be the highest states  in
such representations. 
Immediate questions arise as to the identification of the algebra of extended conformal
symmetry and under
what  conditions on
$\bm{s}$, $\bm{w}$  and $\bm{s}',\bm{w}'$ do the states $|\bm{\psi}_{p,\bm{s}}(\bm{w})\rangle$ 
and $|\bm{\psi}_{p,\bm{s}'}(\bm{w}')\rangle$ 
belong to the same chiral module. Moreover, what are the selection rules 
 for the admissible values of $\bm{s}$ and $\bar{\bm{s}}$ for
the states \eqref{kjasnaaab3}.
All these points  for the most part remain open to us.  
Here we present some analysis concerning the selection rules.

\subsection{General quantization condition}
The relations \eqref{prod1a} yield an important consequence.
It is obtained by making the   specializations $(\ell,\ell')\mapsto (\ell+1,\ell)$ and  $(\ell,\ell')\mapsto (\ell,\ell+1)$
therein, where $\ell$ is taken to be modulo $r$,
  and considering the ratio. The l.h.s. of the result is expressed in terms of the eigenvalues ${\cal K}^{(\ell)}$ and ${\cal K}^{(\ell+1)}$ \eqref{asj21gh} of
the quasi-shift operators.
Keeping in mind that $\log\big((-1)^{\tt w}{\cal K}^{(\ell)}\big)=O(N^{-\frac{1}{r}})$ for $r$ odd and 
 $\log\big((-1)^{\tt w}{\cal K}^{(\ell)}\big)=(-1)^\ell\, \frac{\pi}{n}\,s+O(N^{-\frac{1}{r}})$ 
for  $r$ even, one finds
\be\label{qunat1A}
\bigg(\frac{  N}{rN_0}\bigg)^{\frac{4\ri}{r}\,(-1)^\ell\, { s}}\ \ 
\frac{
\mathfrak{C}_{p,\bm{s}}^{(\ell+1)}({ {\boldsymbol w}})}{\mathfrak{C}_{p,\bm{s}}^{(\ell)}({{\boldsymbol w}})}\,
\frac{\bar{\mathfrak{C}}_{{\bar p},\bar{\bm{s}}}^{(\ell)}({\bar {\boldsymbol w}})}{\bar{\mathfrak{C}}_{{\bar p},\bar{\bm{s}}}^{(\ell+1)}({\bar {\boldsymbol w}})}= 
(-1)^\sigma\,\re^{-\frac{2\pi\ri}{r}\,S^z}
\Big(1
+O\big(N^{-\epsilon}\,\big)\Big)\qquad\qquad (\ell\sim\ell+r)\, ,
\ee
where
\be
\sigma=\begin{cases} 0 & {\rm for}\ \ \frac{N}{r}\ {\rm even} \\[0.2cm]
 1 & {\rm for}\ \ \frac{N}{r}\ {\rm odd}
\end{cases}\ . 
\ee
Recall that we always assume  $N$ is divisible by $r$ and, in constructing an RG trajectory $|\Psi_N\rangle$, the parity 
of $N/r$ must be kept fixed.
Thus we conclude that the parameters $\bm{s}$, $\bar{\bm{s}}$ characterizing a low energy Bethe state
are not independent, but obey  $r-1$  conditions
 (the product of \eqref{qunat1A} over $\ell=1,2,\ldots,r$ is satisfied identically).
\medskip

For the case $r=2$, a closed analytic formula exists for the asymptotic
coefficients $\mathfrak{C}_{p,\bm{s}}^{(\ell)}({ {\boldsymbol w}})$ and 
$\bar{\mathfrak{C}}_{{\bar p},\bar{\bm{s}}}^{(\ell)}({\bar {\boldsymbol w}})$, see 
sec.\,3 of ref.\cite{Kotousov:2019nvt}.
Through the study of \eqref{qunat1A} it was found that the 
 space of states of the   conformal field theory underlying the ${\cal Z}_2$
invariant spin chain possesses both a continuous component with $s$ taking any real values
as well as a discrete one, where $s$ belongs to a discrete set of pure imaginary numbers.
When $r$ is odd, $s$ is by definition zero so that the $N$ dependent factor in the l.h.s. of the
above relations is absent. Then, if $p$ and $\bar{p}$ are given, 
  \eqref{qunat1A} constitute $r-1$ independent complex conditions
imposed on the $r-1$ complex numbers $(s_1,\ldots, s_{\frac{r-1}{2}})$ and
 $(\bar{s}_1,\ldots, \bar{s}_{\frac{r-1}{2}})$. They thus play the r\^{o}le of 
a ``quantization condition'' and are expected to yield a discrete set of admissible values for 
 $\bm{s}$ and $\bar{\bm{s}}$. For even $r$, the
 $N$ dependent factor is present in the relations \eqref{qunat1A}. As in the case $r=2$ we expect that
 together with a discrete set of values,
 $s= s_{\frac{r}{2}}\equiv\bar{s}_{\frac{r}{2}}$ may take any real values leading to a continuous
 component in the spectrum. The other $r-2$ complex numbers $(s_1,\ldots s_{\frac{r}{2}-1})$
 and $(\bar{s}_1,\ldots \bar{s}_{\frac{r}{2}-1})$  belong to a certain  discrete set, similar to the
 case of odd $r$. Unfortunately, a comprehensive analysis of the quantization condition \eqref{qunat1A} is hampered by the fact that an
 analytic expression for the asymptotic coefficients is absent. Nevertheless,
 by   computing $\mathfrak{C}_{p,\bm{s}}^{(\ell)}({{\boldsymbol w}})$
 and $\bar{\mathfrak{C}}_{{\bar p},\bar{\bm{s}}}^{(\ell)}({\bar {\boldsymbol w}})$ via the 
 procedure explained in Appendix \ref{AppC}, some numerical checks  were carried out, see 
tab.\,\ref{tab1}.
\begin{table}
\begin{center}
\scalebox{0.845}{
\begin{tabular}{|l|c|c|l|l|l|l|l|}
\cline{4-7}
\multicolumn{3}{c|}{}  & \multicolumn{4}{c|}{} \\[-0.4cm]
\multicolumn{3}{c|}{} & \multicolumn{4}{c|}{$|$l.h.s.\,$-$\ r.h.s.$|$ of relation \eqref{qunat1A}} \\[0.0cm]
\hline
& &  & & & & \\[-0.3cm]
$({\tt L},\bar{\tt L})$ & $\bm{s}=(s_1,s_2),\,
\bm{w}=\{w_\alpha\}_{\alpha=1}^{\tt L}$ & $\bar{\bm{s}}=(\bar{s}_1,\bar{s}_2),\,
\bar{\bm{w}}=\{\bar{w}_\alpha\}_{\alpha=1}^{\bar{{\tt L}}}$ &  $\ell=1$ & $\ell=2$ & $\ell=3$ & $\ell=4$  \\[0.1cm]
\hline
&  & & & & & \\[-0.4cm]
\multirow{2}{*}{$(0,0)$} &$s_1=-0.5482 + 1.6871\, \ri$ &
$\bar{s}_1=-0.4520 - 1.3912\,\ri$ & \multirow{2}{*}{$1.6\times 10^{-4}$}
 &\multirow{2}{*}{$4.6\times 10^{-5}$} &\multirow{2}{*}{$1.8\times 10^{-4}$}
 &\multirow{2}{*}{$1.8\times 10^{-4}$} \\[0.0cm]
& & & & & & \\[-0.4cm]
& $s_2=+0.8181 + 0.5943\,\ri $ &$\bar{s}_2=+0.8017 - 0.5825\,\ri$ & & & & \\[0.0cm]
\hline
&  & & & & & \\[-0.4cm]
\multirow{2}{*}{$(0,1)$} &$s_1=+0.0974 + 1.8226\,\ri $ &
$\bar{s}_1=-2.9344 - 3.7480\,\ri$ & \multirow{3}{*}{$3.1\times 10^{-4}$} &\multirow{3}{*}{$3.9\times 10^{-4}$} &\multirow{3}{*}{$6.9\times 10^{-4}$} &\multirow{3}{*}{$6.8\times 10^{-4}$} \\[0.0cm]
& & & & & & \\[-0.4cm]
& $s_2=+0.3824 + 0.9444\,\ri$ &$\bar{s}_2=+0.7373 + 0.2855\,\ri$ & & & & \\[0.0cm]
& & & & & & \\[-0.4cm]
& & $\bar{w}_1=+1.2595 - 0.9676\,\ri$ & & & & \\[0.0cm]
\hline
&  & & & & & \\[-0.4cm]
\multirow{2}{*}{$(1,0)$} &$s_1=+0.2683 - 0.8257\,\ri$ &
$\bar{s}_1=-0.5075-1.5618\,\ri$ &
 \multirow{3}{*}{$5.2\times 10^{-4}$} &\multirow{3}{*}{$4.4\times 10^{-4}$} &\multirow{3}{*}{$3.9\times 10^{-4}$} &\multirow{3}{*}{$3.9\times 10^{-4}$} \\[0.0cm]
& & & & & & \\[-0.4cm]
& $s_2=+0.4242 + 0.3082\,\ri$ &$\bar{s}_2=+0.9775 -0.7102\,\ri$ &
 & & & \\[0.0cm]
& & & & & & \\[-0.4cm]
& ${w}_1=-0.6077 - 1.8703\,\ri$  & & & & & \\[0.0cm]
\hline
&  & & & & & \\[-0.4cm]
\multirow{4}{*}{$(2,1)$} &$s_1=-4.0464-1.2882\,\ri$ &
$\bar{s}_1=-2.6122-4.0725\,\ri$ & \multirow{4}{*}{$4.8\times 10^{-4}$} &
\multirow{4}{*}{$9.6\times 10^{-4}$} &\multirow{4}{*}{$1.1\times 10^{-3}$} &\multirow{4}{*}{
$1.8\times 10^{-3}$} \\[0.0cm]
& & & & & & \\[-0.4cm]
& $s_2=+0.7113+1.7374\,\ri$ &$\bar{s}_2=+0.6267+0.6401\,\ri$ & & & & \\[0.0cm]
& & & & & & \\[-0.4cm]
& $w_1=+1.3211 - 1.7905\,\ri$ &$\bar{w}_1=+1.2004 - 1.0049\,\ri$ & & & & \\[0.0cm]
& & & & & & \\[-0.4cm]
& $w_2=+1.5367 - 0.9398\,\ri$ & & & & & \\[0.0cm]
\hline
\end{tabular}
}
\caption{\small%
Presented is numerical data obtained from the spin chain with $r=5$, 
$n=2.5$, ${\tt k}=0.1$ and in the sector $S^z=1$
that  was used for the verification of  the quantization condition \eqref{qunat1A}.
The  rows   correspond to different RG trajectories.
For all of them ${\tt w}=0$, while  the chiral levels ${\tt L}$ and $\bar{\tt L}$ 
are shown in the first column. The next two columns list the values of $\bm{s}=(s_1,s_2)$ and
 $\bar{\bm{s}}=(\bar{s}_1,\bar{s}_2)$.
They were obtained from $b_a(N)$ \eqref{ajs12v}
by interpolating  data at finite $N$ to $N=\infty$, see eq.\,\eqref{asnb2vb}.
Also, for the states with non-zero levels, we give
the sets $\bm{w}$ and $\bar{\bm{w}}$, which  satisfy the system of algebraic equations \eqref{aosaasiBBB}
and its barred version, respectively.
The absolute value of the difference between the l.h.s. and the r.h.s. of the four independent relations \eqref{qunat1A}
with $\ell=1,2,3,4$ is displayed in the four last columns. For the r.h.s.,
the correction term $O(N^{-\epsilon})$ was ignored and $\sigma=0$ since the RG trajectories were
built for lattice sizes keeping $N/5$ an even integer.
\label{tab1}
}
\end{center}
\end{table}

\subsection{Quantization condition for the primary Bethe states\label{sec42}}
Let's first discuss some simple facts concerning the pattern of the Bethe roots for the primary Bethe states.
For this purpose, it is useful to make the following observation regarding the solutions to the Bethe Ansatz equations \eqref{bae}
with the inhomogeneities $\eta_J$ as in eq.\,\eqref{zsym1}. Suppose that the number of Bethe roots $M=\frac{N}{2}-S^z$ is divisible by $r$
and assume that $\zeta_j$ have the form
\be\label{ksa712h}
\re^{\frac{2\pi\ri }{r}(\ell-1)}\,\re^{-\frac{2}{r}\,\alpha_c} \qquad\qquad (c=1,2,\ldots, M/r\,;\ \ell=1,2,\ldots r)\,.
\ee
Substituting this into \eqref{bae}, one obtains
\be\label{asoidoi1902fffff}
\bigg[\frac{\cosh(\alpha_c-\frac{\ri\tilde{\gamma}}{2})}{
\cosh(\alpha_c+\frac{\ri\tilde{\gamma}}{2})
}\bigg]^{\frac{N}{r}}=-\re^{2\ri\pi{\tt k}}\,\prod_{b=1}^{\frac{N-2S^z}{2r}}
\frac{\sinh(\alpha_c-\alpha_b-\ri\tilde{\gamma})}{\sinh(\alpha_c-\alpha_b+\ri\tilde{\gamma})}\,.
\ee
These are easily recognized to be the Bethe Ansatz equations for the Heisenberg XXZ spin\,-\,$\frac{1}{2}$ chain \eqref{asiisaias}
of length $N/r$ and
where $\gamma$ is substituted for $\tilde{\gamma}$ with
\be
\tilde{\gamma}=\frac{\pi r}{n+r}\, .
\ee
The equations correspond to the sector with the $z$-projection of the total spin taking the value $S^z/r=0,\frac{1}{2},1,\ldots$\, .
It is well know that  for the vacuum state in this sector, i.e., the state with the lowest eigenvalue
of the XXZ Hamiltonian, the corresponding
solution is such that  all the $\alpha_c$ are real and distinct,
\be
\alpha_1<\alpha_2<\ldots<\alpha_{M/r}\,,
\ee
provided the absolute value of the twist parameter ${\tt k}$ is sufficiently small.
Computing the
eigenvalue of the Hamiltonian $\mathbb{H}$ of the ${\cal Z}_r$ invariant spin chain via eq.\,\eqref{aoisd9819821A},
where the Bethe roots are  taken to be as in \eqref{ksa712h} with this choice of $\alpha_c$,
turns out to yield  the lowest energy  
in the sector with given value of $S^z$.
\medskip

For a generic
 primary Bethe state of the ${\cal Z}_r$ invariant spin chain the pattern of the Bethe roots remains
  qualitatively similar.
Namely, they  are split into $r$ groups such that the argument of  $\zeta_j$ in each group  is approximately
equal to $\frac{2\pi }{r}\,(\ell-1)$  with 
$\ell=1,2,\ldots,r$. However, there occurs the possibility that the number of Bethe
roots  with roughly equal phases is different. In other words,
\be
\{\zeta_j\}_{j=1}^M=\{\zeta_j^{(1)}\}_{j=1}^{M_1}\,\cup\,\{\zeta_j^{(1)}\}_{j=1}^{M_2}\cup\ldots
\cup\{\zeta_j^{(r)}\}_{j=1}^{M_r}
\ee
with
\be
\arg\big(\zeta_j^{(\ell)}\big)\approx \frac{2\pi}{r}\, (\ell-1)\qquad\qquad ({\rm mod}\ 2\pi)
\ee
and the integers $M_\ell$ are not necessarily  the same.
It should be mentioned that at the edges of the Bethe root distribution along each ray, the 
argument of some of the  roots may deviate significantly from $\frac{2\pi}{r}\, (\ell-1)$.
Then the allocation of the roots to the neighboring groups labeled by $\ell$ or $\ell+1$ (mod\,$r$)
could become ambiguous. Ignoring this subtlety, the states may be assigned the set of numbers
\be\label{kajsshjb32}
\mathfrak{m}_1=M_1-\frac{N}{2r}\,,\ \ \ \ \   \mathfrak{m}_2=M_2-\frac{N}{2r}\,,\ \ \ \ \ldots\,, \ \ \ \ 
\mathfrak{m}_r=M_r-\frac{N}{2r}\,,
\ee
which are integers for $N/r$ even and half-integers for $N/r$ odd.
They are not independent, but satisfy 
\be
\mathfrak{m}_1+\mathfrak{m}_2+\ldots+\mathfrak{m}_r=-S^z\le 0\, ,
\ee
since the total number of roots is equal to $\frac{N}{2}-S^z$.
The states where at the  edges of the root distribution the Bethe roots do not deviate much from the rays 
and, in addition, there are no significant gaps along the ray typically correspond to the primary Bethe states.
\medskip

For the solution of the Bethe Ansatz equations of the form \eqref{ksa712h},  corresponding 
to the lowest energy state  ($S^z=0$) of the ${\cal Z}_r$ invariant spin chain, the
eigenvalues of the
quasi-shift operators \eqref{asj21gh} are given by
\be
{\cal K}^{(\ell)}=1\ .
\ee
In turn, this implies that in the scaling limit all the parameters $s_a$ and $\bar{s}_a$ occurring in the ODEs \eqref{hassay} and \eqref{hassayA} are zero.
Then, upon a change of variables, the differential equations coincide with the Schr\"{o}dinger equations which describe
the scaling limit of the primary states for the XXZ spin chain  (see eq.\eqref{asmn1278sdhj}).
\medskip

Consider the quantization condition \eqref{qunat1A} specialized to the primary Bethe states.
In this case, the sets $\bm{w}$   and $\bar{\bm{w}}$ are trivial and the 
relation can be written as
\be\label{mnsanb21198}
\bigg(\frac{ 2^{\frac{r}{n}}\, N}{rN_0}\bigg)^{\frac{4\ri}{r}\,(-1)^\ell\, { s}}\ \ 
\frac{F_{p}(\bm{s}^{(\ell+1)})}{F_{p}(\bm{s}^{(\ell)})}\,
\frac{F_{\bar{p}}(\bar{\bm{s}}^{(\ell)})}{F_{\bar{p}}(\bar{\bm{s}}^{(\ell+1)})}= 
(-1)^\sigma\,\re^{-\frac{2\pi\ri}{r}\,S^z}
\Big(1
+O\big(N^{-\epsilon}\,\big)\Big)\ .
\ee
Here  $\bm{s}^{(\ell)}$ and $\bar{\bm{s}}^{(\ell)}$ denote
\be\label{mnjhd1a}
\bm{s}^{(\ell)}=\big(s_1^{(\ell)},\ldots,s_{[\frac{r}{2}]}^{(\ell)}\big)\,,\qquad\qquad\qquad
\bar{\bm{s}}^{(\ell)}=\big(\bar{s}_1^{(\ell)},\ldots,\bar{s}_{[\frac{r}{2}]}^{(\ell)}\big)\
\ee
with
\bea\label{mnjhd1b}
{ s}^{(\ell)}_a=(-1)^{ar}\ 
\re^{+\frac{\ri\pi a}{r}(2\ell-1)}\ s_a\,,\qquad\qquad
 {\bar { s}}^{(\ell)}_a=(-1)^{ar}\ \re^{-\frac{\ri\pi a}{r}(2\ell-1)}\ {\bar s}_a\ .
\eea
The function 
\be\label{hastst}
F_p(\bm{s})\equiv  F_p\big(s_1,\ldots s_{[\frac{r}{2}]}\big)
\ee
is  a certain connection coefficient for the ODE 
\bea\label{ODEC1asdasdss}
\bigg[\!-\partial_v^2+\re^{rv }+p^2+
\sum_{a=1}^{[\frac{r}{2}]}  
       { s}_a\  \re^{ a v}\bigg]\,\tilde{ \psi}=0\, .
\eea
To define it, one considers a Jost solution to the differential equation, which 
for $\Re e(p)\ge 0$ is uniquely specified by the condition: 
\bea
\tilde{ \psi}_p\to \re^{pv}\ \ \  \ \ \ \ \ {\rm as}\ \ \ \qquad v\to-\infty\ .
\eea
Then the coefficient $F_p(\bm{s})$ occurs  in the asymptotics:
\bea
\tilde{ \psi}_{p}(v)\to F_{p}({\boldsymbol s})\ 
\exp\bigg(-\frac{rv}{4}+  \frac{s v}{2}+\frac{2}{r}\ \re^{\frac{rv}{2}}\bigg)
\qquad {\rm as}\qquad v\to+\infty\,.
\eea
Recall that $s$, which appears in the above equation along with \eqref{mnsanb21198}, is identically zero
for $r$ odd, while $s\equiv {s}_{\frac{r}{2}}\equiv\bar{s}_{\frac{r}{2}}$ for even $r$.
Clearly, $F_{p}({\boldsymbol s})$ is an entire function of all the variables 
$(s_1,\ldots,{s}_{[\frac{r}{2}]})$ so that it is unambiguously defined when its arguments
are any complex numbers.
\bigskip

Taking the logarithm of  both sides of \eqref{mnsanb21198} one obtains
\be\label{amsnhg21v32}
\frac{4}{r}\,(-1)^{\ell-1}\, { s}\log\bigg(\frac{ 2^{\frac{r}{n}}\, N}{rN_0}\bigg)+\ri
\log\Bigg[\frac{F_{p}(\bm{s}^{(\ell+1)})}{F_{p}(\bm{s}^{(\ell)})}\,
\frac{F_{\bar{p}}(\bar{\bm{s}}^{(\ell)})}{F_{\bar{p}}(\bar{\bm{s}}^{(\ell+1)})}\Bigg]= 
2\pi\,\Big(\mathfrak{n}_\ell+\frac{S^z}{r}\,\Big)\,
+O\big(N^{-\epsilon}\,\big)\, .
\ee
Here $\mathfrak{n}_\ell$ with $\ell=1,\ldots,r$ are some numbers, such that
\be
\mathfrak{n}_\ell\in \begin{cases}
\mathbb{Z} & \ \ {\rm for} \ \ N/r\ {\rm even} \\ 
\mathbb{Z} +\frac{1}{2}& \ \ {\rm for} \ \ N/r\ {\rm odd}
\end{cases}\ .
\ee
In order to assign them a precise meaning, it is necessary to fix the branch of the  logarithm  containing the functions
$F_p$ and $F_{\bar{p}}$. An evident requirement is that the sum over $\ell=1,2,\ldots,r$ of the l.h.s. 
vanishes. This implies the condition
\be
\mathfrak{n}_1+\ldots +\mathfrak{n}_r=-S^z\, .
\ee
One may expect that, with the branch of the  logarithm being suitably chosen,
 the (half-)integers $\mathfrak{n}_\ell$ are simply related to $\mathfrak{m}_\ell$
labeling the primary Bethe states  from \eqref{kajsshjb32}.
To explore this further, let's consider the cases $r=2,3$.

\medskip
For $r=2$ the connection coefficient $F_{p}({\boldsymbol s})$  admits the simple analytical expression:
\be\label{asn1879}
F_{p}( s)=2^{\frac{ s}{2}-p-\frac{1}{2}}\ 
\frac{\Gamma(1+2p)}{\Gamma(\frac{1}{2}+p+\frac{ s}{2})}\ .
%\qquad\qquad (r=2)\ .
\ee
This is because the corresponding ODE \eqref{ODEC1asdasdss} reduces to
\bea\label{ODEC1}
\Big[-\partial_v^2+\re^{2v }+p^2+
s  \,\re^{ v}\Big]\,\tilde{ \psi}=0\qquad\qquad (r=2)\,,
\eea
which may be brought to the form of the  confluent hypergeometric equation. 
The relations \eqref{amsnhg21v32} specialized to $r=2$ consist of only one independent equation. Setting $\ell =2$ 
and $\ell=1$ therein and
taking the difference,  one finds
\be\label{askjh217832jhd}
-4s\log\Bigg[\frac{N\,2^{\frac{2}{n}}\Gamma(\frac{3}{2}+\frac{1}{n})}{\sqrt{\pi}
\Gamma(1+\frac{1}{n})}\Bigg]-2\ri\, \log\Bigg[2^{-2\ri s}\,\frac{\Gamma(\frac{1}{2}+p+\frac{\ri s}{2})}{%
\Gamma(\frac{1}{2}+p-\frac{\ri s}{2})}
\frac{\Gamma(\frac{1}{2}+{\bar p}+\frac{\ri s}{2})}{\Gamma(\frac{1}{2}+{\bar p}-\frac{\ri s}{2})}
\Bigg]=2\pi \,(\mathfrak{n}_2-\mathfrak{n}_1)+O\big(N^{-\epsilon}\big)
\ee
where 
 $N_0$ was substituted for its  expression  \eqref{iaisissu} in terms of the parameter $n$.
This is  the quantization condition studied in 
refs.\cite{Ikhlef:2011ay,Bazhanov:2019xvy,Bazhanov:2019xvyA} in the context of the 
${\cal Z}_2$ invariant spin chain with $\mathfrak{n}_2-\mathfrak{n}_1$
replaced by  $\mathfrak{m}\equiv\mathfrak{m}_2-\mathfrak{m}_1\in\mathbb{Z}$.\footnote{%
Our parameter $(-\frac{s}{2})$  coincides with $s$, which is used in those works, see also Appendix \ref{AppA}.}
The branch of the logarithm in this case is chosen such that the l.h.s. is a continuous function
for real $s$, which vanishes at $s=0$. Taking into account 
that  $\mathfrak{n}_1+\mathfrak{n}_2=\mathfrak{m}_1+\mathfrak{m}_2=-S^z$,
one concludes that
\be
\mathfrak{n}_\ell=\mathfrak{m}_\ell \qquad\qquad \qquad(r=2)\, .
\ee
\smallskip

For $r\ge 3$ a simple analytic formula for  $F_p(\bm{s})$, similar to \eqref{asn1879}, does not exist.
Nevertheless this function can be analyzed within the perturbation theory  and WKB approximation applied to eq.\eqref{ODEC1asdasdss}.
For $r=3$ the ODE becomes
\bea\label{ODEC1a}
\Big[-\partial_v^2+\re^{3v }+p^2+
s_1 \, \re^{ v}\Big]\,\tilde{ \psi}=0\qquad\qquad (r=3)
\eea
and a perturbative calculation yields the first terms of the Taylor expansion
\bea\label{susuasuhhss}
\log F_p(s_1)=\log\bigg[\frac{1}{2\sqrt{\pi}}\  3^{\frac{1}{2}+\frac{2p}{3}}\ 
\Gamma\big(1+\tfrac{2p}{3}\big)\,\bigg]+
 f_1\big(\tfrac{p}{3},\tfrac{1}{3}\big)\, C\ s_1 -
f_2\big(\tfrac{p}{3},\tfrac{1}{3}\big)\, C^2\ s_1^2
+O(s^3_1)\ .
\eea
Here the functions $f_1(h,g)$ and $f_2(h,g)$ are quoted in Appendix \ref{AppB}, while
\bea
C=\frac{3^{\frac{2}{3}}}{\Gamma^2(-\frac{1}{3})}\ .
\eea
On the other hand, the WKB approximation gives that as $|\arg(s_1)|<\pi$  and $|s_1|\to\infty$,
\bea
\label{asssi}
\log F_p(s_1)=\log\bigg(\frac{\Gamma(1+2p)}{2\sqrt{\pi}}\bigg)-p\,\log(s_1)+
\frac{ \Gamma^2(\frac{1}{4})}{3 \sqrt{\pi}}\ s_1^{\frac{3}{4}}
+\frac{ \Gamma^2(\frac{3}{4})}{8 \sqrt{\pi}}\  (1-8p^2)\, s_1^{-\frac{3}{4}}
+O\big(s_1^{-\frac{3}{2}}\big)
\eea
(in the r.h.s. of this equation and \eqref{susuasuhhss} $\log(1)=0$).
The zeroes of the entire function $F_p(s_1)$ accumulate on the negative $s_1$ axis, which is a Stokes line.
As $s_1\to-\infty$, it is possible to show that
\bea
 F_p(s_1)&=&\frac{\Gamma(1+2p)}{\sqrt{\pi}}\ |s_1|^{-p}\ \exp\bigg(-|s_1|^{\frac{3}{4}}\ \frac{ \Gamma^2(\frac{1}{4})}{3 \sqrt{2\pi}}
-\frac{ \Gamma^2(\frac{3}{4})}{8 \sqrt{2\pi}}\  (1-8p^2)\, |s_1|^{-\frac{3}{4}}
+O\big(|s_1|^{-\frac{3}{2}}\big)\bigg)\nonumber\\[0.2in]
&\times&\cos\bigg(\pi p-|s_1|^{\frac{3}{4}}\ \frac{ \Gamma^2(\frac{1}{4})}{3 \sqrt{2\pi}}
+\frac{ \Gamma^2(\frac{3}{4})}{8 \sqrt{2\pi}}\  (1-8p^2)\, |s_1|^{-\frac{3}{4}}
+O\big(|s_1|^{-\frac{3}{2}}\big)\,\bigg)\, .
 \eea
This way,  the function $\log F_p(s_1)$ can be fixed 
at small 
$s_1$ via formula \eqref{susuasuhhss} and then  extended by continuity to
the wedge  $|\arg(s_1)|<\pi-\delta$,
where $0<\delta\ll 1$ is chosen in such a way so that the zeroes of $F_p(s_1)$ are excluded from the domain.
\medskip

The quantization condition \eqref{amsnhg21v32} for $r=3$ 
can be written as
\bea\label{skajb12b32}
&&\ri\log\Big(
F_p\big(\re^{+\frac{2\pi\ri}{3}\ell}\,s_1\big)\Big)-
\ri\log\Big(
F_p\big(\re^{+\frac{2\pi\ri}{3}(\ell-1)}\,s_1\big)\Big)+\ri\log\Big(
F_{\bar{p}}\big(\re^{-\frac{2\pi\ri}{3}(\ell-1)}\,\bar{s}_1\big)\Big)-\ri\log\Big(
F_{\bar{p}}\big(\re^{-\frac{2\pi\ri}{3}\ell}\,\bar{s}_1\big)\Big)\nonumber \\[0.2cm]
&& = 
2\pi\,\Big(\mathfrak{n}_\ell+\frac{S^z}{r}\,\Big)\,
+O\big(N^{-\epsilon}\,\big)\, .
\eea
We found that if  the logarithm of $ F_p(s_1)$  is fixed as above, then
 the (half-)integers $\mathfrak{n}_\ell$ 
are related to
$\mathfrak{m}_\ell$ from eq.\,\eqref{kajsshjb32} as
\be
\mathfrak{n}_\ell=\mathfrak{m}_{\ell-1}\qquad\qquad \qquad(\ell\sim\ell +r\,;\ r=3)\, .
\ee
Furthermore, by means of the  asymptotic formula for the logarithm of $F_p(s_1)$ \eqref{asssi}, it is straightforward to
 show
\bea\label{asmbv123}
|s_1|&=&\pi^2  \, \bigg[\frac{3}{\Gamma^2(\frac{1}{4})}\bigg]^{\frac{4}{3}}\ 
{\mathfrak m}^{\frac{4}{3}}\ \bigg(1+\frac{1-8{\bar p}^2}{9\pi\, 
{\mathfrak m}^2}+o\big({\mathfrak m}^{-2}\big)\bigg)\nonumber\\[-0.2cm]
\\[-0.3cm]
|{\bar s}_1|&=&\pi^2  \, \bigg[\frac{3}{\Gamma^2(\frac{1}{4})}\bigg]^{\frac{4}{3}}\ 
{\mathfrak m}^{\frac{4}{3}}\ \bigg(1+\frac{1-8{ p}^2}{9\pi\, 
{\mathfrak m}^2}+o\big({\mathfrak m}^{-2}\big)\bigg)\ .\nonumber
\eea
Here $\mathfrak{m}$ is given in terms of the triple $(\mathfrak{m}_1,\mathfrak{m}_2,\mathfrak{m}_3)$ as
\be\label{asmbv123aaa}
{\mathfrak m}=\sqrt{\frac{{\mathfrak m}_j^2+{\mathfrak m}_k^2}{2}}\,,
\ee
where $\mathfrak{m}_j,\,\mathfrak{m}_k$ are defined via the condition
\be\label{ask32iu}
\qquad \mathfrak{m}_j\mathfrak{m}_k\ge 0\, .
\ee
Formula \eqref{asmbv123} is valid as $\mathfrak{m}\gg 1$ with $p$ and $\bar{p}$
 being kept fixed. Note that
 if all $\mathfrak{m}_1$, $\mathfrak{m}_2$ and $\mathfrak{m}_3$ have the same sign,
their absolute value must be smaller than $S^z=p+\bar{p}\ge 0$. As a result, \eqref{asmbv123}
becomes inapplicable. Also, 
for all the RG trajectories we were able to construct, where the (half-)integers
 $(\mathfrak{m}_1,\mathfrak{m}_2,\mathfrak{m}_3)$
could be defined unambiguously, at least two of them were smaller or equal to zero so that 
$\mathfrak{m}_j,\mathfrak{m}_k\le 0$.
\medskip

\begin{table}
\begin{center}
\scalebox{0.9}{
\begin{tabular}{|c|c|c|c|c|}
\hline
& \multicolumn{2}{l|}{} & \multicolumn{2}{l|}{} \\[-0.4cm]
 & \multicolumn{2}{c|}{$|s_1|/\mathfrak{m}^{\frac{4}{3}}$} & 
 \multicolumn{2}{c|}{$\arg(s_1)$}\\[0.0cm]
\hline
& & & &  \\[-0.4cm]
$(\mathfrak{m}_1,\mathfrak{m}_2,\mathfrak{m}_3)$ & Bethe Ansatz & formula \eqref{asmbv123} & Bethe Ansatz  &  formula \eqref{akjshg12aa24} \\[0.0cm]
\hline
& & & &  \\[-0.4cm]
$(-2,-1,1)$ &1.265280 &1.280089 & 2.534721 &   \\[0.0cm]
\cline{1-4}
& & & &  \\[-0.4cm]
$(-4,-2,4)$ &1.351599 &1.352403 & 2.523989 &$+2.523396$  \\[0.0cm]
\cline{1-4}
& & & &  \\[-0.4cm]
$(-6,-3,7)$ &1.365639 &1.365794& 2.523508&\\[0.0cm]
\hline
\end{tabular}
}
\bigskip

\scalebox{0.9}{
\begin{tabular}{|c|c|c|c|c|}
\hline
& \multicolumn{2}{l|}{} & \multicolumn{2}{l|}{} \\[-0.4cm]
 & \multicolumn{2}{c|}{$|\bar{s}_1|/\mathfrak{m}^{\frac{4}{3}}$} & 
 \multicolumn{2}{c|}{$\arg(\bar{s}_1)$}\\[0.0cm]
\hline
& & & &  \\[-0.4cm]
$(\mathfrak{m}_1,\mathfrak{m}_2,\mathfrak{m}_3)$ & Bethe Ansatz & formula \eqref{asmbv123} & Bethe Ansatz  &  formula \eqref{akjshg12aa24} \\[0.0cm]
\hline
& & & &  \\[-0.4cm]
$(-2,-1,1)$ &1.185686  &1.194405 &$ -2.522902$ &   \\[0.0cm]
\cline{1-4}
& & & &  \\[-0.4cm]
$(-4,-2,4)$ &1.330539 &1.330982 & $-2.523408$ & $-2.523396$  \\[0.0cm]
\cline{1-4}
& & & &  \\[-0.4cm]
$(-6,-3,7)$ &1.356189 &1.356273& $-2.523398$ &\\[0.0cm]
\hline
\end{tabular}
}
\end{center}

\caption{\small%
\label{tab2}In order to demonstrate the accuracy of the asymptotic formulae \eqref{asmbv123} and \eqref{akjshg12aa24}, primary Bethe states 
$|\Psi_N\rangle$ were constructed for the ${\cal Z}_3$ invariant spin chain that are characterized by different triples
$(\mathfrak{m}_1,\mathfrak{m}_2,\mathfrak{m}_3)$. The values of $s_1,\bar{s}_1$ 
were then obtained by computing $b_1, b_{-1}$ \eqref{ajs12v} from the eigenvalues of the quasi-shift operator
 and then taking the scaling limits 
 $s_1= C_1 \slim_{N\to\infty}b_1$, 
 $\bar{s}_1=-C_1 \slim_{N\to\infty}b_{-1}$   with
the coefficient $C_1$ as in eq.\,\eqref{asnb2vbAA}.
The result was used to generate the numbers listed in the columns  ``Bethe Ansatz''. Recall that $\mathfrak{m}$ is given in
terms of $(\mathfrak{m}_1,\mathfrak{m}_2,\mathfrak{m}_3)$ by eq.\,\eqref{asmbv123aaa} with $\mathfrak{m}_j$, $\mathfrak{m}_k$
being  defined through the conditions \eqref{ask32iu} and \eqref{askjjh12}. Note that for the
 three primary Bethe states considered, the asymptotic
formula \eqref{akjshg12aa24}
 yields the same result  for $\arg(s_1)$, which is given in the last column, and similarly for $\arg(\bar{s}_1)$.
The parameters were taken to be $n=2.5$, ${\tt k}=0.05$, while for all the states ${\tt w}=0$, $S^z=2$.
}
\end{table}
The condition \eqref{ask32iu} does not fix $\mathfrak{m}_j,\mathfrak{m}_k$ taken from the triple
$(\mathfrak{m}_1,\mathfrak{m}_2,\mathfrak{m}_3)$ unambiguously. This is not important for the
description of the asymptotics of $|s_1|$ and $|\bar{s}_1|$ since \eqref{asmbv123} involves only
 $\mathfrak{m}$, which is invariant under the permutation $\mathfrak{m}_j\leftrightarrow\mathfrak{m}_k$.
Turning to the large $\mathfrak{m}$
asymptotic formula for the arguments of $s_1,\bar{s}_1$ we supplement \eqref{ask32iu} with the requirement 
\be\label{askjjh12}
|\mathfrak{m}_j|\geq|\mathfrak{m}_k| \ .
\ee
Then we found that 
\bea\label{akjshg12aa24}
&&\arg(s_1)=
\frac{2\pi}{3}\,\mu_\ell+(-1)^{\#({\cal P})}\ 
\frac{4}{3}\,\arctan\Big(\frac{{\mathfrak m}_j-{\mathfrak m}_k}{{\mathfrak m}_j+{\mathfrak m}_k}\Big)+
o\big(|{\mathfrak m}|^{-2}\,\big)\nonumber\\[0.3cm]
&&\arg(s_1)+\arg(\bar{s}_1)=o\big(|{\mathfrak m}|^{-2}\,\big)\,,
\eea
where $(-1)^{\#({\cal P})}$ stands for the parity of the permutation ${\cal P}:\,(1,2,3)\mapsto (j,k,\ell)$, while
\bea
\mu_\ell=\begin{cases}
0 & {\rm for} \ \ \ \ell=1\\[0.1cm]
-1 & {\rm for} \ \ \ \ell=2\\[0.1cm]
+1 & {\rm for} \ \ \ \ell=3\\
\end{cases}\, .
\eea
 Note that for the case  $\mathfrak{m}_j=\mathfrak{m}_k$, the arctangent in  \eqref{akjshg12aa24}
vanishes and $\arg(s_1)=\frac{2\pi}{3}\,\mu_\ell+o(|{\mathfrak m}|^{-2}\,)$ and similarly for $\arg({\bar s}_1)$.
Also, the condition \eqref{ask32iu}  guarantees the inequality
\be
-\frac{\pi}{3}\,<\, \frac{4}{3}\, \arctan\bigg(\frac{{\mathfrak m}_j-{\mathfrak m}_k}{{\mathfrak m}_j+{\mathfrak m}_k}\bigg)\,
\le\,
\frac{\pi}{3}\ .
\ee
In order to illustrate the accuracy of \eqref{asmbv123} and \eqref{akjshg12aa24}, 
some numerical 
data for $s_1$ and $\bar{s}_1$ obtained from the solution of the Bethe Ansatz equations corresponding
to the primary Bethe states is compared   with the predictions coming from  these asymptotic formulae  in tab.\,\ref{tab2}.

\section{The spectral determinants at large $E$}

 In the context of the ODE/IQFT correspondence, an important r\^{o}le belongs to the study of the large $E$
 asymptotic expansion of the spectral determinants. The first few terms in the expansion 
 were already presented in \eqref{askb1287ashhg}-\eqref{askj38921}. 
 Let's denote the r.h.s.
 of the last of these formulae by
\bea\label{askj287jhasd}
C^{(\ell)}(\theta)\equiv{\mathfrak C}^{(\ell)}_{p,{\boldsymbol s}}({\boldsymbol w})\ \exp\Bigg[\,\frac{N_0}{\cos(\frac{\pi r}{2n})}\ \re^\theta+ \bigg(\,
(-1)^{\ell -1}\, \ri s
 -\frac{2 n p}{n+r}\bigg)
\, \frac{\theta}{r}\, \bigg]\, .
\eea
Recall that the integer $\ell=1,2,\ldots r$  labels the different wedges
 $\arg\big((-1)^{r-1}\,E\,\big)\in\big(\frac{2\pi}{r}(\ell-1),\frac{2\pi}{r}\ell\,\big)$ $({\rm mod}\ 2\pi)$
 in the complex $E$ plane.
An analysis of the ODE \eqref{hasas} shows that the full asymptotic expansion may be represented in the form
\bea\label{askjh312}
D_p(E\,|\,\bm{w})\asymp C^{(\ell)}(\theta)\ \exp\Big(\,S^{(\ell)}(\theta)+I(\theta)+{\tilde H}^{(\ell)}(\theta)\,\Big)\, ,
\eea
where  the formal asymptotic series $S^{(\ell)}(\theta)$, $I(\theta)$ and ${\tilde H}(\theta)$ are described as follows. 

\smallskip
The largest corrections to \eqref{askj38921}  come from $S^{(\ell)}(\theta)$. In particular,
\bea\label{askj98123}
S^{(\ell)}(\theta)=\frac{1}{2 n\sqrt{\pi}}\ \sum_{a=1}^{[\frac{r-1}{2}]} \, 
%(-1)^{ar}\ \re^{\frac{\ri\pi a}{r}(2\ell-1)}
(\eta_\ell)^a\ 
\Gamma\big(\tfrac{1}{2}-\tfrac{2a-r}{2n}\big)\,\Gamma\big(\tfrac{2a-r}{2n}\big)
\ s_a\ \, \re^{-(1-\frac{2a}{r})\,\theta} +S_1^{(\ell)}(\theta)
\eea
with ${S}_1^{(\ell)}(\theta)$ decaying faster  as $\theta\to\infty$  than the explicitly displayed terms.
The latter linearly depend on the parameters $s_a$ entering into the differential equation \eqref{hasas}.
Within the usual interpretation $s_a$ are the eigenvalues of certain Integrals of Motion  (IM)
${\bf S}_a$.
The subscript  $a$ coincides with the  charge of these operators
w.r.t. the ${\cal Z}_r$ symmetry transformations. Furthermore, their conformal dimensions are given by
\bea 
{\bf S}_a\,:  \ \ \ \ \ (\Delta,{\bar \Delta})=\big(1-\tfrac{2a}{r},0\big) \,  \ \ \ \ \ \ \ \ \ 
\big(a=1,\ldots,[\tfrac{r}{2}]\big)\, . 
\eea
Similarly the large $\bar{E}$ asymptotic of $\bar{D}_{\bar{p}}(\bar{E}\,|\,\bar{\bm{w}})$ involves
the eigenvalues of the operators $\bar{{\bf S}}_a$, whose ${\cal Z}_r$ charges are  $(-a)$
and conformal dimensions read as
\bea 
\bar{{\bf S}}_a\,: \ \ \ \ \  (\Delta,{\bar \Delta})=\big(0,1-\tfrac{2a}{r}\big)  \,  \ \ \ \ \ \ \ \ \ 
\big(a=1,\ldots,[\tfrac{r}{2}]\big)\, . 
\eea
The formal asymptotic series  ${S}_1^{(\ell)}({\theta})$ in \eqref{askj98123} has the general structure
\bea\label{askasj821}
S_1^{(\ell)}({\theta})
=\sum_{m=1}^\infty\sum_{a=1}^{[\frac{r}{2}]}\ (\eta_\ell)^a\ S_{m,a}\ \re^{-(2m+1-\frac{2a}{r})\,\theta}\ .
\eea
Again, the coefficients $S_{m,a}$  of the expansion are expressed in terms of the eigenvalues
of certain operators whose ${\cal Z}_r$ charges are $a$ and which have
   conformal dimensions $\Delta=2m+1-\frac{2a}{r}$ and $\bar{\Delta}=0$.
\medskip

Among the   coefficients $S_{m,a}$, those 
with $a=\frac{r}{2}$ for even $r$ deserve special attention:
\bea\label{askasj821aaa}
S_{m,\frac{r}{2}}\equiv  I_{2m}\ \ \ \  \ (m=1,2,\ldots;\ \ r-{\rm even})\, .
\eea
They are  the
eigenvalues of certain operators ${\bf I}_{2m}$ with ${\cal Z}_r$ charges  $\frac{r}{2}\,({\rm mod}\ r)$
and which have integer conformal dimensions 
\bea
{\bf I}_{2m}\,: \ \ \ \ \ 
(\Delta,{\bar \Delta})=\big(2m,0\big) \, .
\eea
The operators ${\bf I}_{2m}$ are expected to be local IM, i.e., they admit an expression of the form
\be \label{askj7812jh}
{\bf I}_{2m}=\int_0^{2\pi}\frac{\rd u}{2\pi}\, T_{2m+1}(u)\,,
\ee
where $T_{2m+1}(u)$ are chiral local fields of Lorentz spin $2m+1$ such that 
\be
\bar{\partial}T_{2m+1}(u)=0\, .
\ee
Similarly the expansion of the spectral determinant $\bar{D}_{\bar{p}}(\bar{E}\,|\,\bar{\bm{w}})$ 
yields the eigenvalues of
\be
\bar{{\bf I}}_{2m}=\int_0^{2\pi}\frac{\rd \bar{u}}{2\pi}\  \bar{T}_{2m+1}(\bar{u})\,,\qquad\qquad
\partial\bar{T}_{2m+1}(\bar{u})=0\, .
\ee
These have the same ${\cal Z}_r$ charge as ${\bf I}_{2m}$, i.e., $\frac{r}{2}\  ({\rm mod}\ r)$.
\medskip

While ${\bf I}_{2m}$ and ${\bar{\bf I}}_{2m}$ appear only when $r$ is even, it is expected that for generic
positive integer $r$
the theory possesses the additional local IM ${\bf I}_{2m-1}$ and $\bar{{\bf I}}_{2m-1}$. For the former ones
\be\label{askj81972sa}
{\bf I}_{2m-1}=\int_0^{2\pi}\frac{\rd u}{2\pi}\, T_{2m}(u)\,,\qquad\qquad
{\bar{\partial}{T}}_{2m}({u})=0\,,
\ee
where the  chiral local density $T_{2m}$ has Lorentz spin $2m$ 
and similarly for $\bar{{\bf I}}_{2m-1}$.
These local IM are all  ${\cal Z}_r$ invariant. The eigenvalues of ${\bf I}_{2m-1}$ 
appear  in   $I(\theta)$ in the formula \eqref{askjh312}. In particular, the leading term reads explicitly as
\bea
I(\theta)=-\frac{1}{\sqrt{\pi}}\ 
\Gamma\big(-\tfrac{r}{2n}\big)\,\Gamma\big(\tfrac{3}{2}+\tfrac{r}{2n}\big)\ 
I_1\  \re^{-\theta}+\ldots\ ,
\eea
where
\be\label{kjaskj12}
I_1=\frac{p^2}{n+r}+\frac{s^2}{4n}-\frac{r}{24}+{\tt L}\qquad\qquad (s\equiv 0\ {\rm for}\ r\ {\rm odd};
\ s\equiv s_{\frac{r}{2}} \ {\rm for}\ r\ {\rm even})\, .
\ee
The corresponding operator  is of the form
\be\label{askj12}
{\bf I}_1=\int_0^{2\pi}\frac{\rd u}{2\pi}\, \big((\partial\varphi)^2+\ldots\big)\, .
\ee
Here $\partial\varphi$ is the chiral component of the current associated
with the global ${\rm U}(1)$ symmetry of the spin chain.\footnote{%
The current $\partial\varphi$ is normalized  through the operator product expansion
$\partial\varphi(u_1)\,\partial\varphi(u_2)=-\frac{1}{2\,(u_1-u_2)^2}+O(1)$.
} The ellipses stand for the contribution of other degrees of freedom. Note that
formula \eqref{askj12}  fixes any ambiguity in the normalization of ${\bf I}_1$,
which is chosen in such a way that its eigenvalues are given by \eqref{kjaskj12}.
This follows since the eigenvalue of the zero mode
 $\int_0^{2\pi} \frac{{\rm d} u}{2\pi}\, \partial \varphi$  is $P=\frac{p}{\sqrt{n+r}}$.
The CFT Hamiltonian coincides with 
 \be
{\bf  H}_{\rm CFT}={\bf I}_1+\bar{{\bf I}}_1\ .
 \ee
\smallskip

The full series for $I(\theta)$ takes the form 
\bea
I(\theta)=
\sum_{m=1}^\infty \frac{(-1)^{m-1} (n+r)^m}{2n\sqrt{\pi}\, m!}\  \Gamma\big(-(m-\half )\, \tfrac{r}{n}\big)\,  \Gamma\big( (m-\half )\,\tfrac{n+r}{n}\big)\ \ I_{2m-1}\ \re^{-(2m-1)\theta} \ .
\eea
Here the  coefficients $I_{2m-1}$ depend on $p$ such that as $p\to\infty$, 
\bea
I_{2m-1}= \bigg(\frac{p^2}{n+r}\bigg)^m+O\big(\,p^{2m-2}\,\big)\ .
\eea
These are the eigenvalues of the local IM ${\bf I}_{2m-1}$
from \eqref{askj81972sa}.
\medskip

 Finally, the term ${\tilde H}^{(\ell)}(\theta)$ from \eqref{askjh312} stands for the formal asymptotic series
\bea
{\tilde H}^{(\ell)}(\theta)=\sum_{m=1}^\infty {\tilde H}^{(\ell)}_m\ \re^{-\frac{2n m}{r}\,\theta}\, .
\eea
It involves the eigenvalues of the ``dual non-local'' IM, whose Lorentz spin depends on $n$
and is given by $\frac{2nm}{r}$ with $m=1,2,3,\ldots\ $. For the case $r=1$, such operators were first discussed in the
context of the  quantum KdV theory  in ref.\cite{Bazhanov:1996dr}. Also note that the coefficient
${\mathfrak C}^{(\ell)}_{p,{\boldsymbol s}}({\boldsymbol w})$ in \eqref{askj287jhasd} can be 
interpreted as the eigenvalue of the simplest non-local IM, which is related to
the so-called reflection operators \cite{Zamolodchikov:1995aa,Kotousov:2019nvt}.

\section{Conclusion}
The main result of the paper is the class of second order linear differential equations and the quantization condition
 that describe the scaling limit of the ${\cal Z}_r$ invariant
spin chain in the critical regime with anisotropy parameter $q=\re^{\frac{\ri\pi}{n+r}}$ and $n>0$.
We can not claim to have developed an ODE/IQFT correspondence for the model as 
the field theory description lies beyond the scope of this work. We believe
 that further studying the theory is worthwhile  since even a superficial analysis reveals many
  interesting features of the CFT underlying the critical behaviour.
   Among them  is an infinite degeneracy of the ground state 
 (as well as all conformal primary states) and the presence of a continuous component in the spectrum
 which occurs in the case of even $r$.
To conclude the paper, we would like to mention two possible directions which, in our opinion,
may help to better understand the critical behaviour of the model.

\medskip
The first concerns the algebra of extended conformal symmetry. A way of  proceeding in its study
is provided by the densities for the local IM \eqref{askj7812jh},\,\eqref{askj81972sa}
$(T_j,\bar{T}_j)$ with $j=2,4,6,\ldots$ for odd $r$ and $j=2,3,4,\ldots$ for even $r$.
 Since such fields are periodic, e.g., $T_j(u+2\pi)=T_j(u)$, and occur inside an integral,
 they are defined up to a total
derivative:
\be
T_j\mapsto T_j+\partial {\cal O}_{j-1}\,,\qquad\qquad
\bar{T}_j\mapsto \bar{T}_j+{\bar \partial }\bar{{\cal O}}_{j-1}\ ,
\ee
where ${\cal O}_{j-1}$  $\big(\bar{\cal O}_{j-1}\big)$ are local chiral fields 
of Lorentz spin $j-1$  $(1-j)$.
Based on the experience gained from the study of the cases $r=1,2$ we expect that
the densities can be chosen such that they generate a closed 
${\cal W}$-algebra.
The explicit construction of the algebra of extended conformal symmetry 
would be an important step for describing the CFT underlying
the critical behaviour of the  lattice system.
\medskip

The second direction is the study of the limit $n\to\infty$.
As is the case for $r=2$, one 
expects that it can be interpreted as
 a classical limit, where the CFT admits a Lagrangian description. If so, this  would  certainly yield valuable
 insights into the physical content of the theory.

 \medskip

\section*{Acknowledgments}
GK acknowledges discussions with Holger Frahm. 
\medskip

\noindent
The research of GK is supported
by the Deutsche Forschungsgemeinschaft under grant
No. Fr 737/9-2.
The final part of this work was carried out during GK's visit to the NHETC at Rutgers University. 
He is grateful for the support and hospitality he received during the stay.
\medskip

\noindent
The research of SL is partially funded by the NSF under grant number NSF-PHY-2210187.

\appendix
\section{Appendix\label{AppA}}
The special cases $r=1,2$ for the proposed ODE \eqref{hassay}
have already been discussed in the literature. For $r=1$,  the differential equation reads as
\be
\bigg[-\partial_z^2
+\frac{p^2-\frac{1}{4}}{z^2}-\frac{1}{z}
+ E^{-n-1}\ z^{n-1}
\bigg]\, \Psi=0\qquad\qquad\qquad (r=1)\, .
\ee
Via the change of variables:
\be
z=\tfrac{1}{4}\,E_{\rm XXZ}
\,x^2\,,\qquad\qquad \Psi=\sqrt{x}\,\Psi_{\rm XXZ}
\ee
and parameters
\be
p=\sqrt{\alpha+1}\,(\ell+\tfrac{1}{2})\,,\qquad n=\alpha\,,\qquad 
E=2^{-\frac{2\alpha}{\alpha+1}}\,E_{\rm XXZ}
\ee
this ODE becomes the Schr\"{o}dinger equation
 for the anharmonic oscillator appearing in refs.\,\cite{Voros:1994,Dorey:1998pt,Bazhanov:1998wj}:
\be\label{asmn1278sdhj}
\bigg[-\partial_x^2+\frac{\ell(\ell+1)}{x^2}+x^{2\alpha}-E_{\rm XXZ}\bigg]\Psi_{\rm XXZ}=0\, .
\ee
It describes the scaling behaviour of the XXZ spin chain with anisotropy parameter $q=\re^{\frac{\ri\pi}{\alpha+1}}$
see, e.g., refs.\cite{Kotousov:2019ygw,Bazhanov:2019xvyA}.
\medskip

When $r=2$, eq.\,\eqref{hassay} becomes 
\be
\bigg[-\partial_z^2
+\frac{p^2-\frac{1}{4}}{z^2}-\frac{s_1}{z}-1
+ E^{-n-2}\ z^{n}
\bigg]\, \Psi=0\qquad\qquad\qquad (r=2)\, .
\ee
The substitution
\be
 z\mapsto \ri z\,,\qquad\qquad s_1\mapsto -2s\,,\qquad\qquad E\mapsto \ri\lambda
\ee
brings it to the form of the differential equation discussed in ref.\cite{Bazhanov:2019xvy} in the context of the scaling limit
of the ${\cal Z}_2$ invariant spin chain.

\section{Appendix\label{AppB}}
Here we present the explicit formulae for $f_1$, $f_2$, which enter into eqs.
\eqref{akjshgj873287} and \eqref{susuasuhhss}. We use the same notation
as in ref.\cite{Bazhanov:2019xvyA}, where these functions have previously appeared.

\medskip

The function $f_1$ is defined as
\bea
f_1(h,g)&=&\frac{\pi\Gamma(1-2g)}{\sin(\pi g)}\ \frac{\Gamma(g+2h)}{\Gamma(1-g+2 h)}\ .
\eea
As for $f_2$, it is more complicated and is given by the integral
\be\label{jassusau000}
f_2(h, g)=2^{1-4g}\,\frac{\Gamma^2(1-g)}{\Gamma^2(\frac{1}{2}+g)}\
\frac{\Gamma(2g+2 h)}{\Gamma(1-2g+2 h)}\,\int_{ -\infty}^\infty\frac{\rd x}{2\pi}\,\frac{S_1(x)}{x+\ri h}\ \ \ \ \ \  
\big(0<g<\tfrac{1}{2},\ \Re e(h)>0\big)
\ee
with
\be\label{Sdef1a}
S_1(x)=\sinh(2\pi x)\,\Gamma(1-2g+2\ri x)\,\Gamma(1-2g-2\ri x)\,\big(\Gamma(g+2\ri x)\Gamma(g-2\ri x)\big)^2\ .
\ee
For $\tfrac{1}{2}<g<1$ one has
\bea\label{jassusau}
f_2(h,g)&=&2^{1-4 g}\,\frac{\Gamma^2(1-g)}{\Gamma^2(\frac{1}{2}+g)}\
\frac{\Gamma(2g+2 h)}{\Gamma(1-2g+2h)}\,\Bigg(\int_{-\infty}^\infty\frac{\rd x}{2\pi}\,
\frac{S_1(x)}{x+\ri h}  \\[0.2cm]
&-&
\frac{\sin(2\pi g)\Gamma(3-4 g)\Gamma^2(1-g)\Gamma^2(3g-1)}{(2h+1-2g)
(2h-1+2 g)}\Bigg)\qquad
 \qquad \big(\,\tfrac{1}{2}<g<1, \ \Re e(h)>0\, \big)\ .\nonumber
\eea

\section{Appendix\label{AppC}}

The coefficient ${\mathfrak C}^{(\ell)}_{p,{\boldsymbol s}}({\boldsymbol w})$,
appearing in the large $E$ 
 asymptotic expansion of the spectral determinant \eqref{askj38921}, may be expressed in terms of the connection
coefficients of the differential equation
\bea\label{ODEC1}
\bigg[\!-\partial_v^2+\re^{rv }+p^2+
\sum_{a=1}^{[\frac{r}{2}]}  
       { s}^{(\ell)}_a\  \re^{ a v}+\sum_{\alpha=1}^{{\tt L}}\bigg(\frac{2}{\big(1+w^{(\ell)}_\alpha \re^{-v}\big)^2}+
       \frac{n_\alpha}{1+w^{(\ell)}_\alpha\re^{-v}}\bigg)
                         \bigg]\,{\tilde \psi}=0\ ,
\eea
where
\bea\label{sajhg1278hgsd}
{ s}^{(\ell)}_a=(-1)^{ar}\ \re^{+\frac{\ri\pi a}{r}(2\ell-1)}\ s_a\ \ \ \qquad{\rm and}\qquad  \ \ \ 
{ w}^{(\ell)}_\alpha=(-1)^{ r}\ \re^{-\frac{\ri\pi }{r}(2\ell-1)}\ w_\alpha\ .
\eea
This ODE is obtained from  \eqref{hasas} by changing variables,
\bea\label{sajhhg21}
z=(-1)^{r-1}\ \re^{\frac{\ri\pi}{r}(2\ell-1)} \ \re^{v}\ ,\ \ \ \ \ \ \ \Psi=\re^{\frac{v}{2}}\ {\tilde \psi}\,,
\eea
and  setting $E$ to infinity therein.
Assuming that $\Re e(p)\ge 0$ we consider a solution of \eqref{ODEC1}
such that
\bea
{\tilde \psi}_p\to \re^{pv}\ \ \  \ \ \ \ \ {\rm as}\ \ \ \ v\to-\infty\ .
\eea
A straightforward WKB analysis yields that when $v\to+\infty$,
\bea\label{samnnb1321}
{\tilde \psi}_{p}(v)\to C^{(\ell)}_{p,{\boldsymbol s}}({\boldsymbol w})\ 
\exp\bigg(-\frac{rv}{4}- \ri \,(-1)^{\ell} \  \frac{s v}{2}+\frac{2}{r}\ \re^{\frac{rv}{2}}\bigg)
\eea
with
\bea
{s}\equiv \begin{cases}
\,0\ \ \ \ &{\rm for \ \ odd}\ \ r\\
\, {s}_{\frac{r}{2}}\ \ \ \ &{\rm for\ \ even}\ \ r
 \end{cases}\  .
\eea
One can show that ${\mathfrak  C}^{(\ell)}_{p,{\boldsymbol s}}({\boldsymbol w})$
is expressed in terms of the connection coefficient  $C_{p,\bm{s}}^{(\ell)}(\bm{w})$
as 
\bea\label{sanb21bbv3}
{\mathfrak  C}^{(\ell)}_{p,{\boldsymbol s}}({\boldsymbol w})
=\sqrt{\frac{4\pi}{n+r}}\ \   (n+r)^{-\frac{2p}{n+r}}\ 2^{-\ri\, (-1)^{\ell}\, \frac{ s}{n}}\  
\frac{ C^{(\ell)}_{p,{\boldsymbol s}}({\boldsymbol w})}
{\Gamma\big(1+\tfrac{2p}{n+r}\big)}\,.
\eea
\smallskip

The computation of $\bar{{\mathfrak  C}}^{(\ell)}_{\bar{p},\bar{\boldsymbol s}}(\bar{\boldsymbol w})$,
which occurs in the large $\bar{E}$ asymptotic formula  \eqref{askj38921} of the spectral determinant
$\bar{D}_{\bar{p}}(\bar{E}\,|\,\bar{\bm{w}})$ is analogous.
 The relevant ODE
would formally coincide with \eqref{ODEC1} upon the substitution of the variables $(v,\tilde{\psi})\mapsto 
(\bar{v},\tilde{\bar{\psi}})$ and  parameters 
$(p,s_a^{(\ell)},w_\alpha^{(\ell)},n_\alpha,{\tt L})\mapsto 
(\bar{p},\bar{s}_a^{(\ell)},\bar{w}_\alpha^{(\ell)},\bar{n}_\alpha,\bar{\tt L})$, where together with \eqref{sajhg1278hgsd}
we  use the notation 
\bea\label{sajhg1278hgsde}
\bar{ s}^{(\ell)}_a=(-1)^{ar}\ \re^{-\frac{\ri\pi a}{r}(2\ell-1)}\ \bar{s}_a\ \ \ \qquad{\rm and}\qquad  \ \ \ 
\bar{ w}^{(\ell)}_\alpha=(-1)^{ r}\ \re^{+\frac{\ri\pi }{r}(2\ell-1)}\ \bar{w}_\alpha\ .
\eea
The  connection coefficient is extracted from the $\bar{v}\to+\infty$ asymptotic   of the solution 
$\tilde{\bar{\psi}}_{\bar{p}}$, which is defined by the condition 
$\tilde{\bar{\psi}}_{\bar{p}}\to \re^{\bar{p}\bar{v}}$ as $\bar{v}\to-\infty$. Namely,
\bea
\tilde{\bar{\psi}}_{\bar{p}}(\bar{v})\to \bar{C}^{(\ell)}_{\bar{p},\bar{{\boldsymbol s}}}(\bar{\boldsymbol w})\ 
\exp\bigg(-\frac{rv}{4}+ \ri \,(-1)^{\ell} \  \frac{s v}{2}+\frac{2}{r}\ \re^{\frac{rv}{2}}\bigg)
\eea
(recall that $s\equiv \bar{s}_{\frac{r}{2}}$ for $r$ even
and is identically zero for $r$ odd).
Then the barred version of  \eqref{sanb21bbv3} reads as 
\bea
\bar{\mathfrak C}^{(\ell)}_{\bar{p},\bar{\boldsymbol s}}(\bar{\boldsymbol w})
=\sqrt{\frac{4\pi}{n+r}}\ \   (n+r)^{-\frac{2\bar{p}}{n+r}}\ 2^{+\ri\, (-1)^{\ell}\, \frac{ s}{n}}\  
\frac{ \bar{C}^{(\ell)}_{\bar{p},\bar{\boldsymbol s}}(\bar{\boldsymbol w})}
{\Gamma\big(1+\tfrac{2\bar{p}}{n+r}\big)}\,.
\eea

\bigskip

The following comments are in order here.
For the case when there are no apparent singularities,
the coefficients $C_{p,\bm{s}}^{({\ell})}$   and $\bar{C}_{\bar{p},\bar{\bm{s}}}^{({\ell})}$  
are expressed in terms of the function  $F_p(\bm{s})=F_p(s_1,\ldots,s_{[\frac{r}{2}]})$ \eqref{hastst}
used in sec.\,\ref{sec42}:
\be
C_{p,\bm{s}}^{({\ell})}(\bm{\emptyset})=F_p(\bm{s}^{(\ell)})\, ,\qquad\qquad
\bar{C}_{\bar{p},\bar{\bm{s}}}^{({\ell})}(\bm{\emptyset})=F_{\bar{p}}(\bar{\bm{s}}^{(\ell)})\,,
\ee
where $\bm{s}^{(\ell)}=\big(s_1^{(\ell)},\ldots,s_{[\frac{r}{2}]}^{(\ell)}\big)$,
$\bar{\bm{s}}^{(\ell)}=\big(\bar{s}_1^{(\ell)},\ldots,\bar{s}_{[\frac{r}{2}]}^{(\ell)}\big)$ with 
  $s_a^{(\ell)}$ and  $\bar{s}_a^{(\ell)}$ being defined in eqs.\,\eqref{sajhg1278hgsd} and \eqref{sajhg1278hgsde}, respectively.
\medskip

The above procedure for calculating the asymptotic coefficients, say, 
 ${\mathfrak  C}^{(\ell)}_{p,{\boldsymbol s}}({\boldsymbol w})$
works under the assumption that $\Re e(p)\ge 0$.  Nevertheless, it
may be extended to any $2p\ne -1,-2,\ldots$
as follows. One notes that
$\Psi_p(z)=\re^{\frac{v}{2}}\,\tilde{\psi}_p$ with $z$ related to $v$ as in \eqref{sajhhg21}
solves the differential equation
\bea
\Big(-\partial_z^2+t_0(z)+t_1(z)
\Big)\, \Psi_p=0
\eea
and satisfies
\be
\Psi_p\big(\re^{2\pi\ri}z\big)=-\re^{2\pi\ri p}\,\Psi_p(z)\ .
\ee
Here the l.h.s. is understood as an analytical continuation along a small contour wrapping around $z=0$ in
the counter-clockwise direction. Such a solution $\Psi_p(z)$ may be defined for any $2p\ne -1,-2,\ldots\, .$
By  reverting to the original variables $v$, $\tilde{\psi}_p$ and  using \eqref{samnnb1321},
one extracts  the connection coefficient  $C^{(\ell)}_{p,{\boldsymbol s}}({\boldsymbol w})$.
In turn, ${\mathfrak  C}^{(\ell)}_{p,{\boldsymbol s}}({\boldsymbol w})$ 
is obtained via formula \eqref{sanb21bbv3}.

\end{document}